\documentclass{aa}  

\usepackage{graphicx}
\usepackage{txfonts}
\usepackage{xcolor}
\usepackage{lipsum}
\usepackage{hyperref}
\usepackage{placeins}           
\newcommand{\m}{$^{-1}$}
\newcommand{\mm}{$^{-2}$}
\newcommand{\mum}{$\mu$m~}
\newcommand{\Wmm}{W~m$^{-2}$~}
\newcommand{\mucdm}{$\mu$cd~m\mm~}

\newcommand{\ob}[1]{ {\bf#1}~}
\newcommand{\obr}[1]{\textcolor{red}{$\mathbf{#1}$}} 

\newcommand{\obl}[1]{{#1}~} 

\begin{document}

   \title{Large or bright satellite constellations}
   \subtitle{Effects on observations, including on the background sky brightness}


%

   \author{O. R. Hainaut\inst{1}\fnmsep\thanks{Corresponding author: ohainaut@eso.org}
        } 

   \institute{
   European Southern Observatory, Karl-Schwarzschild-Stra{\ss}e 2, 85748 Garching-bei-M\"unchen, Germany
   }

   \date{Received 2026 Mar. 30 - Reviewed 2026 May 05}

 
\abstract
    {
        Large satellite constellations impact astronomy through both visible trails and the generation of diffuse and atmospherically scattered light. Quantification of this cumulative sky-background component is essential to assess the full ramifications  of extant and proposed systems. 
    }
    {
        This study evaluates the effect of proposed constellations --- ranging from current deployments to mega-constellations and very bright reflector concepts --- on direct trail losses, diffuse background, and scattered sky brightness. 
    }
    {
        The methodology employs a numerical model for Mie and Rayleigh scattering in the $V$ band, adapted from moonlight sky-brightness calculations and validated against observations of moonlight and stellar background light. This is combined with the SatConAnalytic  package to quantify scattered light, diffuse light from undetected satellites, and direct losses from detected trails. 

    } 
    {
        Constellations comprising  $\sim60\,000$ satellites that adhere to the $V_\mathrm{550~km} \ge 7$ recommendation exert  a negligible effect on sky brightness, contributing only about $10^{-4}$ of the natural dark sky. Conversely,  mega-constellations with $10^6$ satellites render  trails pervasive, thereby affecting the majority of long exposures. Bright satellites, such as spacecraft analogous to AST SpaceMobile, significantly impact saturating detectors even when their number is moderate. Extremely bright satellites pose a far more severe threat: a 5000-satellite Reflect Orbital-like constellation could elevate the scattered sky background by 20\%--30\%, and a population of $50\,000$  could increase it by 200\%--300\%. 
    }
    {
        The constellations currently proposed for launch, which exceeds 1\,700\,000 objects and including satellites brighter than $V_\mathrm{550~km} = 7$, would substantially degrade astronomical observations. Maintaining satellite brightness below $V_\mathrm{550~km} = 7$ is important for all instruments, but critical for safeguarding saturating instruments, such as the Rubin LSST camera and for limiting sky-background pollution. Even under this constraint, the total satellite population should remain below  $\sim100\,000$ satellites to ensure that field-of-view losses do not exceed typical technical downtime.
    }


   \maketitle

\nolinenumbers

\section{Introduction}
The darkness of the sky background is one of the key parameters determining the quality of an observatory: for any background-dominated observation, the exposure time required to reach a given signal-to-noise ratio (S/N) scales linearly with sky brightness. This is also critical for $\gamma$- and cosmic-ray observations based on Cherenkov radiation from cascades of secondary particles produced when the primary particle interacts with atoms in the atmosphere, for which a stable, dark background is essential.

Because of this linear dependence of exposure time on sky brightness, critical observations in the visible can be carried out only from sites with the darkest possible skies, during moonless nights. Moonlight precludes such observations, and telescopes typically switch instead to infrared instruments, for which moonlight is not an issue.
Light pollution from cities and industries can raise sky brightness by orders of magnitude over distances of up to hundreds of kilometres \citep{Cinzano2001}. Professional observatories have therefore retreated to remote sites in search of \obl{the darkest skies. The best sites have a level of light pollution  below 1\% of the natural dark-sky brightness.
Light pollution at the 10\% level is incompatible with state-of-the-art observatories \citet{DQS1}.} The effects of artificial light at night (ALAN) and possible mitigation methods are well documented \citep[for a review, see][]{Barentine05}.

New satellite constellations are being announced, some extremely large and others extremely bright. In this work, we evaluate their contribution to scattered and diffuse light in the optical regime and discuss their direct effects on observations.

\subsection{Large satellite constellations}

\begin{figure}[ht]
    \centering
    \includegraphics[width=1\linewidth]{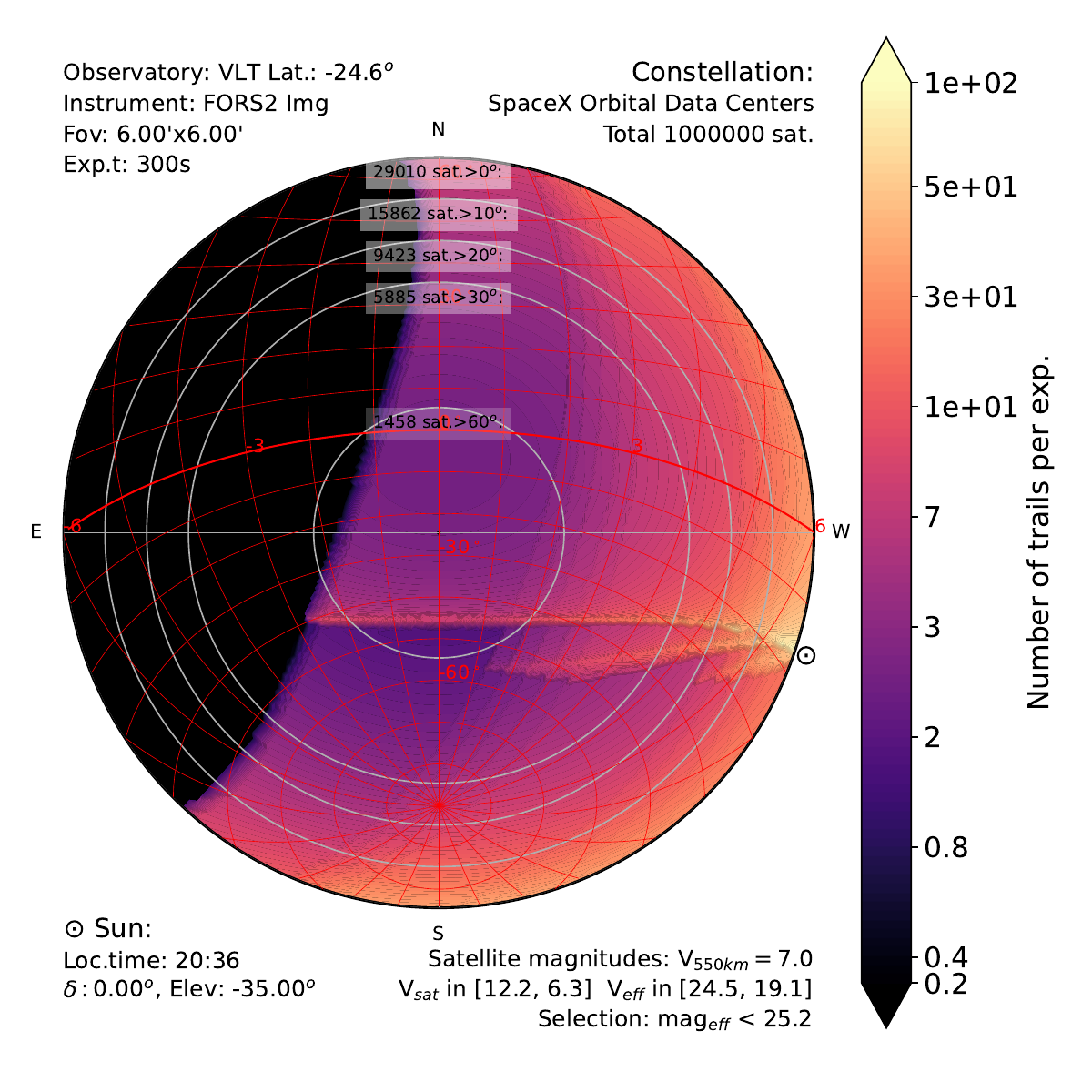}
    \caption{Example of the effect of SpaceX's one-million-satellite Orbital Data Center constellation on a 300-second FORS2 exposure at Paranal. Even 1.3~h after the end of astronomical twilight, more than half the sky is still affected, with an average of more than eight satellite trails per image at zenith. In regions corresponding to the constellation cusps, more than 20 trails cross each image. The effect on a 15-second exposure with the $3^\circ$ LSST camera at the Rubin observatory is similar.
    The figure is a map of the sky above the observatory (indicated at top left), with the zenith at the centre and the horizon along the outer circle.
    The dark region toward the east corresponds to the part of the sky where the satellites are already in the Earth's shadow and are therefore invisible. The $\odot$ symbol on the horizon indicates the Sun's azimuth; its position is given at bottom left. The sky maps in this paper correspond to solar elevations $e_\odot \le -21^\circ$, i.e. well after astronomical twilight.
    The red grid marks hour angle and declination. The grey circles correspond to elevations of 10$^\circ$, 20$^\circ$, 30$^\circ$, and 60$^\circ$ above the horizon (most observations are made above 30$^\circ$). The colour scale on the right gives, in this example, the number of satellite trails per exposure. The annotations at top right and bottom right give details of the satellite constellation and the characteristics of the individual satellites.
    }
    \label{fig:1Msat}
\end{figure}

Since the spectacular launch of the first batch of SpaceX's Starlink satellites in May 2019, the number of artificial satellites orbiting the Earth has increased from $\sim$2000 to over 14\,000, dominated by the deployment of the Starlink telecommunication constellation and a few other smaller constellations. That number rises further to over 32\,000 when including dead satellites and debris (from Jonathan McDowell's satellite statistics web page\footnote{\url{https://planet4589.org/space/stats/acdec.html}}, Feb.~2026). 

Early spectacular satellite ``photo-bombing'' of wide-field images\footnote{e.g. this example from DECam: \url{https://noirlab.edu/public/images/iotw1946a/}} triggered the astronomical community to react. This resulted in a series of publications investigating the number and period of visibility of the constellation satellites \citep[e.g.][]{mcd20}, as well as quantifying their impact on astronomical observations \citep[e.g.][]{hw20,BHG22}. In parallel, international conferences were organized --- SatCon-1 \citep{SatCon1} and -2 \citep{SatCon2}, Dark and Quiet Skies 1 \citep[D\&QS, ][]{DQS1} and -2 \citep{DQS2} --- where the optical and radio astronomy communities met with the satellite industry and relevant policy-makers. One of the key recommendations resulting from the SatCon conferences is that satellites must be fainter than magnitude $V=7$ in absolute terms and, for satellites with altitudes higher than 550~km, their magnitudes reduced to 550~km,  $V_\mathrm{550~km}$, must also be fainter than 7 \citep[Priority No. 1 recommendation][p.~1237]{DQS2}. This ensures that the satellites are not visible to the naked eye (matching its limiting magnitude in optimal conditions) and that they do not saturate the detector of the  NSF-DOE Vera C. Rubin Observatory (see Sect.~\ref{sect:instr} for more details on this).

Large satellite constellations also have a significant impact on radio astronomy. This is obvious if a satellite's directional antenna directly illuminates a radio telescope pointing at that satellite. There is also a cumulative impact through the side lobes of the satellites and the telescopes, and through unintended emissions by the satellites. The SatCon and D\&QS reports also discuss these matters. In what follows, we focus on the optical regime.  

Since the first launches, a series of mitigation methods have been implemented to reduce the brightness of the satellites. Some are operational (such as adjusting the satellite attitude or the solar panel angle), some are at the design level (e.g. covering the nadir-pointing face of the satellite with a specular coating that reflects sunlight away from Earth). SpaceX makes their findings available \citep{starlink}. Mitigation techniques are also being developed to minimize the number of satellite trails in observations \citep{BHG22}, and to detect trails \citep{Stoppa+2024,Chen2025} so they can be masked \citep{Hasan2022}. 

The numerous launches required to put a large number of satellites in orbit, their operation, and the eventual return of these satellites burning in the upper atmosphere also raise a series of concerns beyond the interference with optical and radio astronomy: sustainability of low-Earth orbit populated by many thousands of satellites, the associated generation of space debris, atmospheric pollution caused by the launches and re-entries, etc. \citep[see][for a summary]{law+22}. 

\cite{Kocifaj+21} investigated the effect of an increased number of space debris associated with a large satellite population on the diffuse sky background. Using the available debris size distribution and scaling their number to the satellite population, they concluded that these debris could contribute 16.2~\mucdm (24.6~magnitude per square arcsecond, or MpSA), and could even reach 21.1~\mucdm (24.3~MpSA) at the end of astronomical twilight, i.e. about 10\% of the luminance of the dark sky (the magnitude and illuminance, surface brightness and luminance units, and their conversions are discussed below and detailed in Appendix~\ref{Appendix}). While this study should be refined, it shows that space debris have the potential to increase the sky brightness to a level that would interfere with astronomical observations.

\cite{BHG22} considered the impact of $\sim$60\,000 satellites with $V_\mathrm{550~km} = 7$ and concluded that, in the optical domain, mitigation was possible but implied work at all stages, from the satellites and constellation design and operation, through observation scheduling and execution, to the data processing. \\

Since then, a series of developments has taken place:
\begin{itemize}
    \item The number of foreseen satellites has dramatically increased: McDowell currently lists 1\,766\,623 planned satellites\footnote{\url{https://www.planet4589.org/space/con/conlist.html} as of Feb.~2026}, dominated by SpaceX's Orbital Data Centers (one million satellites), E-Space's Cinnamon (over three hundred thousand satellites), and China's CTC-1 and -2 constellations (close to one hundred thousand satellites each), over a total of 33 large constellations. 
    
    \item Satellite operators are now launching very large satellites, whose size is dominated by a large antenna enabling direct broadband communication with standard smartphones. For instance, AST SpaceMobile launched their ``BlueWalker'' prototypes with a 64~m$^2$ antenna and $V=0$--4~mag \citep[$V=1.4$ on average,][]{mallama+22}, and they are now deploying their  ``BlueBird'' satellites with a 223~m$^2$ antenna. 
    \citet{MallamaCole2025} report their brightness around $V_\mathrm{550~km} = 3.12$, i.e. $\sim 36 \times$ brighter than the recommended limit. Their apparent magnitudes \citep{ColeMallama2025} are in the 0--7 range, i.e. up to $660\times$ brighter than the recommended limit. Other operators are also considering launching large satellites. For instance,  SpaceX's ``V3'', of which they plan to manufacture 5\,000 per year, is described as being ``each the size of roughly a 737 [Boeing plane]\footnote{\href{https://uk.pcmag.com/networking/158321/elon-musk-outlines-ambitious-plan-to-produce-10k-starlink-satellites-per-year}{PCmag.com}, retrieved Feb.~2026}''. 
    No magnitude measurement is available yet, but they could easily breach the $V>7$ IAU recommendation \citep{IAUmag7}.
    
    \item A new type of constellation aims at providing reflected sunlight to specific locations during the night, either for direct illumination or to extend the working hours of solar power stations. For that purpose, Reflect~Orbital plans\footnote{\url{https://www.reflectorbital.com/}} to launch over 5000 satellites by 2030, providing light for 2~h per night, then over  50\,000 satellites by 2035, providing illumination ``24/7''. They indicate that the large mirror-like satellites would illuminate a region 5~km across. The exact specifications of the satellites are not yet available, but it is suggested the satellites could be $54\times54$~m, or $\sim 3000$~m$^2$. While Reflect~Orbital indicates that they will avoid directly illuminating observatories with the specular reflected sunlight beam, it is to be expected that the surface of the reflector will not be perfect, and that part of the light will be diffused. 
    
\end{itemize}

The direct effect of these satellites can be estimated using existing methods, such as the \citet{BHG22} analytical model. As an example, Fig.~\ref{fig:1Msat} illustrates that, even well into the astronomical night, each and every observation will be affected by numerous satellite trails over most of the sky. Even if the satellites are fainter than $V_\mathrm{550~km}$, all of them will result in well-detected trails. These direct effects will be discussed below.

\subsection{Satellite scattered and diffused light as a source of pollution}

\begin{figure}[t]
    \centering
    {\bf a}\includegraphics[width=.98\linewidth]{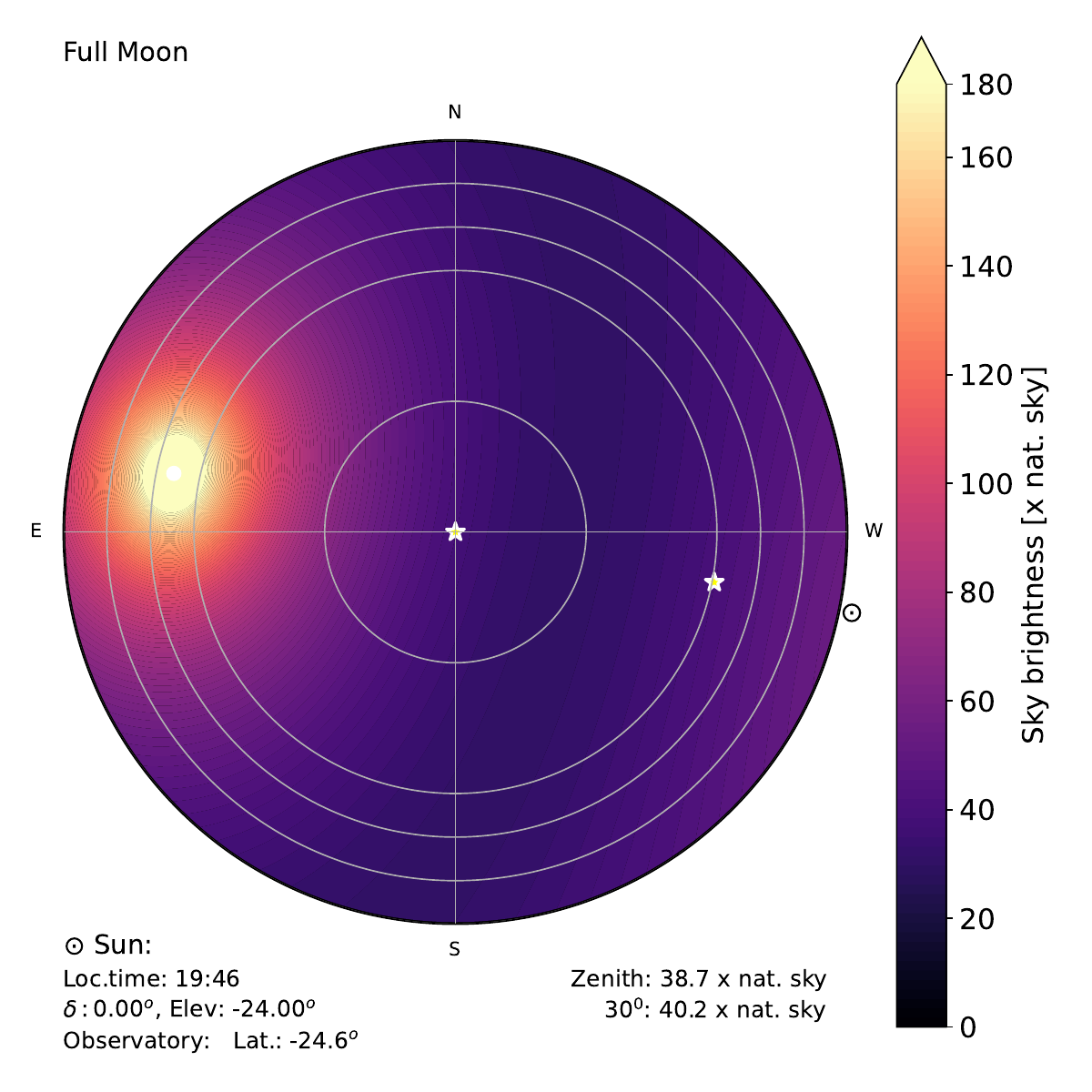}
    {\bf b}\includegraphics[width=.98\linewidth]{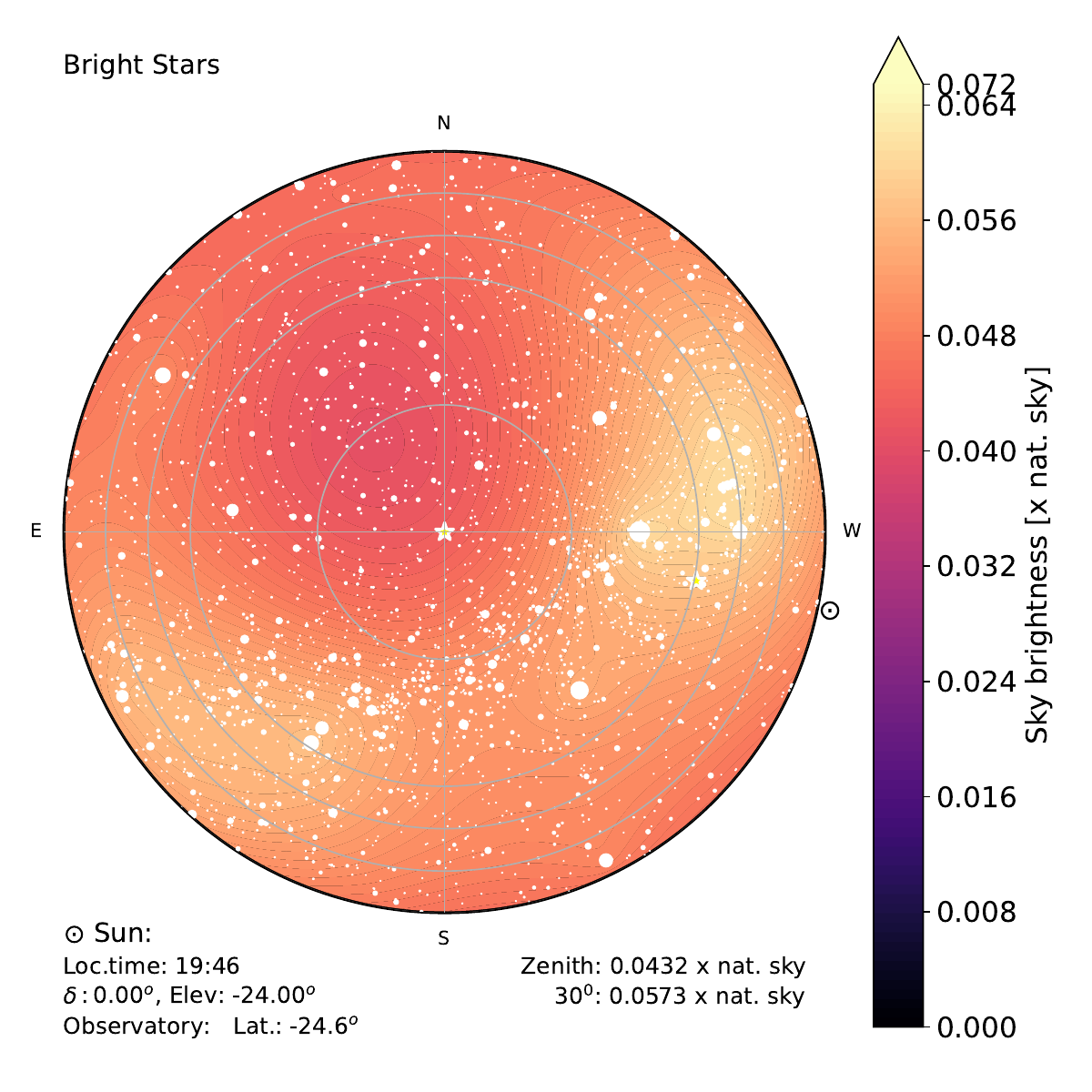}
    \caption{Example of the scattered-light model for the full Moon ({\bf a}) and for stars ({\bf b}, marked by white dots scaled by magnitude), used for validation over the whole sky (see Fig.~\ref{fig:1Msat} for a general explanation of these figures).
    The colour scale is in nanolamberts in panel~a and in fractions of the reference dark sky ($V=22$~MpSA) in panel~b.}
    \label{fig:fullMoon}
\end{figure}

A completely different impact, not considered until now, is the effect of light scattered by the atmosphere. 
In the optical domain, the main component of the dark sky luminance is the airglow, i.e. the emission by upper-atmosphere molecules, followed by the zodiacal light, i.e. sunlight scattered by interplanetary dust, the integrated starlight, and the extragalactic background light. In their reference paper on the night sky, \citet{Leinert+1998} detail these different sources in the optical domain, from ultraviolet to infrared.  
The relative contribution of these components varies by orders of magnitude. The airglow varies with atmospheric conditions; \citet{Barentine2022} provide a detailed discussion of these contributions and their variations. The zodiacal light depends on the ecliptic latitude and the difference in longitude with the Sun. The stellar background changes with galactic latitude and longitude. \cite{Masana2021} show the seasonal variations of these components in their Fig.~11. \cite{Patat2003} reports their variations as measured from Paranal, Chile. 
Based on these results, the contributions to an optimal dark sky (i.e. measured away from the Milky Way and the zodiacal light, and not including the individual contributions from stars) are: about 60\% for the airglow, 30\% for the zodiacal light, 4\% for the scattered starlight, the remainder corresponding to galactic and extragalactic diffuse light. Based on \cite{Leinert+1998} and \cite{Patat2003}, we use $V = 22 $~MpSA as the reference for the surface brightness of a pure, dark sky at zenith. This surface brightness is equivalent to a luminance of 174\mucdm; see Appendix \ref{Appendix} for details, and \citet{Bara2020} for a discussion. Any additional light source illuminating the atmosphere --- natural or artificial --- contributes to the scattered light background. 

Finally, satellites may be too faint to be detected as point sources or trails. In that case, they contribute to the sky background as a source of diffuse light. This is not relevant for images obtained with large telescopes, whose limiting magnitudes are much fainter than the effective magnitudes of all LEO satellites. For example, a combined 300-second exposure in $V$ with FORS2, under median conditions, will result in a $5\sigma$ detection of an object with $V=25.7$, while the faintest LEO satellite in our simulation (see Sect.~\ref{Sect:SLOWGWAK}) has $V_\mathrm{eff}=11.6$: all satellites are detected as trails. It will, however, be relevant for some wide-field (e.g. all-sky) cameras or even for naked-eye observations. It is also critically relevant for observations with brighter limiting magnitudes, such as most spectroscopic observations \citep[see][for a discussion of the spectroscopic case]{HM24}.

We introduce a simple model of atmospheric scattering and validate it. We then use it to quantify the effect of existing and upcoming satellite constellations on the sky background via scattered and diffused light pollution, and discuss the consequences. 

\section{Simulation methods}

The evaluation of the scattered and diffuse light pollution, as well as the satellite densities, trail densities, and losses is computed using our Python implementation of the algorithm described in \citet{BHG22}, expanded for this paper. The Python code and the accompanying configuration files are available on GitHub.\footnote{SatConAnalytic package, \url{https://github.com/ohainaut/SatConAnalytic}}

\subsection{Observatories and instruments}\label{sect:instr}

The simulations are performed for the FOcal Reducer/low dispersion Spectrograph~2 \citep[FORS2;][]{FORS94} in imaging mode. Typical combined exposures have a field-of-view (FoV) of $6\times6'$ for an exposure time of 300~s, with a limiting magnitude $V \sim 25.8$ ($5\sigma$). FORS2 is mounted on one of the 8.2-meter Unit Telescopes of ESO's Very Large Telescope, on Paranal, Chile (latitude $24.62^\circ$ S). We use FORS2 as representative of traditional imaging cameras on large telescopes.

We also performed simulations for the  NSF-DOE Vera C. Rubin Observatory  Legacy Survey of Space and Time (LSST) camera. This camera combines a huge FoV of $3.5^\circ$ diameter covered by 189 individual CCDs assembled in 21 rafts of $3\times3$ detectors. The camera is fed by the large, 8.4-meter diameter mirror of the telescope. Rubin observatory is located on Cerro Pach\'on, Chile (latitude $30.24^\circ$ S). A typical 15-second exposure has a limiting magnitude $V \sim 24.9$ ($5\sigma$).
\obl{The observatory is finalizing the commissioning of the system; it is likely that some of the operational parameter will change when going into full operations. For instance, the exposure time is increased from 15 to 30~s, which would doubles the counts from the following simulations.}
This combination makes the LSST a unique instrument in terms of sensitivity and FoV. These characteristics also make the LSST particularly sensitive to satellite trails. 
\obl{Furthermore, because of the high-density electronics required to control the CCDs, a satellite trail bright enough to saturate the detector will not only cause a broad trail, but will also cause a series of parallel ghost trails resulting from cross-talk between the CCD controllers. These ghost trails appear in the affected CCD and in the other CCDs from the same raft, resulting in a dramatically multiplied loss of FoV. These effects are discussed in detail by \citet{Tyson2020}.} The potential of a satellite to trigger this cross-talk effect depends on its visual magnitude, but also on its apparent angular velocity (a slower, higher satellite will remain longer on a given pixel), and on the apparent angular size of the satellite (a higher satellite will appear smaller, and its light will be concentrated on fewer pixels). Following \citet{Tyson2020}, these parameters can be simplified to a threshold $V_\mathrm{550~km} \le 7$, which converts to an effective magnitude (accounting for trailing) $V_\mathrm{eff} \le 18.3$ \citep[see][for a discussion of the effective magnitude]{BHG22}. This effect is one of the origins of the IAU recommendation $V_\mathrm{550~km} > 7$; the other one, $V>7$, ensures that the satellites are not visible to the naked eye. The fact that the numerical values are equal is a welcome coincidence. We use the LSST camera as representative of the imaging cameras suffering from this type of effects (which we will label as {\em dramatic saturation}) -- LSST, because of its large field of view on a large telescope, is probably the instrument most sensitive to bright satellite trails. 

Spectrographs, even mounted on very large telescopes, have much brighter limiting magnitudes. It is therefore common that a spectrograph does not register a satellite trail \citep[see][for a detailed discussion of the effects on spectrographs]{HM24}. In that case, the trail contributes to increasing the background; this is the "diffuse light pollution".

\subsection{Observation geometry}
\obl{For a satellite to affect the observations, it must be above the horizon and illuminated by the Sun. These conditions depend on the latitude of the observatory, the position of the Sun, and the position of the satellite. In the following simulations, we use the position of the FORS2 and LSST instruments. We either scan the full night, or focus on the time when the Sun is at an elevation around $-24^\circ$; this corresponds to $\sim$1h into the astronomical night, and is therefore representative of near-twilight observations. Using the \citet{BHG22} method, we either scan the whole sky, or focus on the zenith (the most favourable observing direction) and on $30^\circ$ elevation (corresponding to the typical limit for most observations) towards the direction of the Sun (where the number of illuminated satellites will be the highest), thereby bracketting most conditions.}

\subsection{Direct contamination}

We first evaluate the direct effect of the satellites on the observations, as described in detail by \citet{BHG22}. A rigorous quantification of the losses caused by satellite trails depends on the science case for which the data are acquired. For instance, in the case of a survey aiming at counting distant galaxies, the critical effect of satellite trails is to reduce the useful area of the field-of-view (FoV). In the case of targeted observations, however, a satellite trail missing the observed object will not be noticed directly, but may affect the quality of the sky subtraction. To generalize the concept of data losses, we use the simple metric of the fraction of the FoV compromised by the trails. For this purpose, each trail is considered to cross the entire FoV, and its width is set to 5$''$, accounting for the intrinsic size of the satellite, the width of the point-spread-function (PSF) and its wings including the defocusing effects \obl{\citep[see][for details]{Tyson2020}}. To account for the wings of very bright satellite, that value is doubled for a satellite 100$\times$ brighter. We define the loss metric as the total fraction of FoV caused by all the trails crossing an exposure. This neglects the superposition of trails, and can possibly result in a loss fraction larger than 100\%, indicating that each pixel is on average affected by more than one trail.
For the case of the LSST camera, we also account for the saturation effect described above.

\subsection{Scattered light}\label{Sect:scatteredLight}

We aim to evaluate the contribution of scattered light from sources illuminating the atmosphere, and to compare it to the optimal reference dark sky with $V=22$~MpSA or $L=174$~\mucdm. We use the method described by \cite{KS91} (hereafter KS91) to evaluate the contribution of a satellite to the sky brightness. They developed this method to compute the contribution of the Moon to the sky brightness and validated it on ``a dozen of observations" of the moonlit sky over Mauna~Kea. It also reproduces well extensive measurements performed on Paranal \citep{Patat2003}. More detailed models of the Paranal night sky have been developed \citep[e.g.]{Noll2012}; similarly, a much more advanced scattered-moonlight model is now available \citep{Jones2013}. However, the accuracy of the original KS91 method is sufficient for our purposes.

Following KS91 (their Eq.~15), the surface brightness $V$ caused by the Moon is
\begin{equation}
    V = f(\rho) ~ I^* ~ 10^{-0.4~ k ~X(Z_{\mathrm{Source}}) }
     \left( 1 - 10^{-0.4 ~k ~ X(Z)} \right)                   \label{Eq:sbMoon}
\end{equation}
where $f(\rho)$, the scattering function, is the intensity of the scattered light as a function of the scattering angle $\rho$ between the line of sight and the source,
$X(Z)$ is the length of the optical path in airmasses (i.e. 1 at zenith) as a function of the zenith distance $Z$. $Z_{\mathrm{Source}}$ is the airmass of the source, i.e. the Moon, and $Z$ the airmass of the direction of observation.
$k$ is the atmospheric extinction, in magnitudes per airmass. We use $k=0.125$, corresponding to a typical value in the $V$ band (see, for instance, the FORS2 data quality control\footnote{\url{https://www.eso.org/observing/dfo/quality/FORS2/qc/photcoeff.html}}).
$I^*$ is the illuminance of the source outside the atmosphere, so that the term  
$I^* ~ 10^{-0.4~ k ~X(Z_{\mathrm{Source}})} $  is the illuminance after atmospheric extinction.

For the airmass function, we use their Eq.~3:
\begin{equation}
    X(Z) = \sqrt{1- 0.96 \sin^2 Z},   \label{Eq:airmass}
\end{equation}
which they found to represent the effect of scattering better than either the simplified $X = \sec Z$ (valid only far from the horizon) or more complex relations used for photometric measurements very close to the horizon. 

The scattering function $f(\rho)$ is the sum of the Rayleigh component $f_\mathrm{R}$ (caused by air molecules) and the Mie component $f_\mathrm{M}$ (accounting for atmospheric aerosols). For the Rayleigh scattering, they use
\begin{equation}
    f\mathrm{R}(\rho) =  10^{5.36} \left( 1.06 + \cos^2 \rho \right),   \label{Eq:Rayleigh}
\end{equation}
and for the Mie scattering,
\begin{equation}
    f\mathrm{M}(\rho) = 10^{6.15 - \rho/40},        \label{Eq:Mie}
\end{equation}
where $\rho$ is in degrees, and the numerical factors were obtained empirically from their measurements, using foot-candles (fc) as the unit for the Moon illuminance and nano lamberts (nL) for the luminance. KS96 remark that $f_\mathrm{M}$ is underestimated for $\rho < 10^\circ$. As we are concerned with effects at large $\rho$, this is not an issue.

Combining Equations \ref{Eq:sbMoon}, \ref{Eq:airmass}, \ref{Eq:Rayleigh}, and \ref{Eq:Mie}, we obtain the luminance (in nL) of the scattered light caused by a source with illuminance $I$ (in fc). The transformations described in Appendix~\ref{Appendix} convert between these non-standard units and standard photometric units.

To validate our implementation, we reproduced the surface brightness of the scattered moonlight, using $V=-12.74$ for the magnitude of the full Moon and setting the coordinates of the Moon directly opposite to the Sun. An example is displayed in Fig.~\ref{fig:fullMoon}.a. Both the spatial structure and the surface brightness of the moonlight are well reproduced, with $V\sim 18$~MpSA as expected.

To validate our model in the case of multiple sources, we computed the scattered light from stars. We used the stars from the Yale Bright Star Catalogue \citep[HR;][]{Hoffleit1991} as sources and, for each star, computed its contribution to the scattered background. As the HR stars are limited to $V \le 8$, this model misses the contribution from the numerous fainter stars. To correct for this, we computed a scaling factor using the magnitude distribution from Gaia DR3. Accounting for all stars down to $V \le 21$, the flux from the HR stars must be multiplied by 2.2. An example is displayed in Fig.~\ref{fig:fullMoon}.b; far from the Milky Way, the scattered light represents 4\% of the $V=22$~MpSA dark sky, increasing to 6\% in the Orion region (right part of the plot).

Equipped with the scattered-light model described above, we can now evaluate the contribution of satellites to the background sky. Each constellation is defined by a collection of shells, each described by an altitude, an orbital inclination, a number of planes, and a total number of satellites. For each constellation, the satellites are characterised by their standard magnitude at zenith at 550~km, $V_\mathrm{550~km}$. The satellites are distributed among orbital planes and, within each plane, along the orbit. At any time, the positions of the satellites around the Earth are computed using simple Keplerian motion. The position of the Sun is computed for the time of observation, and the satellites in the shadow of the Earth are discarded. Similarly, the satellites below the observatory horizon are discarded. The coordinates of the remaining satellites are converted into topocentric azimuth and elevation. This simple mechanism aims to produce a representative distribution of the satellites in the sky, {\em not} to be ephemeris-accurate. Following \cite{BHG22}, the magnitude of an illuminated satellite is simply scaled for its distance to the observer, without solar phase correction. For the Starlink satellites, this reproduces the observed magnitudes within $\sim 1$~mag, which is sufficient for our purposes \citep{BHG22}.

One issue with this method is that it relies on a single realisation of the satellite positions, which leads to local variations near individual satellites. A straightforward way to average out this effect would be to repeat the computation for several realisations and average the results. Instead, we work with the apparent density of satellites (in satellites per square degree) computed analytically using the \citet{BHG22} method for each surface element of the sky. As demonstrated in that paper, this density is a rigorous average of the number of satellites in a given sky element over all possible realisations of their positions. The total flux reflected by the satellites is integrated over each element, yielding a total magnitude that is then used as input to the scattering model. An additional advantage of this approach is that the processing time is independent of the number of satellites (which is particularly beneficial when treating the mega-constellation of Sect.~\ref{Sect:SXODC}).

\subsection{Diffuse light}\label{Sect:diffuseLight}

A satellite can be too faint to be recorded as a trail, either because it is intrinsically too faint or because its apparent angular motion is too fast; in other words, because its effective magnitude \citep[see][for the detailed definition]{BHG22} is fainter than the limiting magnitude of the instrument. In that case, its light still contributes to the diffuse sky background. This effect is critical for very wide-angle systems, such as all-sky cameras or naked-eye observations, but also for most spectroscopic observations where the spectral dispersion of the light results in relatively bright limiting magnitudes \citep{HM24}. 

The contribution of a constellation to the diffuse sky background is computed by integrating the flux from the satellites in a sky surface element. In turn, this is obtained by multiplying the satellite number density in that part of the sky by the angular area considered and by the flux corresponding to one satellite. This calculation is performed for each constellation shell to account for the different magnitudes of the satellites in different shells, and the contributions of all shells are summed, resulting in the luminance of the sky surface element.

\subsection{Simulated quantities and qualitative evaluation}

\begin{table*}[tbp]
\begin{tabular}{lrr|llllllc}
\hline
Constellation & N$_\mathrm{sat}$ & Mag
    & \multicolumn{2}{c}{Satellite Density}
    & \multicolumn{2}{c}{Diffuse light pollution}
    & \multicolumn{2}{c}{Scattered light pollution} 
    & Fig.\\
 & & $V_\mathrm{550~km}$& \multicolumn{2}{c}{[sat/deg$^2$]}
    & \multicolumn{2}{c}{[fraction of dark sky]}
    & \multicolumn{2}{c}{[fraction of dark sky]} 
    &\\
 & & & Zenith & $30^\circ$ & Zenith & $30^\circ$ & Zenith & $30^\circ$ &\\
\hline
\hline
\multicolumn{5}{l}{\bf  Base scenario: sixty thousand satellites} \\
BHG2022 & 64\,526 & 7 & $9.0 \times 10^{-3}$ & $0.053$ & $3.4 \times 10^{-4}$ & $1.6 \times 10^{-3}$ & $1.1 \times 10^{-4}$ & $1.8 \times 10^{-4}$ & \ref{fig:SLOWGWAK}.acd\\
\hline
\multicolumn{5}{l}{\bf Mega-constellations} \\
SXODC-7 & 1M & 7 &$0.58$&$1.6$ & $5.8 \times 10^{-3}$ & \ob{0.015} & $1.1 \times 10^{-3}$ & $1.5 \times 10^{-3}$  & \ref{fig:ap.SXODC}.acd\\
SXODC-6 & " & \ob{6}  & " & " & \ob{0.015} & \ob{0.038} & $2.7 \times 10^{-3}$ & $3.8 \times 10^{-3}$  \\
SXODC-5 & " & \ob{5}  & " & " & \ob{0.037} & \ob{0.095} & $6.8 \times 10^{-3}$ & $9.4 \times 10^{-3}$ \\
\hline
\multicolumn{5}{l}{\bf Very bright satellites: Direct-to-cell} \\
AST   & 243   & \obr{2} & $4.6 \times 10^{-5}$ & $2.6 \times 10^{-4}$ & $2.0 \times 10^{-4}$ & $3.3 \times 10^{-4}$ & $3.2 \times 10^{-5}$ & $4.7 \times 10^{-5}$ & \ref{fig:ap.AST-243}.acd\\
D3000 & 3000  &  \obr{"}  & $4.2 \times 10^{-4}$ & $2.4 \times 10^{-3}$ & $2.4 \times 10^{-3}$ & $3.9 \times 10^{-3}$ & $4.0 \times 10^{-4}$ & $5.6 \times 10^{-4}$ & \ref{fig:ap.AST-3000}.acd\\
\hline
\multicolumn{5}{l}{\bf Extremely bright satellites: Sunlight as a service}   &  \\
RO-2027& 36 & \obr{-4.3} & $5.3 \times 10^{-6}$ & $3.2 \times 10^{-5}$ & $8.6 \times 10^{-3}$ & \ob{0.014} & $1.9 \times 10^{-3}$ & $2.7 \times 10^{-3}$ & 
                                   \ref{fig:SpaceOrbitalScatter}.a \\
RO-2030           & 5000 & \obr{"}  & $7.8 \times 10^{-4}$ & $4.6 \times 10^{-3}$ & n/a &  & \obr{0.27} & \obr{0.37} & 
                                   \ref{fig:SpaceOrbitalScatter}.b\\
RO-2035        & 50\,000 & \obr{"} & $7.8 \times 10^{-3}$ & $0.046$ & n/a &  & \obr{2.7} & \obr{3.7} & 
                                   \ref{fig:SpaceOrbitalScatter}.c\\
\hline
\end{tabular}

~\\

\begin{tabular}{lrr|lllllc}
\hline
Constellation & N$_\mathrm{sat}$ & Mag & Inst.
    & \multicolumn{2}{c}{Trails}
    & \multicolumn{2}{c}{FoV loss} 
    & Fig.\\
 & & $V_\mathrm{550~km}$&  & \multicolumn{2}{c}{per frame}
    & \multicolumn{2}{c}{[fraction of FoV]} &\\
 & & & & Zenith & $30^\circ$ & Zenith & $30^\circ$& \\
\hline
\hline
\multicolumn{5}{l}{\bf  Base scenario: sixty thousand satellites} \\
BHG2022  & 64\,526 &  7     & FORS & $0.145$ & $0.804$                              
                                & $1.36 \times 10^{-3}$ & $5.98 \times 10^{-3}$  
                                    & \ref{fig:SLOWGWAK}.b\\
\hline              
\multicolumn{5}{l}{\bf Mega-constellations} \\
SXODC-7 & 1M & 7              & FORS & $4.27$ & $11.5$                                  & 
                                \ob{0.0593} & \ob{0.160 } &
                                    \ref{fig:ap.SXODC}.b\\
   "    & " & "                      & LSST & $11.6$ & $31.3$                                  & 
                                 $0.0108$ & $0.0290$ & 
                                    \ref{fig:1Msat}\\
SXODC-6 & " & \ob 6                           &   "   & " & "                                           
                                & \obr{ 0.390} & \obr{1.62} &
                                    \ref{fig:SXODC_LSST}.a \\
SXODC-5 & " & \ob 5                             &   "   & " & "                                         
                                & \obr{ 11.6} & \obr{19.5}  \\
\hline
\multicolumn{5}{l}{\bf Very bright satellites: Direct-to-cell} \\
AST & 243 & \obr{2}        & FORS & $7.88 \times 10^{-4}$ & $2.54 \times 10^{-3}$    
                                & $1.09 \times 10^{-5}$ & $3.52 \times 10^{-5}$ & 
                                    \ref{fig:ap.AST-243}.b\\
&&&&&\\
D3000 & 3000 &  \obr{2}    & FORS & $8.19 \times 10^{-3}$ & $0.0265$                 
                                & $1.14 \times 10^{-4}$ & $3.68 \times 10^{-4}$ & 
                                    \ref{fig:ap.AST-3000}.b\\
 " &  "    &     \obr{"}              & LSST & $0.0160$ & $0.0614$                              
                                & $0.0160$   & \ob{0.0614} & 
                                    \ref{fig:AST-faint-vs-bright_elev},
                                    \ref{fig:ap.AST-faint-vs-bright}.d
                                     \\
\hline
\multicolumn{5}{l}{\bf Extremely bright satellites: Sunlight as a service}   &  \\
RO-2027 & 36 & \obr{-4.3}   & FORS & $1.10 \times 10^{-4}$ & $3.65 \times 10^{-4}$    
                                & $1.53 \times 10^{-6}$ & $5.06 \times 10^{-6}$ \\
                        & & & LSST & $2.12 \times 10^{-4}$ & $8.30 \times 10^{-4}$    
                                & $2.12 \times 10^{-4}$ & $8.30 \times 10^{-4}$  & 
                                    \ref{fig:SpaceOrbitalLSST}.a\\
&&&&&\\
RO-2030& 5000 & \obr{"} & FORS & $0.0157$ & $0.0518$                              
                                & $2.18 \times 10^{-4}$ & $7.20 \times 10^{-4}$ \\
                            & & & LSST & $0.0306$ & $0.119$                               
                                & $0.0306$ & \ob{0.119} & 
                                \ref{fig:SpaceOrbitalLSST}.b,  \ref{fig:SpaceOrbitalScatterElevation}.a\\
&&&&&\\
RO-2035 & 50\,000 & \obr{"} & FORS & $0.157$  & $0.518$                               
                                & $2.18 \times 10^{-3}$ & $7.20 \times 10^{-3}$ \\
 & &                        & LSST & $0.306 $ & $1.19$                                
                                & \obr{0.306} & \obr{1.19} 
                                    & \ref{fig:SpaceOrbitalLSST}.c,
                                    \ref{fig:SpaceOrbitalScatterElevation}.b\\
\hline
\end{tabular}
\caption{Results of the simulations. All simulations are performed at solar elevation $e_\odot = -21^\circ$, i.e. well into the astronomical night (e.g. $\sim 1$h40 after sunset for the VLT).
{\bf Constellations}: BHG2022: realistic upcoming set, from \citet{BHG22}, 
            see Sect.~\ref{Sect:SLOWGWAK};
        SXODC: SpaceX Orbital Data Centre, see Sect.~\ref{Sect:SXODC};
        AST: AST Space-Mobile, see Sect.~\ref{sect:AST}; AST-3000 is an extrapolation of their proposed constellation;
        RO: Reflect Orbital, with the year of foreseen deployment, see Sect.~\ref{sect:RO}.
{\bf N}$_\mathrm{sat}$ is the total number of satellites in the constellation.
{\bf Mag} is the magnitude of a satellite, reduced to a distance of 550~km.
In the upper table:
{\bf Satellite density} is the average number of satellites per square degree.
{\bf Diffuse light pollution} is the contribution of the satellites to the sky background for instruments with a limiting magnitude brighter than the effective magnitude of the satellite (e.g. spectrographs).
{\bf Scattered light pollution} is the contribution of all the light from illuminated satellites, scattered (Mie and Rayleigh) by the atmosphere. 
Both types of pollution are given as a fraction of the dark sky ($V=22$~MpSA, 174\mucdm).
The lower table lists instrument-specific effects.
{\bf Inst.} specifies the instrument: FORS is a $6\times6'$ imager taking 300~s exposures, and LSST is a 3.9$^\circ$ camera taking 15~s exposures; see Sect.~\ref{sect:instr} for more details.
{\bf Trail per frame} gives the average number of detected satellite trails in an individual exposure.
{\bf FoV loss} refers to the average fraction of the field of view contaminated by the satellite trails.
{\bf Fig.} points to figures illustrating the values.
The quantities are evaluated at zenith, and at $30^\circ$ elevation in the direction of the solar azimuth; these positions are marked by a star symbol on the sky maps. Values in boldface are above the acceptable threshold (1\% for light pollution, 3\% for losses); values in red are above the disastrous threshold (10\% for light pollution, 30\% for losses). 
}
\label{table:results}
\end{table*}

\begin{figure*}
    \centering
    {\bf a}\includegraphics[width=.485\linewidth]{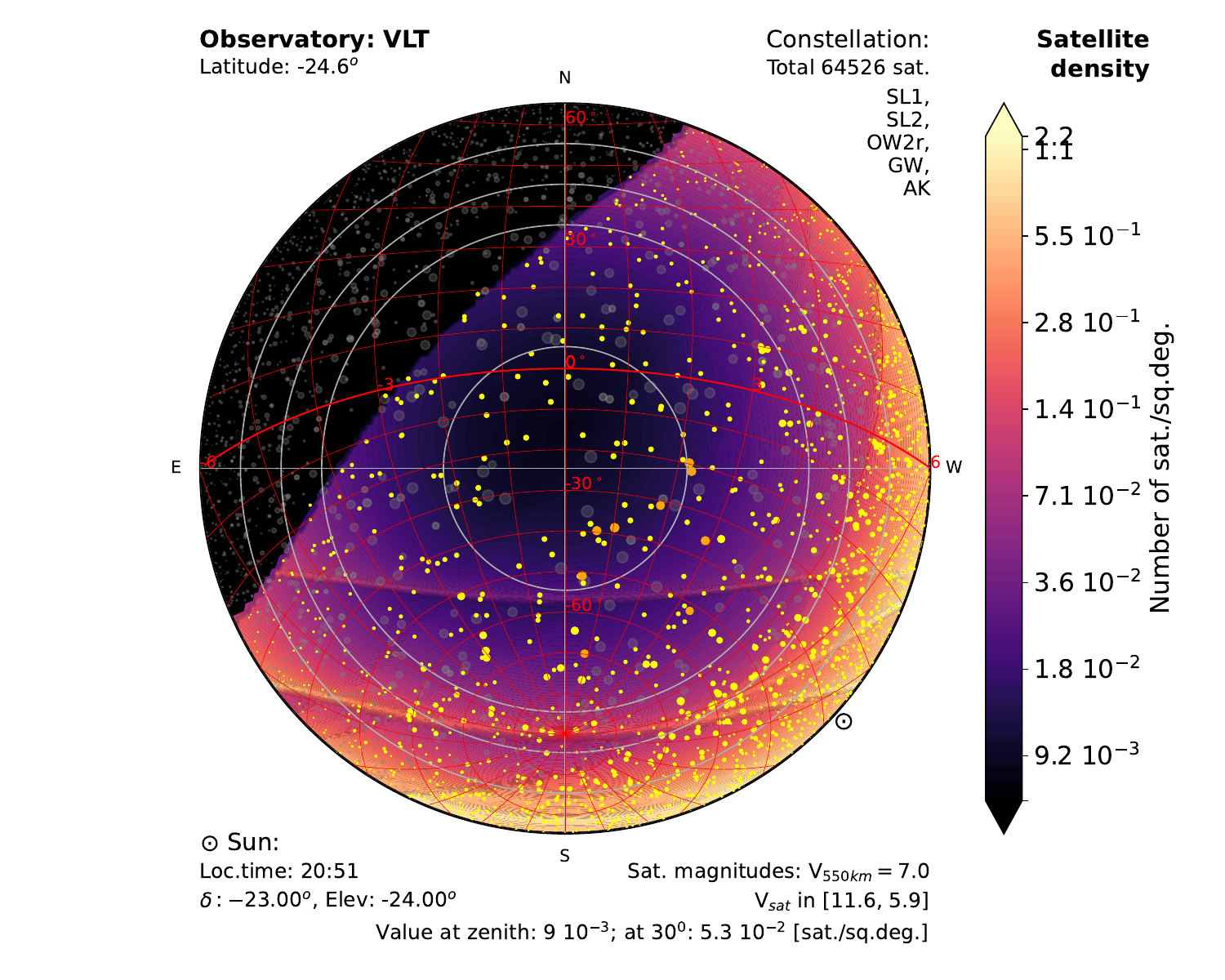}
    {\bf b}\includegraphics[width=.485\linewidth]{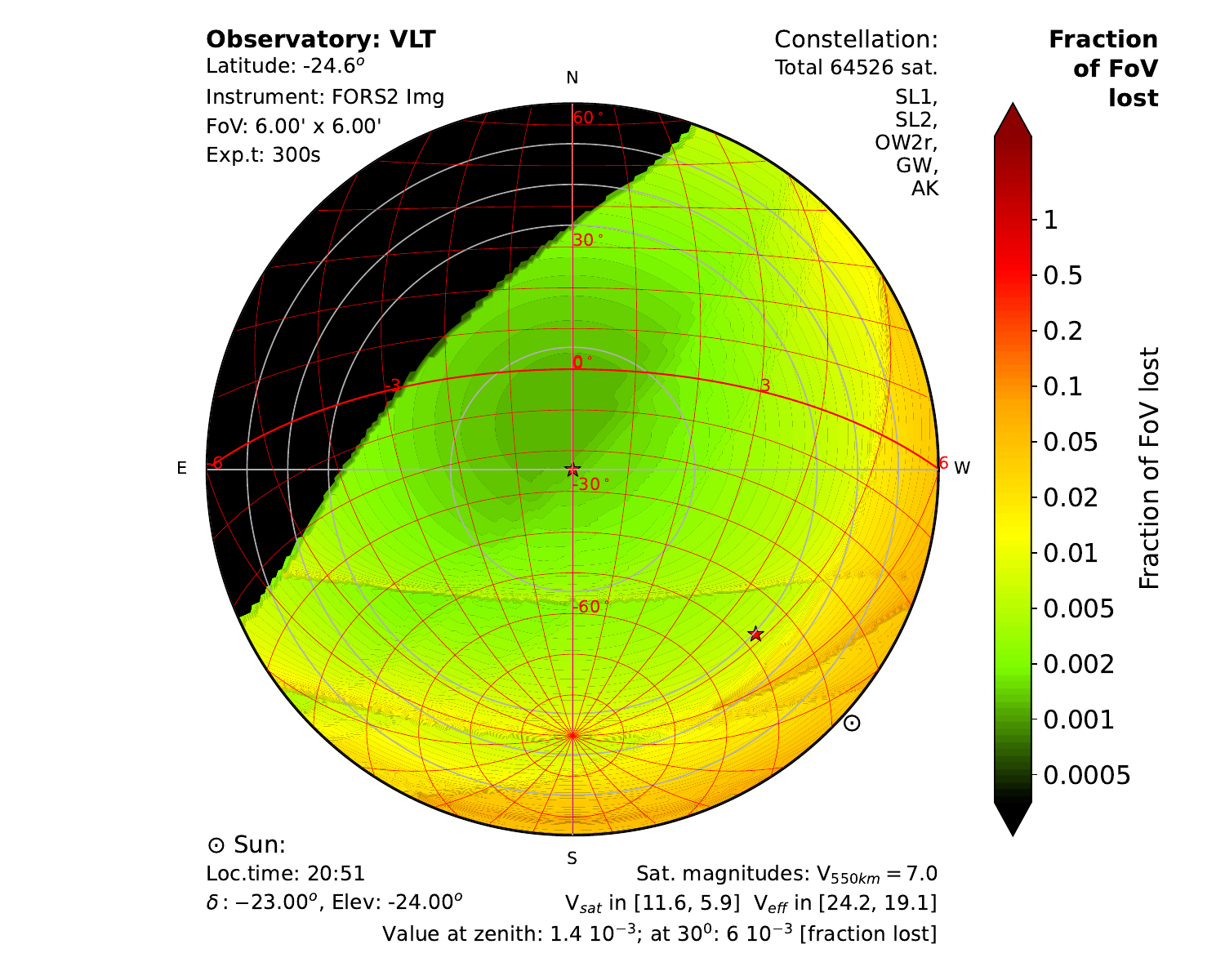}
    {\bf c}\includegraphics[width=.485\linewidth]{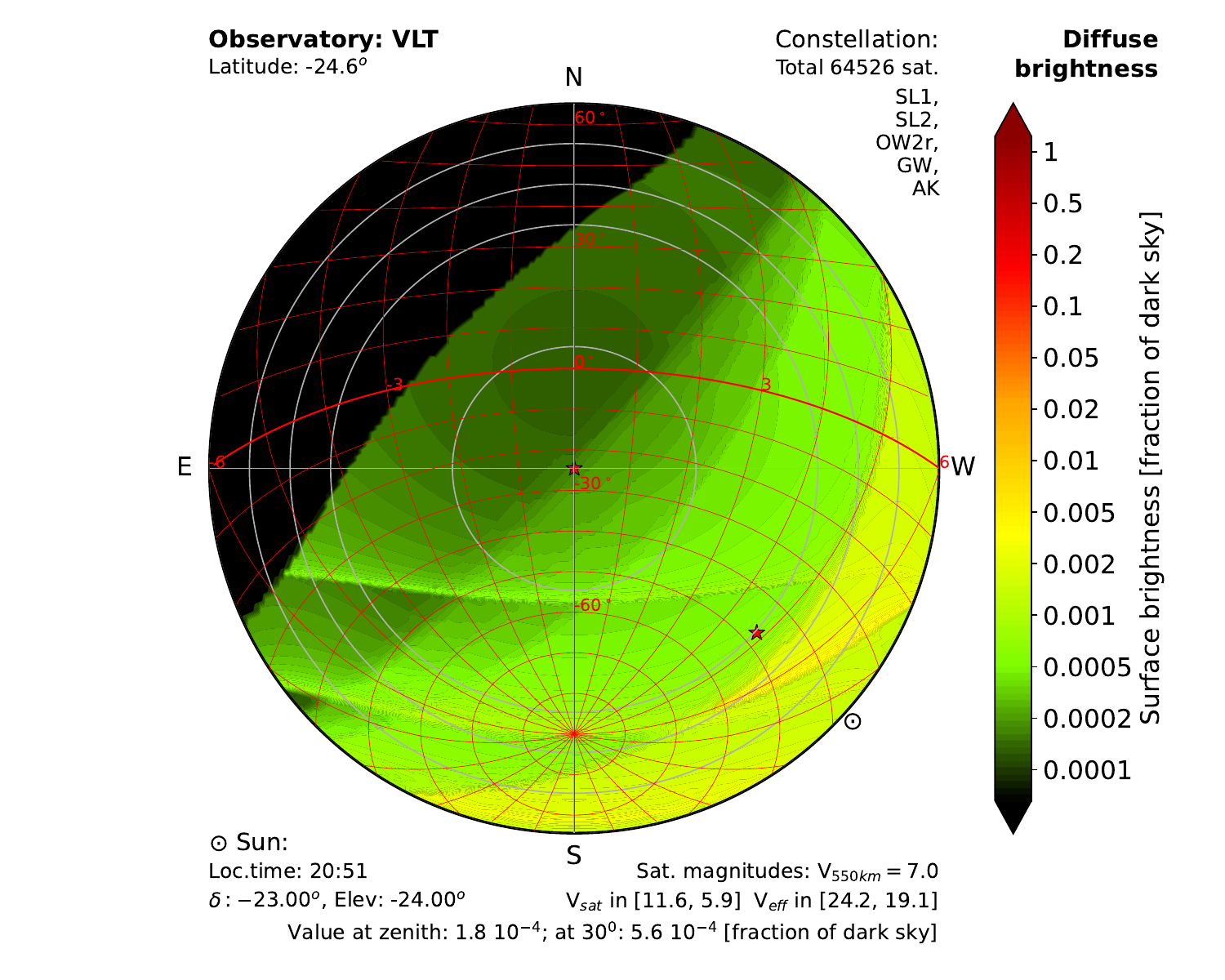}
    {\bf d}\includegraphics[width=.485\linewidth]{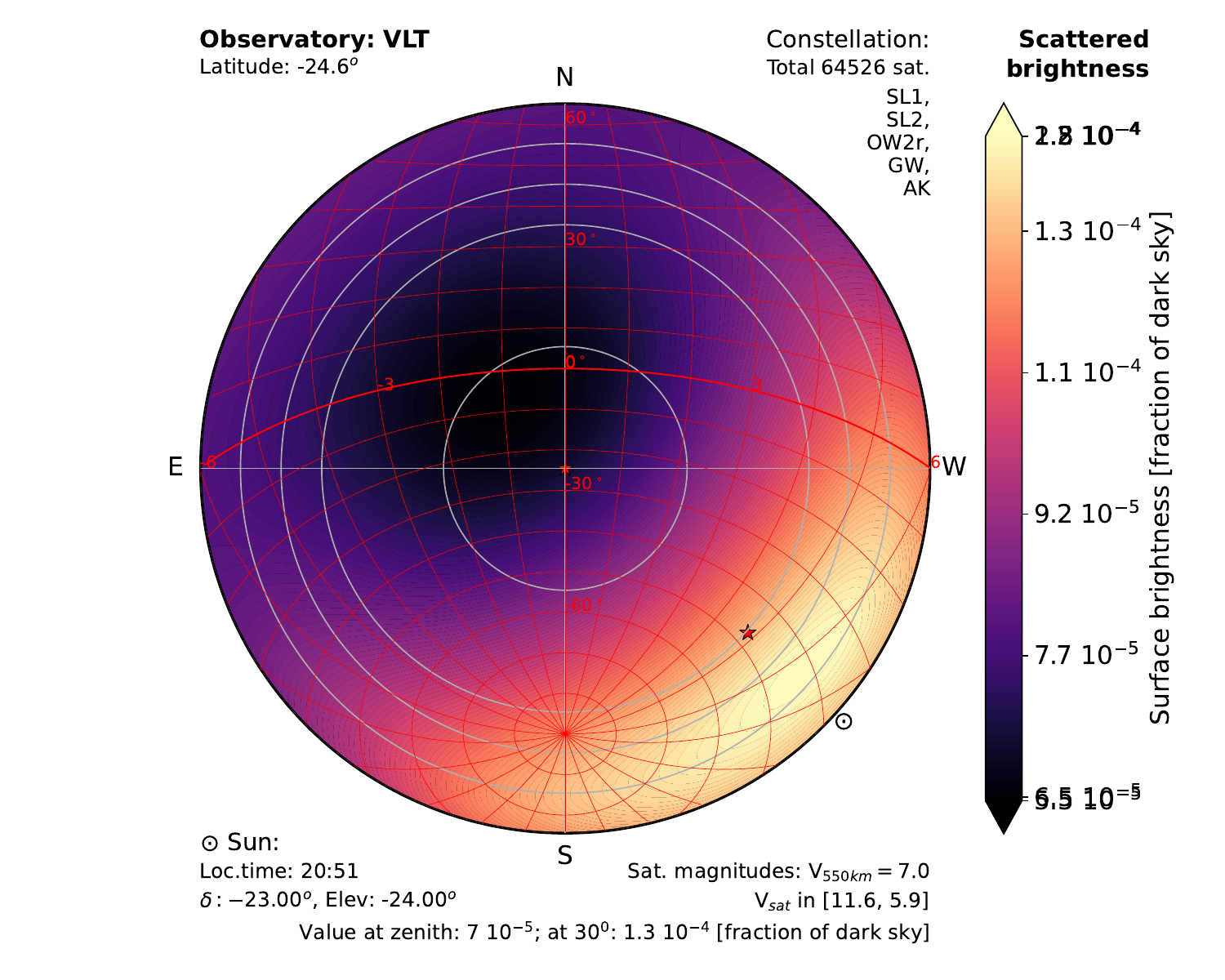}
    \caption{For constellations with 64\,526 satellites, representative of the near future (BHG2022 configuration):
    {\bf a} average satellite density [sat/sq.~deg.]; the dots show an example of the satellite positions (grey for satellites in the Earth's shadow, yellow for illuminated satellites, and orange for satellites with $V<7$);
    {\bf b} fraction of the FoV lost to satellite trails in 300-second ESO VLT FORS2 exposures;
    {\bf c} diffuse sky brightness, that is, the direct contribution of all satellites to the background of instruments unable to detect the individual satellites, as a fraction of the natural dark sky with $V=22$~MpSA;
    {\bf d} scattered sky brightness, including Mie and Rayleigh scattering from all illuminated satellites, as a fraction of the natural dark sky.
    A general description of these sky maps is given in Fig.~\ref{fig:1Msat}.
    }
    \label{fig:SLOWGWAK}
\end{figure*}
\begin{figure*}
    \centering
    \includegraphics[width=.98\linewidth]{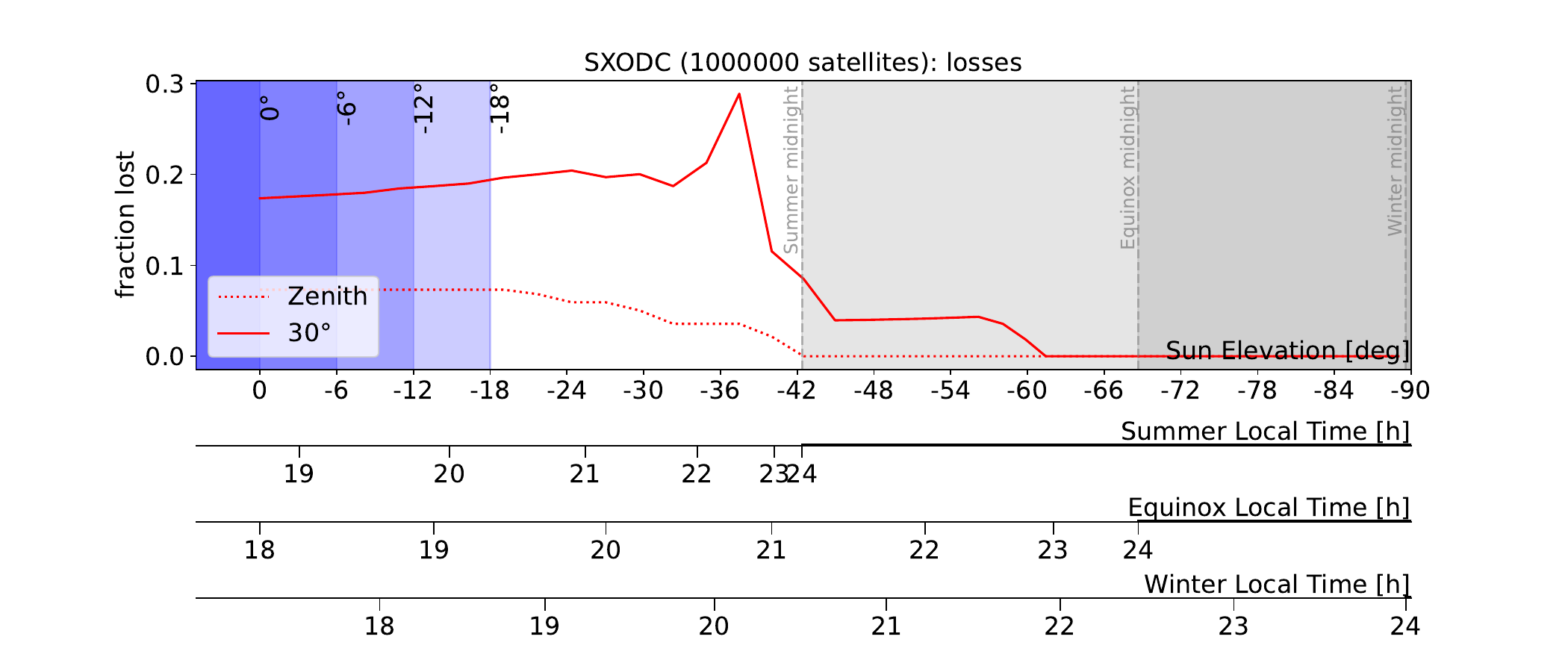}
    \caption{For one million satellites ($V_\mathrm{500~km}=7$), representing the SpaceX Orbital Data Center constellation:
    fraction of the FORS2 field of view lost as a function of solar elevation for observations at zenith and at $30^\circ$ elevation toward the Sun. 
    \obl{The peak at $\sim-36^\circ$ corresponds to the time when the measuring point crosses the cusp of the constellation.}
    The secondary scales convert solar elevation into local solar time for the equinoxes and solstices. Twilights are shown in blue, and inaccessible elevations in grey.
    Twilights are shaded in blue. The corresponding local times are given for the solstices and equinoxes, and inaccessible elevations are shaded in grey.
    The corresponding sky maps are shown in Fig.~\ref{fig:ap.SXODC}.
    }
    \label{fig:SXODCelev}
\end{figure*}

\begin{figure*}
    \centering
    {\bf a.}\includegraphics[width=.48\linewidth]{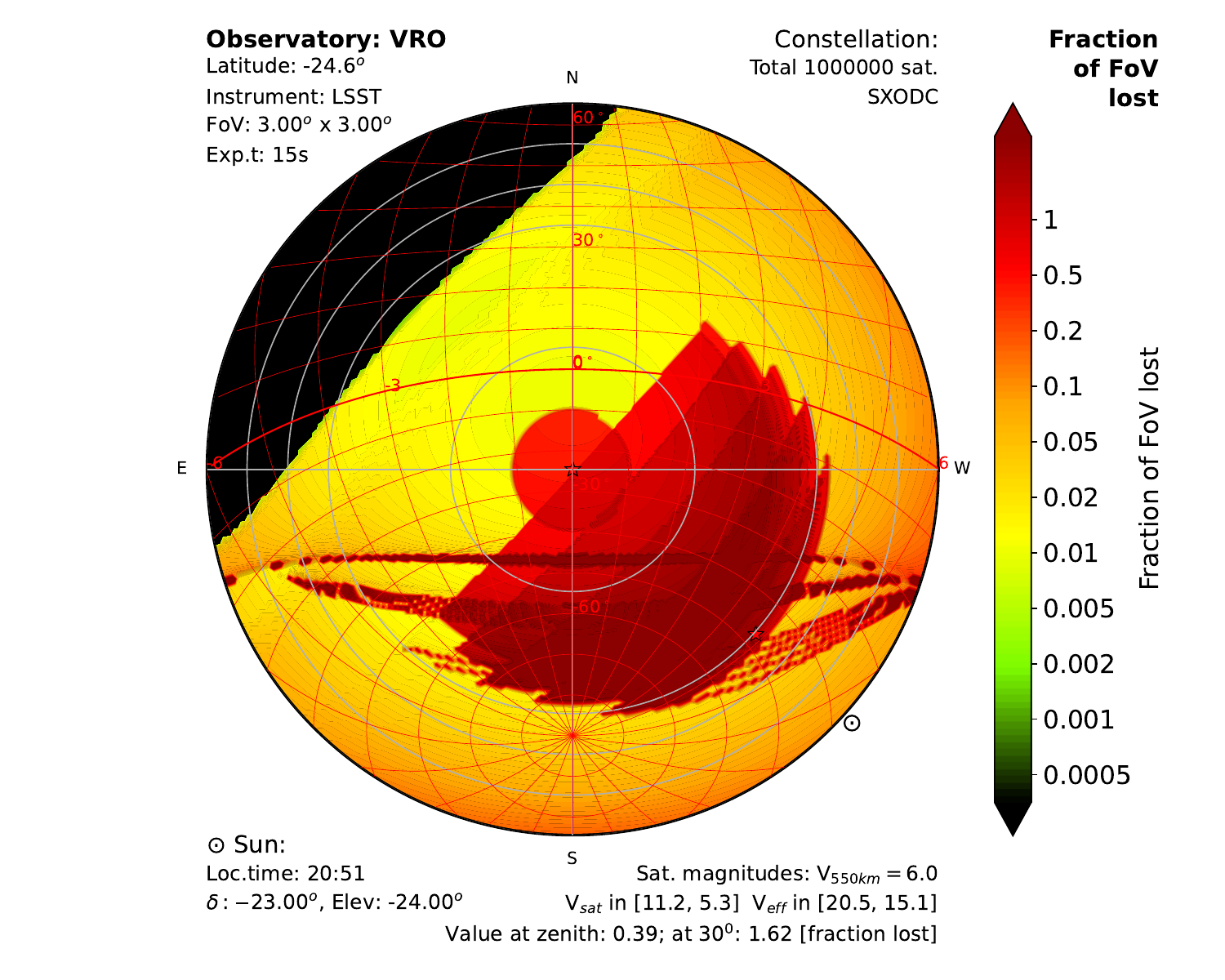}
    {\bf b.}\includegraphics[width=.48\linewidth]{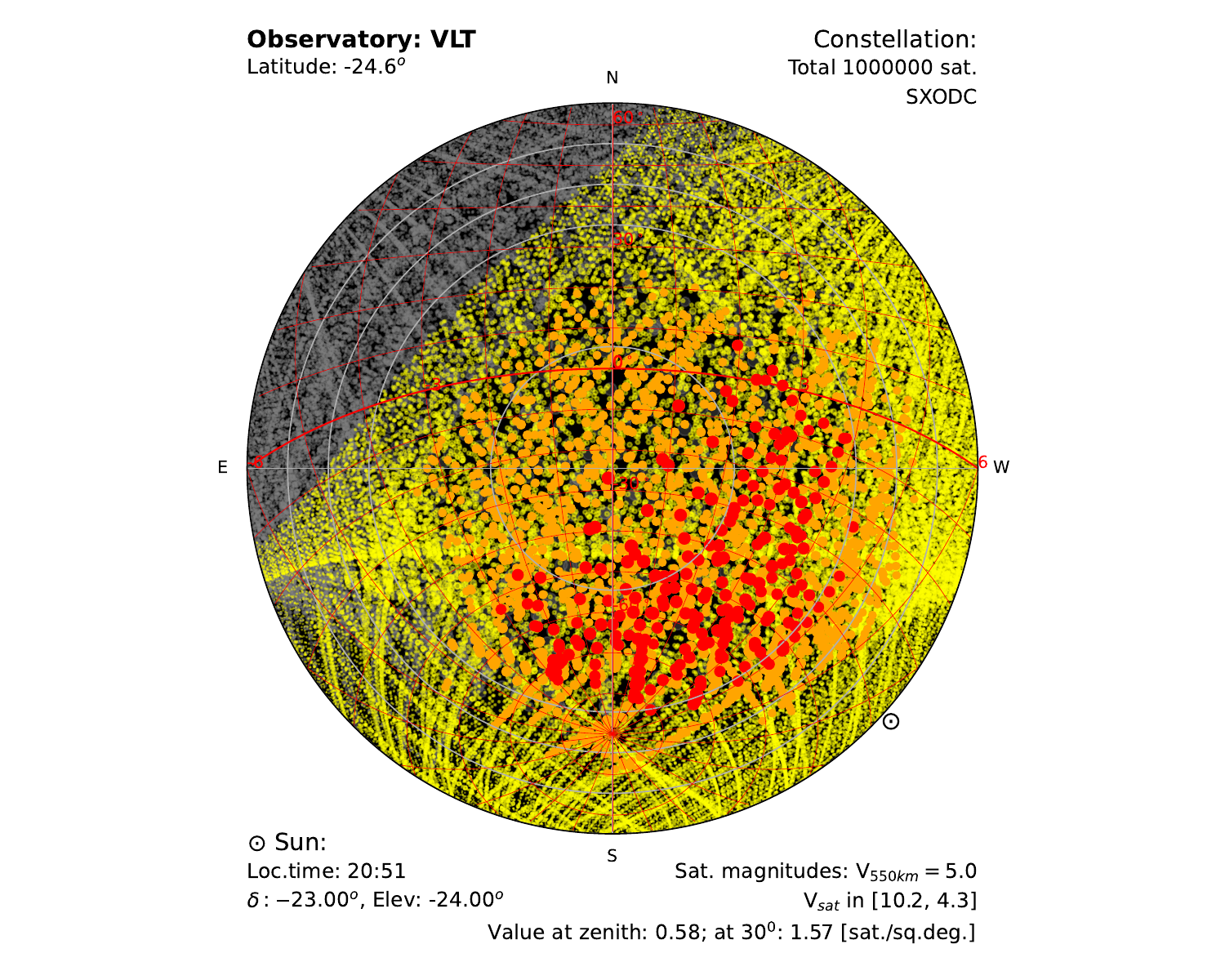}
    \caption{For one million satellites ($V_\mathrm{500~km}=6$), representing the SpaceX Orbital Data Center constellation in slight violation of the $V>7$ recommendation:
    {\bf a.} fraction of the field of view lost for the LSST camera. Values above 1 indicate that, on average, more than one satellite trail saturates a given pixel. The satellites' apparent angular velocity decreases at the constellation cusps, lowering their effective magnitudes and increasing their density, which results in higher losses in these regions. The red ``half circles'' correspond to the lowest and brightest satellites, for which the Earth's shadow has already hidden part of the orbit.
    {\bf b.} Example positions of the individual satellites. The grey dots are satellites in the Earth's shadow and therefore invisible, whereas the coloured dots are illuminated satellites. The 1867 orange dots correspond to satellites brighter than $V=7$, and the 247 red dots to satellites brighter than $V=5$. For comparison, about 3000--4000 stars are this bright (see Fig.~\ref{fig:fullMoon}.b).
    }
    \label{fig:SXODC_LSST}
\end{figure*}

For a given instrument, we set the observing conditions to a solar elevation $e_\odot = -21^\circ$, i.e. $1.5$~h after sunset (the actual time varies with the season and latitude), well into the astronomical night but at an instant when most of the satellites are still illuminated. The simulation is performed for the whole sky above the observatory, using the analytical method described in \citet{BHG22}, resulting in a sky map like the example in Fig.~\ref{fig:1Msat}. The values at zenith and at elevation $e=30^\circ$ (at the azimuth of the Sun) are recorded, corresponding to the range of elevations at which most observations are performed.

For critical cases, the simulation is repeated for solar elevations from $0^\circ$ (sunset or sunrise) to $-90^\circ$ (nadir), corresponding to the full range accessible from Paranal Observatory.

The quantities computed are:
\begin{itemize}
    \item Satellite density: the number of satellites per square degree;
    \item Number of trails crossing the instrument's FoV during an exposure;
    \item FoV losses: average value of the fraction of the FoV lost because of satellite trails. A value of 0.1, for instance, means that, on average, an exposure will be crossed by trails damaging 10\% of its pixels (and {\em not} that one image out of 10 is lost, the others intact). 
    \item Diffuse light pollution: the contribution of satellites below the detection threshold to the background brightness.
    \item Scattered light pollution: the contribution of the scattered light from the satellites to the background brightness.
\end{itemize}

For the FoV losses $f$, we set the following qualitative thresholds:
\begin{itemize}
    \item $f<0.3$\% is considered "good": the satellite-induced losses are in line with other losses affecting the data, such as bad columns on the detector, cosmic-ray hits, or bright-star spikes caused by diffraction on the telescope spider, etc.
    \item $0.3\le f <3$\% is "marginal". On the VLT, the technical losses, caused by equipment failure, are at the $<3$\% level, which is considered a good trade-off between the cost of preventive maintenance (more of which could reduce the technical losses) and the cost of the residual losses. Three per cent is therefore the total acceptable losses budget.
    \item $3 \le f <30$\% is "bad", as this is significantly larger than the weather losses (typically $\sim 10$\% for Paranal). At the 30\% level, satellites would be, by far, the largest contribution to data losses.
    \item $f \ge 30$\% is "disastrous". At that threshold, the production of data is severely compromised. In some cases, the losses can reach 100\%, effectively rendering the telescope useless.
\end{itemize}

For sky-brightness pollution $p$, the thresholds are as follows:
\begin{itemize}
    \item $p<0.5$\% is flagged as "good": the satellite-induced pollution is essentially negligible, well below the 1\% limit for observatories.
    \item $0.5\le p <10$\% is "marginal". The light pollution is still below the IAU general recommendation for light pollution, but the sky is no longer "observatory-grade".
    \item $10 \le p <100$\% is "bad". Observations are significantly affected: the faintest objects are not observable any more, and the exposure times required for the others are increased by up to a factor of two.
    \item $p \ge 100$\% is "disastrous": the light pollution is high enough to preclude deep, background-limited observations. 
\end{itemize}

\section{Simulation results}

\subsection{Upcoming constellations: sixty thousand satellites}\label{Sect:SLOWGWAK}

We consider a set of five telecommunication constellations representing Starlink~Gen.~1 and 2, OneWeb, GuoWang, and Amazon Leo (a.k.a. Kuiper), with a total of over 60 thousand satellites on 29 shells. This is the same configuration used by \citet{BHG22}, detailed in their Table~1 \obl{and hereafter referred to as BHG2022}. While the actual architecture of each constellation is likely to evolve, this set is representative of the upcoming population \obl{of telecommunication satellites}, with a mix of low- and high-altitude orbits from 328 to 1200~km. We set their absolute magnitudes to $V_\mathrm{550~km} = 7$, which may be optimistic. This ensures that most satellites respect the IAU recommendation $V>7$; the only exceptions are the few satellites on very low orbits, appearing close to zenith.
Figure~\ref{fig:SLOWGWAK} displays the results; the values at zenith and at 30$^\circ$ elevation are listed in Table~\ref{table:results}.

As discussed by \citet{BHG22}, the fraction of FoV lost with this set of constellations is a fraction of a per cent in the regions of the sky where the satellites are illuminated, which dwindle about one hour after twilight. While this is an annoyance, that level of loss is still in the "good" range.
The diffuse light on all-sky cameras is of the order of $1$--$5\times10^{-4}$ times the natural sky background (i.e. negligible, in the "good" range). Similarly, the scattered light is of the order of $10^{-4}$ times the natural sky (i.e. also negligible). 

In summary, a 60-thousand-satellite population where all satellites adhere to the $V>7$ limit causes a non-zero but manageable effect on observations.

\subsection{Mega-constellations: one million SpaceX orbital data center satellites}\label{Sect:SXODC}

\begin{figure*}
    \centering
    \includegraphics[width=.985\linewidth]{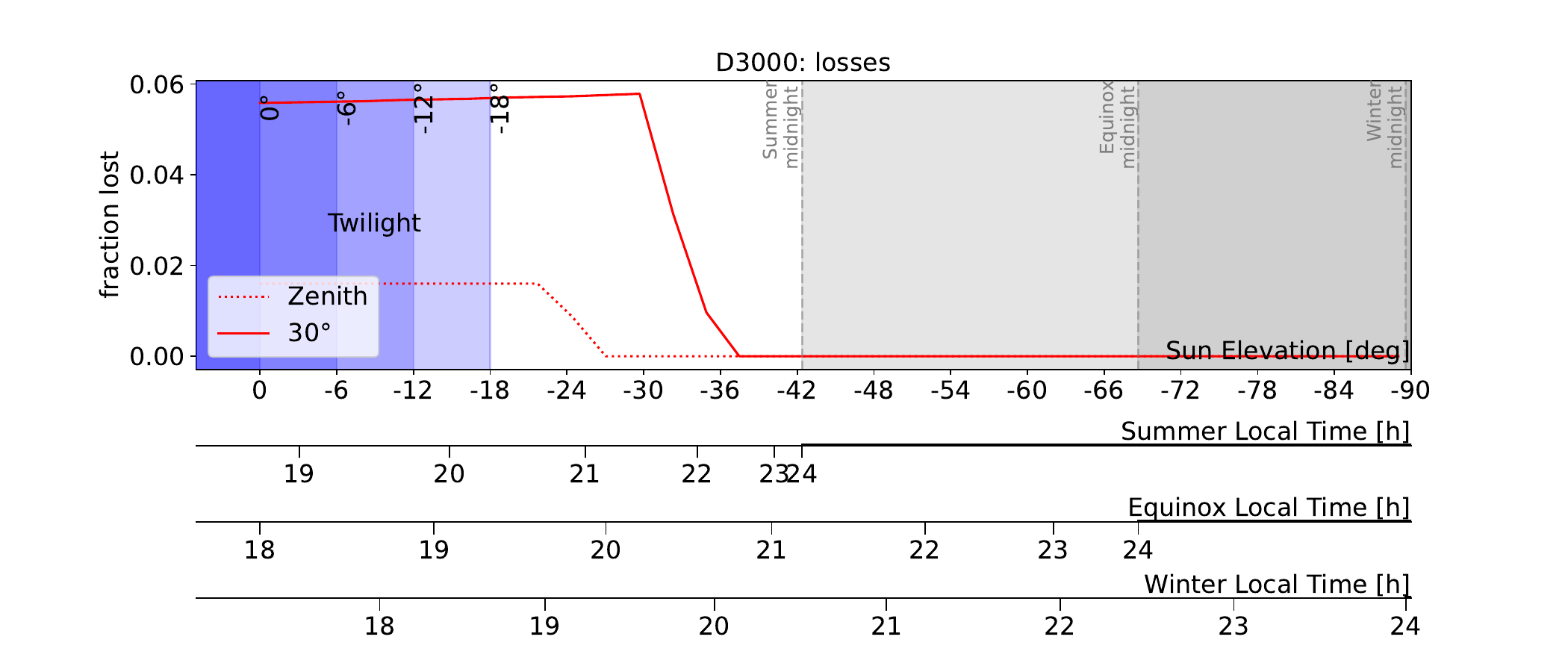}
    \caption{FoV losses for a saturating camera such as LSST, for constellations of 
    3000 BlueBird-like satellites 
    representing a direct-to-cell constellation (code D3000 in the table), 
    as a function of solar elevation.
    Twilights are shaded in blue. The corresponding local times are given for the solstices and equinoxes, and inaccessible elevations are shaded in grey.
    }
    \label{fig:AST-faint-vs-bright_elev}
\end{figure*}

\begin{figure}
    \centering
    \includegraphics[width=1\linewidth]{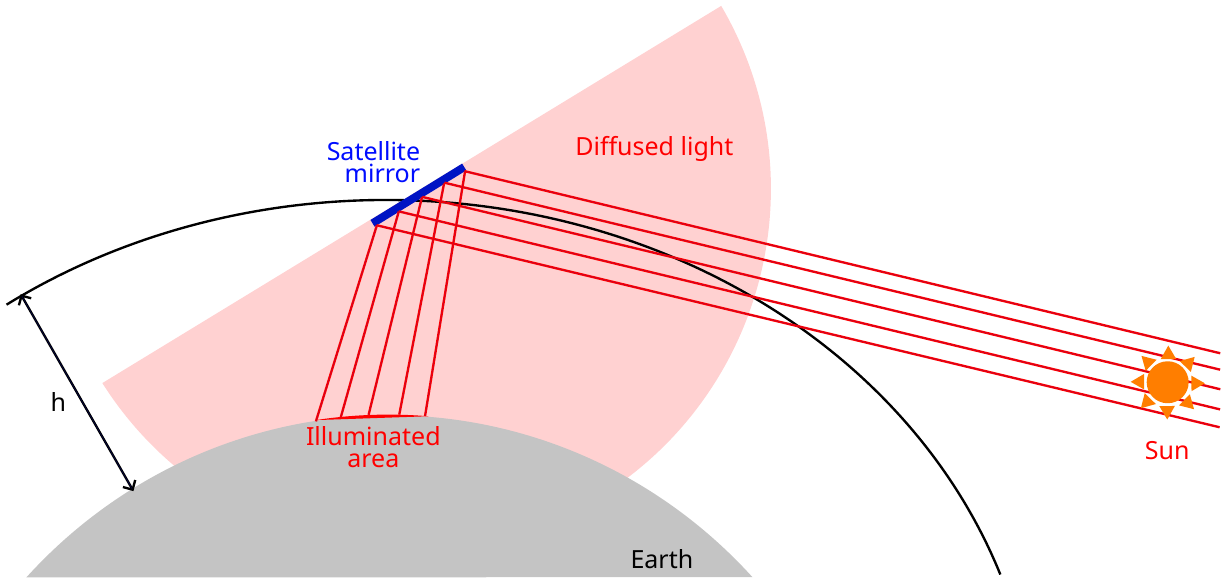}
    \caption{Geometry of a large mirror-like satellite reflecting sunlight toward Earth. Part of the light is reflected onto a small area on the ground, while the remainder is diffused over a hemisphere. This illustration is not to scale.}
    \label{fig:mirror}
\end{figure}

We consider here the SpaceX Orbital Data Center constellation, as described in McDowell's list\footnote{\url{https://planet4589.org/space/con/conlist.html}, retrieved 2026-03-24}, with a total of one million satellites. \obl{From these, 500\,000 are distributed on shells with altitudes between 400 and 1000~km, and the remaining 500\,000 are distributed on sun-synchronous terminator orbits with altitudes between 700 and 1000~km, to benefit from continuous illumination. The latter ones will be above the horizon only during twilight. To evaluate the impact of one million satellites, we spread the sun-synchroneous satellites on traditional Walker pattern. We set their absolute magnitudes to $V_\mathrm{550~km} = 7$, which may be optimistic considering the power requirements of these data centres will require large solar panels and radiators. We therefore also consider  $V_\mathrm{550~km} = 6$ and 5. In the tables and figures, these constellations are labelled SXODC-7, -6, and -5.}

The scattered light pollution (Fig.~\ref{fig:ap.SXODC}.d) remains at the 0.1--0.2\% level of the dark sky background (i.e. in the "good" range), but the diffuse light pollution (Fig.~\ref{fig:ap.SXODC}.c) will be in the 0.5--1.5\% range, i.e. above the 1\% limit for prime astronomical sites, and is therefore flagged as "marginal".

Because of the large number of satellites, the number of trails per image increases dramatically compared to the previous section; consequently, the fraction of the field of view covered by trails also increases dramatically. The example in Fig.~\ref{fig:ap.SXODC}.d indicates that the losses will reach 6--15\% (i.e. "bad") when the Sun is at $-24^\circ$. Figure~\ref{fig:SXODCelev} shows the evolution of the losses with solar elevation and local time, peaking at 28\% when pointing towards the cusps of the constellation. These losses remain above the per cent level for most of the night, except when the Sun is below $-60^\circ$, which happens only during 2~h in spring/autumn nights and 4~h in winter nights --- and not at all in summer. The example of FORS2 is representative of traditional imagers on large telescopes. 

For a saturating detector like Rubin observatory's LSST, it is {\em critical} that the magnitudes of the satellites remain fainter than $V=7$. Should the absolute magnitudes of the satellites reach $V_\mathrm{550~km}=6$, the effects would be devastating, as illustrated in Fig.~\ref{fig:SXODC_LSST}.a. At $V_\mathrm{550~km}=5$, {\em most of the LSST observations would be lost}, firmly in the "disastrous" regime. This sets stringent constraints on the design of the satellites, in terms of cross section and reflective properties, and on their operations, e.g. ensuring that they remain fainter than $V=7$.

Additionally, because of the very large number of satellites, a minor violation of the $V_\mathrm{550~km}>7$ recommendation would result in {\em many} satellites becoming visible to the naked eye. For instance (see Fig.~\ref{fig:SXODC_LSST}.b), should their $V_\mathrm{550~km}=5$, close to 2000 satellites would be visible to the eye (brighter than $V=7$) one hour after sunset. As a comparison, one can typically see four to five thousand stars. The satellites would significantly affect the appearance of the sky.

\subsection{Very bright satellites: AST BlueBirds}\label{sect:AST}

Here, we discuss the case of the bright "BlueBird" satellites by AST-SpaceMobile. Based on their website\footnote{\url{https://ast-science.com}}, the surface of these satellites will be 223~m$^2$. Using the measurements from \citet{MallamaCole2025}, i.e. $V_\mathrm{550~km} = 2$, 
and a total number of 243 satellites (as in their petition to the Federal Communication Commission) on a set of realistic shells, we built the "AST" constellation. \obl{To evaluate the effect of a larger constellation, or the combination of various AST-like constellations, we also built an hypothetical constellation with 3000 BlueBird-like satellites on realistic orbits with altitudes ranging from 550 to 750~km; this hypothetical constellation is "D3000" (for "direct-to-cell").} 

In the case of the 300-satellite constellation, the small number of spacecraft ensures negligible losses on instruments not affected by saturation (Fig.~\ref{fig:ap.AST-243}). In the case of the LSST camera, however, saturation drives the losses to 0.2--0.6\% of the FoV, instead of 0.0002--0.0006\% if the satellites had been fainter (see Fig.~\ref{fig:ap.AST-faint-vs-bright}.a and b). The effect on the sky-background pollution, both diffuse (at the $10^{-4}$ level) and scattered (at the $10^{-5}$ level), is negligible, as expected for a handful of illuminated satellites in the sky.

Even with 3000 bright satellites (Fig.~\ref{fig:ap.AST-3000}), the scattered light pollution remains negligible (4--6 $\times 10^{-4}$), while the diffuse light inches towards the tolerability limit of 1\%: the pollution would represent 0.2--0.4\% of the dark sky. For non-saturating instruments, the FoV loss remains at the $10^{-4}$ level, still driven by the fairly small (compared to previous sections) number of satellites. For a saturating instrument like the LSST, however, the losses become significant, in the 1--6\% range (see Fig.~\ref{fig:AST-faint-vs-bright_elev}). As in the case of the 243-satellite configuration, these losses are completely dominated by the brightness of the satellites: had they respected the $V=7$ limit, the losses would be at the $10^{-5}$ level (see Fig.~\ref{fig:ap.AST-faint-vs-bright}.c and d). All illuminated satellites (about 30--50 in the first and last hours of the night) would be clearly visible to the naked eye as bright sources, ranging from $V\sim 6$ close to the horizon to $V\sim 2$ close to zenith.

Overall, this experiment illustrates that the satellites of a small to moderately sized constellation still must respect the $V>7$ limit, as brighter satellites will have a disproportionate effect on saturating instruments like the LSST Camera.

\subsection{Extremely bright satellites: Reflect Orbital sunlight as a service}\label{sect:RO}

\begin{figure}[tbp]
    \centering
    \includegraphics[width=.98\linewidth]{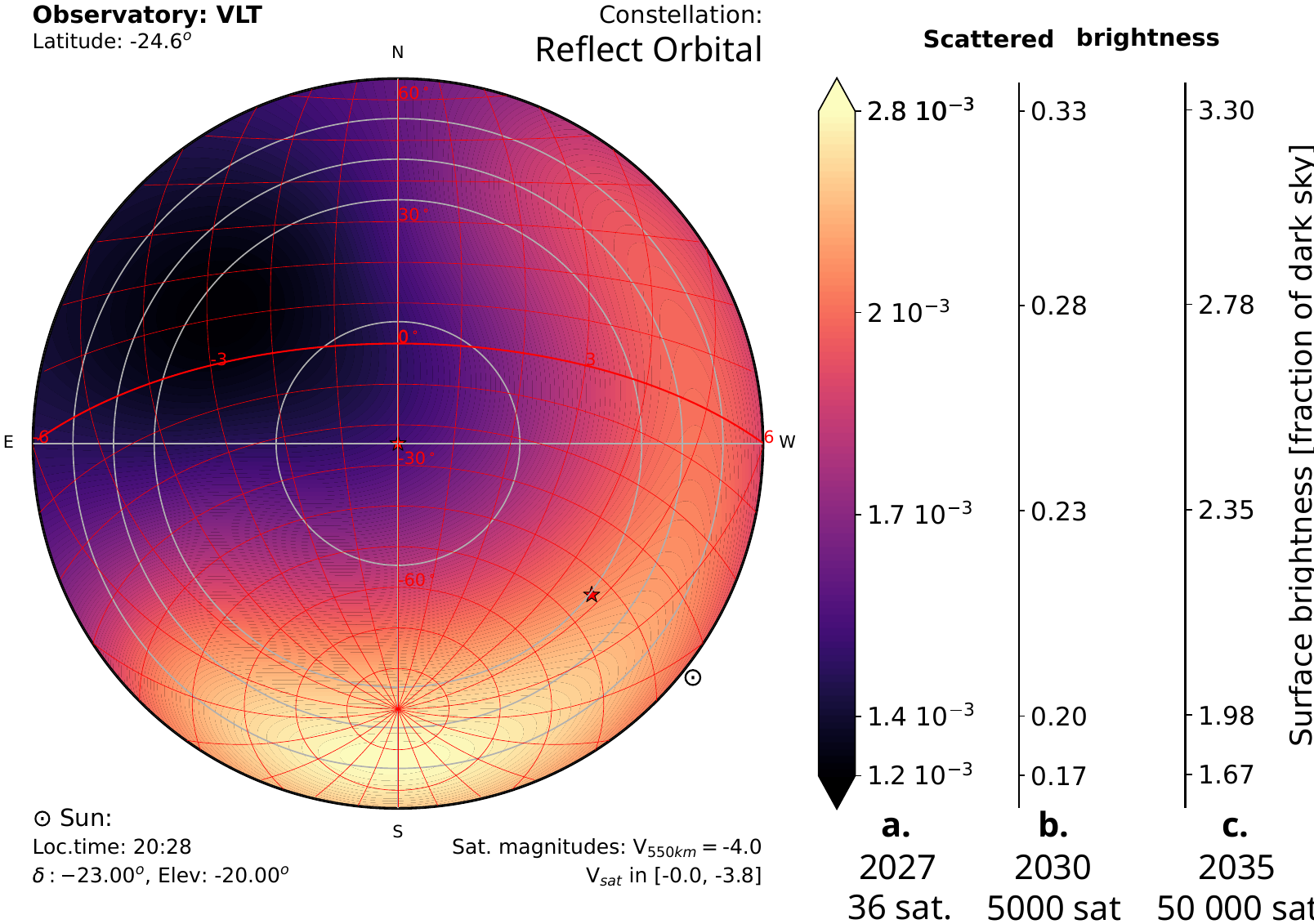}
    \caption{Scattered light, as a fraction of the natural dark sky, for 37 (panel~a), 5000 (b), and 50,000 (c) extremely bright satellites ($V_\mathrm{500~km}=-4$), representing the Reflect Orbital constellations envisaged for 2027, 2030, and 2035, respectively, as seen from outside their illuminating beam.
    For more details, see Fig.~\ref{fig:SLOWGWAK}.
    }
    \label{fig:SpaceOrbitalScatter}
\end{figure}

\begin{figure}[tbp]
    \centering
    {\bf a}\includegraphics[width=.98\linewidth]{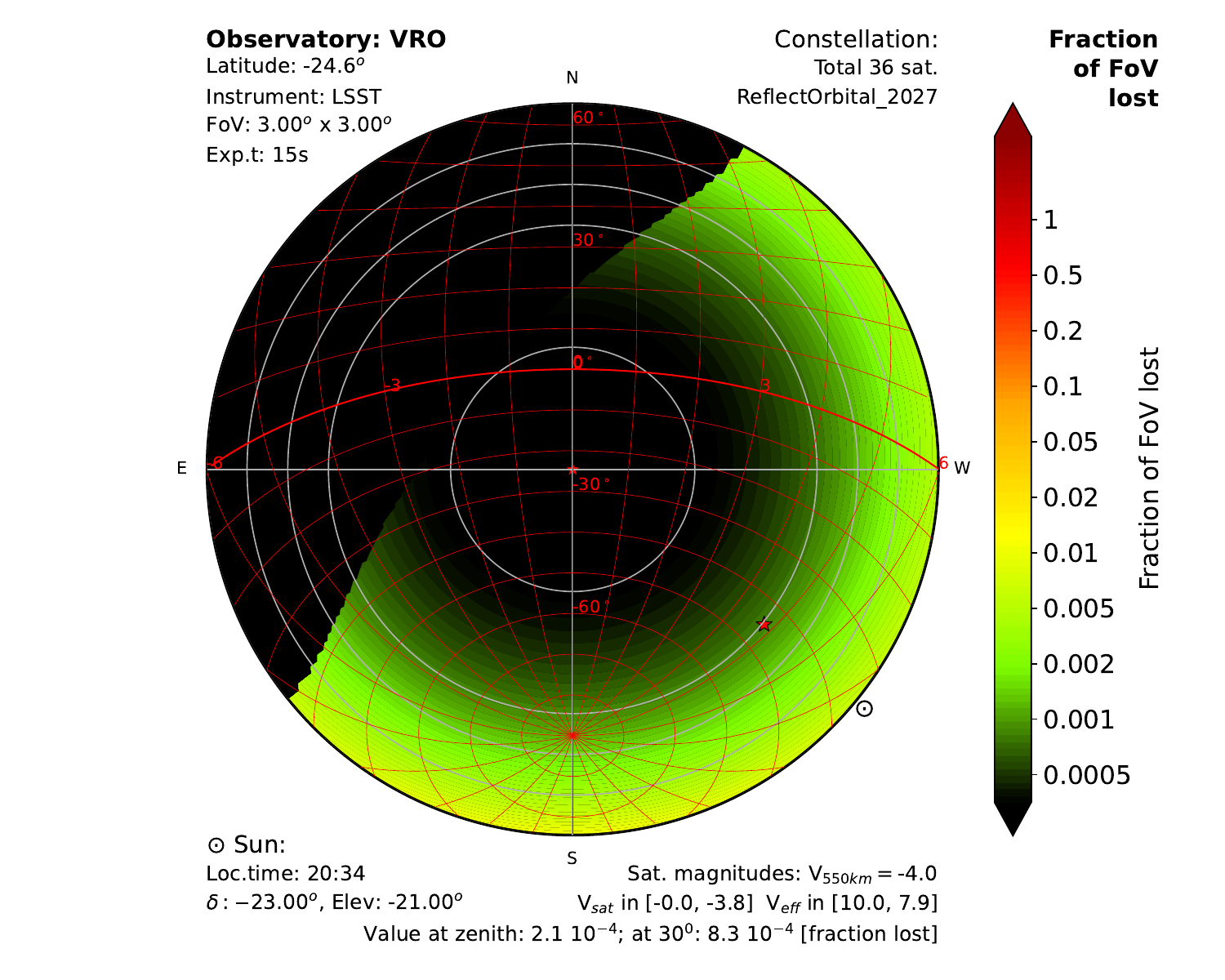}
    {\bf b}\includegraphics[width=.98\linewidth]{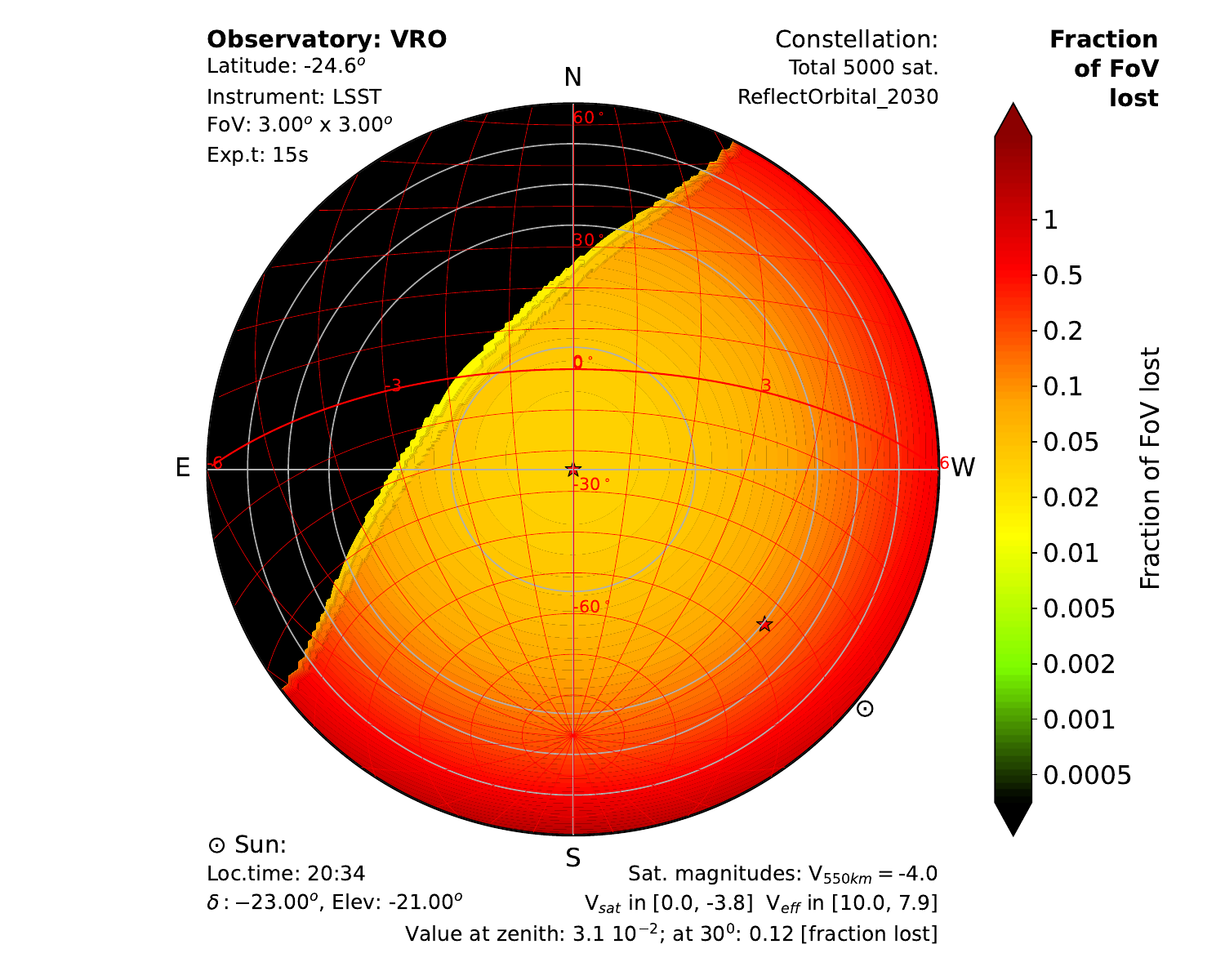}
    {\bf c}\includegraphics[width=.98\linewidth]{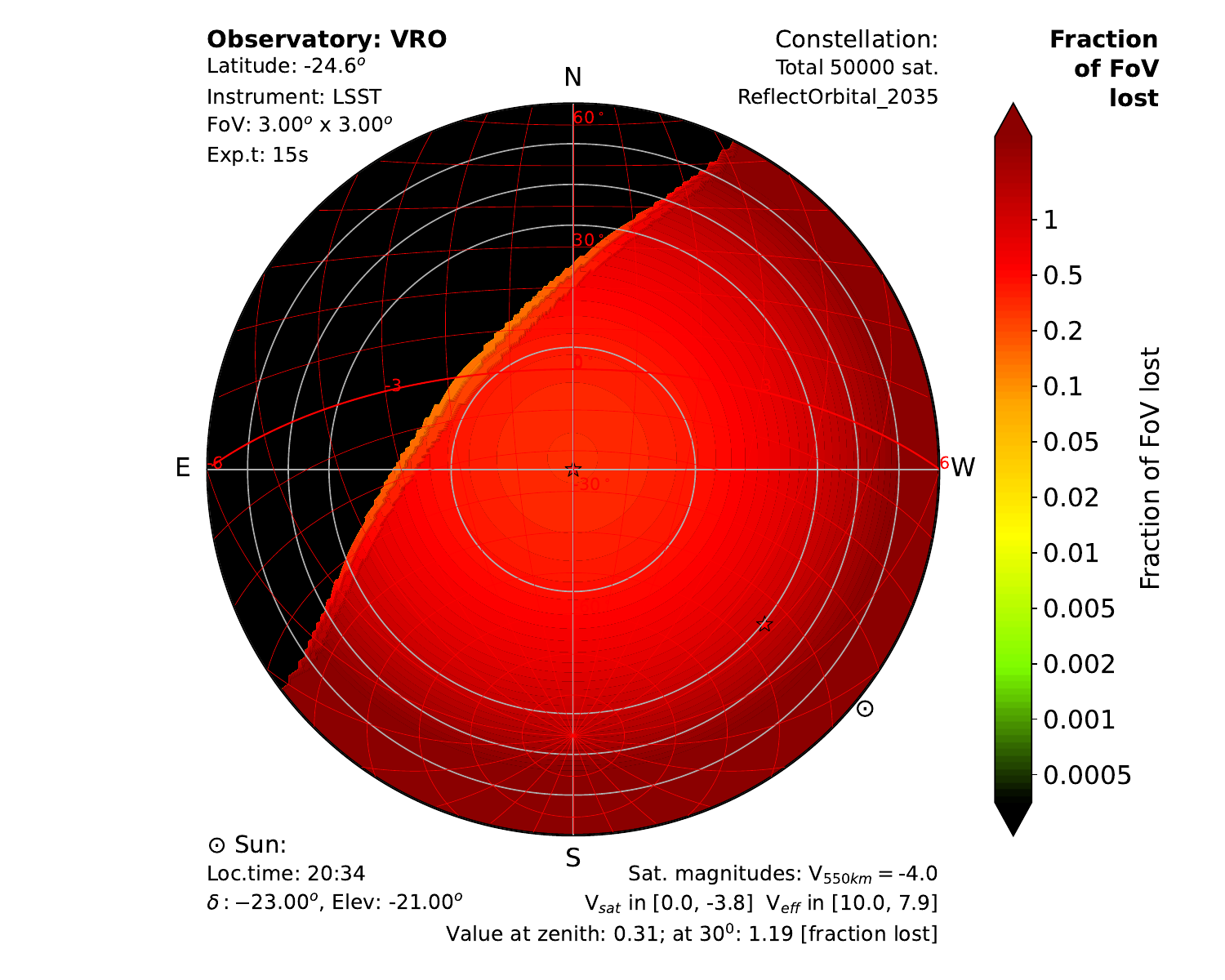}
    \caption{Fraction of exposure lost for an instrument such as the LSST Camera for 36 ({\bf a}), 5000 ({\bf b}), and 50\,000 ({\bf c}) extremely bright satellites ($V_\mathrm{500~km}=-4$), representing the Reflect Orbital constellations envisaged for 2027, 2030, and 2035, respectively, as seen from outside their illuminating beam.
    A loss greater than 1 means that, on average, each pixel is affected by more than one satellite trail.
    For more details on these sky maps, see Fig.~\ref{fig:SLOWGWAK}.
    }
    \label{fig:SpaceOrbitalLSST}
\end{figure}

\begin{figure*}[tbp]
    \centering
    {\bf a}\includegraphics[width=.98\linewidth]{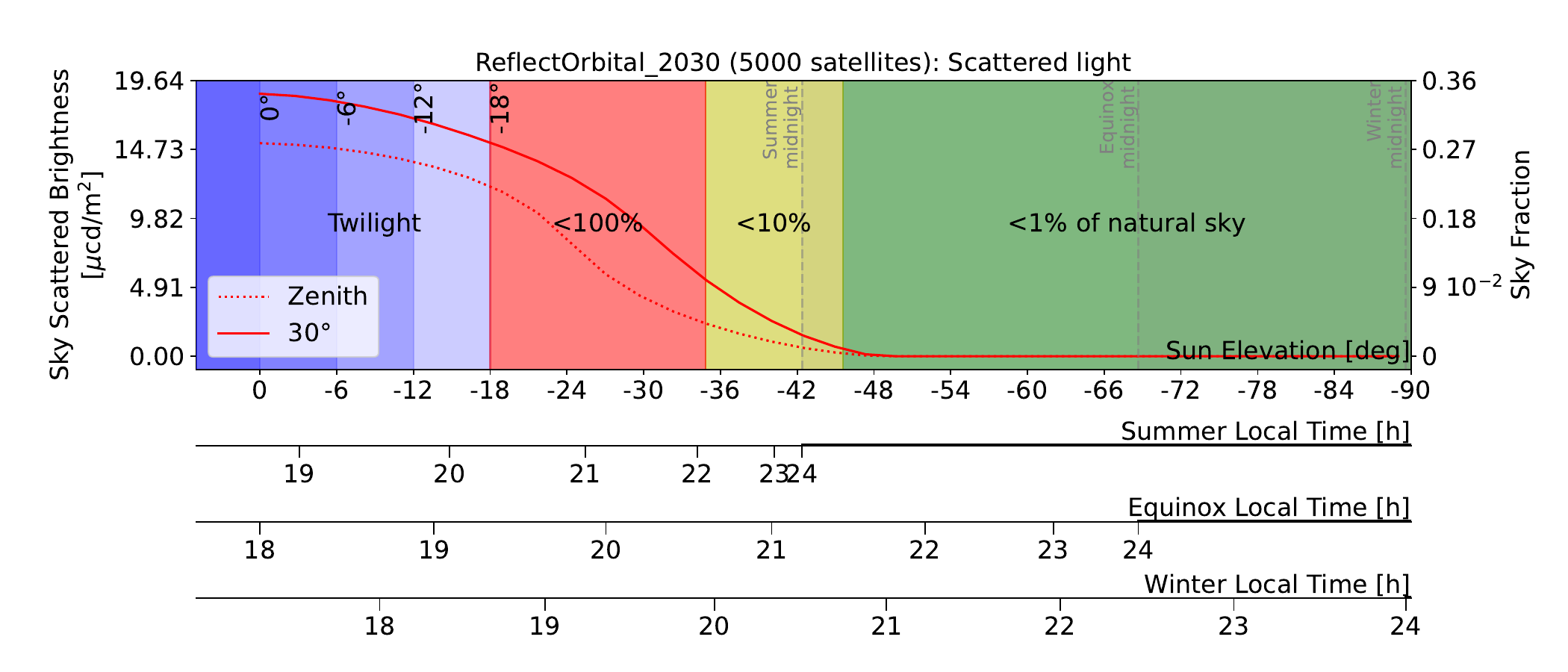}
    {\bf b}\includegraphics[width=.98\linewidth]{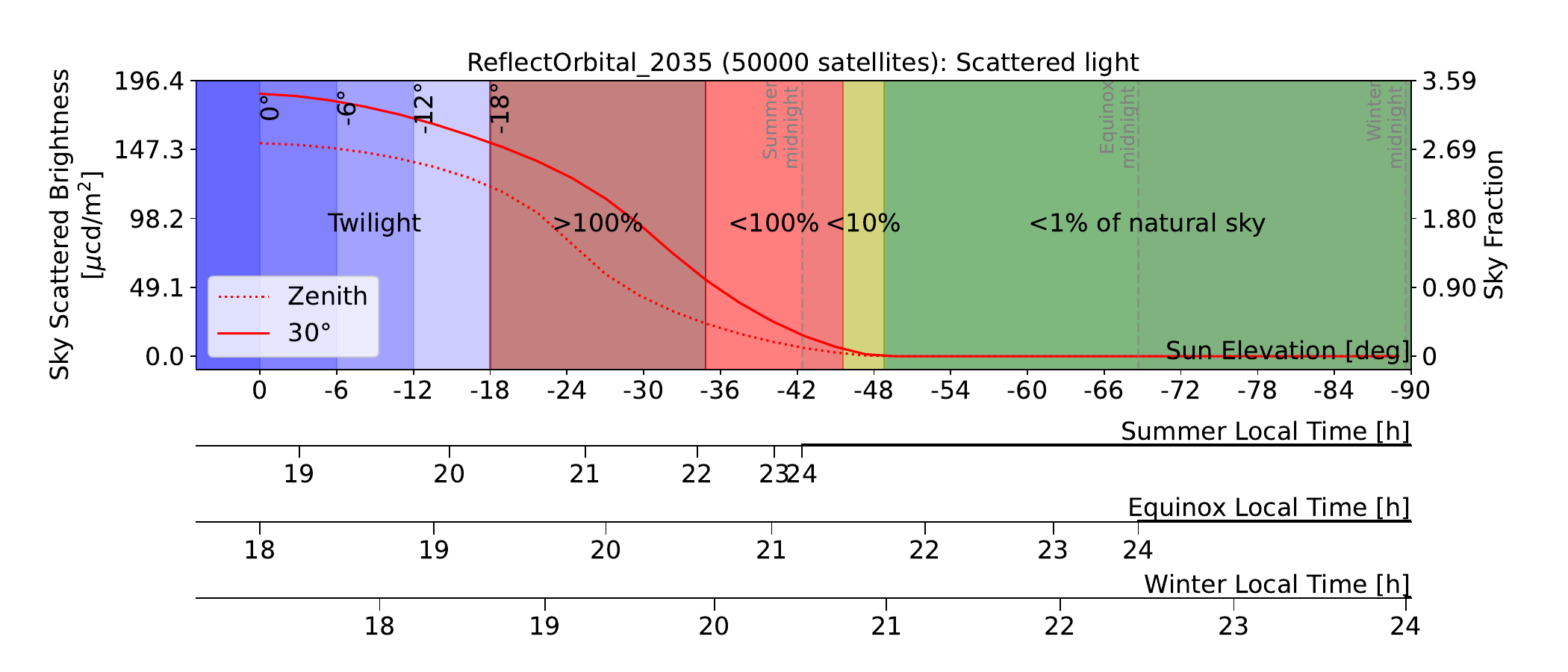}
    \caption{Evolution of scattered light as a function of solar elevation for 5000 (a) and 50,000 (b) extremely bright satellites ($V_\mathrm{500~km}=-4$), representing the Reflect Orbital constellations envisaged for 2030 and 2035, respectively, as seen from outside their illuminating beam.
    Twilights are shaded in blue. The corresponding local times are given for the solstices and equinoxes, and inaccessible elevations are shaded in grey. Observing conditions remain good when the pollution is $<1$\% of the natural sky (green shading), are compromised when the pollution is between 1 and 10\% (yellow), and are jeopardised between 10 and 100\% (red), and impossible above 100\% (dark red).
    }
    \label{fig:SpaceOrbitalScatterElevation}
\end{figure*}

In this section, we tackle the issue of extremely bright objects, such as the very large mirror-like satellites envisioned by Reflect Orbital\footnote{\url{https://www.reflectorbital.com/}}. 
They plan\footnote{\url{https://www.reflectorbital.com}} to launch a single prototype satellite in 2026, soon followed by a modest constellation of 36 satellites in 2027 and a production constellation of 5000 satellites in 2030. They also mention an extension to 50\,000 satellites by 2035. In the table and figures, they are labeled RO-2027, RO-2030, and RO-2035. The company has not made their constellation architecture available, so we distribute the satellites over three Walker shells with altitudes between 600 and 650~km, on orbits at an arbitrary but realistic inclination of 55$^\circ$. For the simulation of the various effects on observations, we assume that no satellite is pointing its reflected beam to or near the observatory, but that all illuminated satellites are in an operational attitude.

To estimate the magnitude of these objects, we consider the mirror area. It is suggested that a Reflect Orbital satellite could be $57\times57 \sim 3\,000$~m$^2$. Following the simple geometry illustrated in Fig.~\ref{fig:mirror}, the mirror is illuminated by the Sun, with an illuminance $V=-26.75$, or $I = 127\,000$~lx $= 164$~W~m\mm in the $V$ band. Assuming that the mirror is oriented at $45^\circ$ with respect to the incident light, it collects $F \sim 348$~kW. A fraction $R$ of that light is reflected towards the Earth, and the remainder is diffused over half a sphere. Formally, part of it is also absorbed. However, assuming that the reflector is made of high-reflectivity coated Mylar, we neglect the absorption.

The diffused light, $(1-R)~F$, is spread over $2\pi$~sr. A unit area on the ground subtends $1/h^2$~sr, with $h=550$~km being the altitude of the satellites, resulting in the diffused-light illuminance $I_\mathrm{diff}= (1-R)I / (2\pi h^2)$ at zenith.

The fraction of reflected to diffused light, $R$, is difficult to estimate. For a carefully optimised dielectric film, \cite{starlink} obtained a ratio of $10^7$ between the reflected and diffused illuminances. An earlier version of the film resulted in a ratio of $10^5$. While a Mylar sheet is likely to also have a very high ratio, it is expected that the unfurling of the mirror will leave some small- and large-scale irregularities. Furthermore, the mirror is expected to degrade with time, resulting in less light being reflected in the main beam. Setting $R=0.1$, we obtain a ratio of $9\times10^{-5}$ between the diffused and reflected illuminances, which results in 
\begin{itemize}
\item illuminance in the reflected beam $I = 1.8\times10^{-3}$W~m\mm$=1.4$~lx, or $V=-14.3$. This is $\sim 4$ times brighter than the full moon, and the illuminance produced is in line with Reflect Orbital's web page. 
\item diffuse illuminance outside the beam $I_\mathrm{diff}= 1.7\times10^{-7}$W~m\mm = $1.3\times10^{-4}$~lx or $V_\mathrm{diff}=-4.3$, i.e. close to the maximum brightness of Venus.
\end{itemize}

Because of the modest number of satellites (compared to those in Sect.~\ref{Sect:SLOWGWAK} and \ref{Sect:SXODC}), the direct losses caused by satellite trails on the data will be small for instruments not suffering from saturation. For the saturating LSST Camera, however, the effects of even a single satellite crossing its field of view will be devastating \citep{Tyson2020}, compromising a very high fraction of the frame affected by that single trail, as illustrated in Fig.~\ref{fig:SpaceOrbitalLSST}. The 2030 constellation (5000 satellites) would ruin 3\% (at zenith) to 12\% (at $30^\circ$ elevation) of the LSST FoV one hour after sunset. With the 2035 constellation (50,000 satellites), the losses would be catastrophic, about 30\% at zenith to 100\% at $30^\circ$: {\em all} LSST images would be entirely unusable. 

Clearly, all the illuminated satellites will be visible as objects brighter than the brightest stars ($V$ in the $-4$ to 0 range). With 5000 satellites, one can expect $\sim 130$ Venus-bright satellites, and with a 50\,000-satellite constellation, about 1300 Venus-like satellites will criss-cross the sky. Furthermore, those directly illuminating the observer will appear as points brighter than the full Moon, which have no natural equivalent. 

These satellites are so bright that most instruments will detect them directly. Only for long exposures with high-resolution spectrographs will their effective magnitudes be fainter than the instrument's limiting magnitude. In that case, the test constellations with 37, 5000, and 50,000 satellites will contribute increases to the sky background by factors of 0.09, 0.7, and $7\times$ (respectively) with respect to the dark sky background.

In terms of scattered light, only the smallest, 36-satellite constellation contributes an acceptable level of light pollution (0.15 to 0.21\% of the dark sky; see Fig.~\ref{fig:SpaceOrbitalScatter}a). With 5000 satellites, the pollution reaches 20 to 30\% (Fig.~\ref{fig:SpaceOrbitalScatter}.b), corresponding to increased exposure times by the same amount, and sky brightnesses  $V = 21.8$ to 21.7~MpSA, or Bortle Class 2.  
Finally, with 50\,000 satellites, the scattered light pollution will reach 2 to 3$\times$ the dark sky, i.e. increasing the total sky brightness by a factor 3 to 4. This corresponds to $V = 20.8$ to 20.5~MpSA, and would push a prime site to Bortle Class 4.5, that is "semi-suburban", unsuitable for dark-time astronomical observations. 

Figure~\ref{fig:SpaceOrbitalScatterElevation} displays the evolution of the scattered light pollution with the elevation of the Sun for the 5000 and 50,000 satellites, showing that good observing conditions ($<1$\% pollution) can be preserved only when the Sun is below $-48^\circ$ elevation. Accounting for seasonal changes, this geometry does not occur at all in summer nights, which are then 100\% compromised. Around the equinoxes, $\sim47$\% of the night is compromised, and at winter solstice, $\sim40$\% of the night is compromised. It is important to note that the change from 5000 to 50,000 satellites does not change these 1\% pollution limits significantly, but it increases the fraction of the night during which the pollution is above 10\% of the dark sky.

\subsection{Generalization}\label{generalization}
  
\begin{figure*}[tbp]
   \sidecaption
    \begin{minipage}{12.5cm}
    {\bf a}\includegraphics[width=12cm]{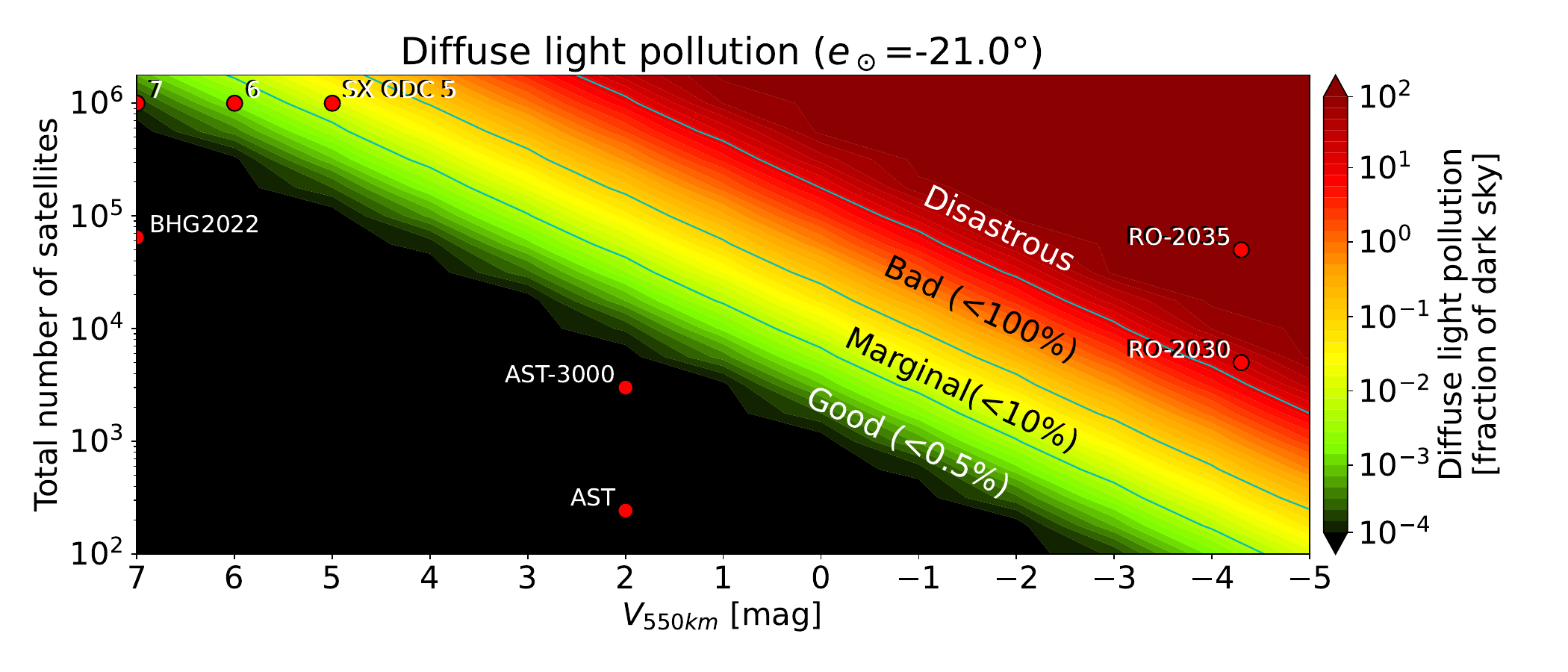}
    {\bf b}\includegraphics[width=12cm]{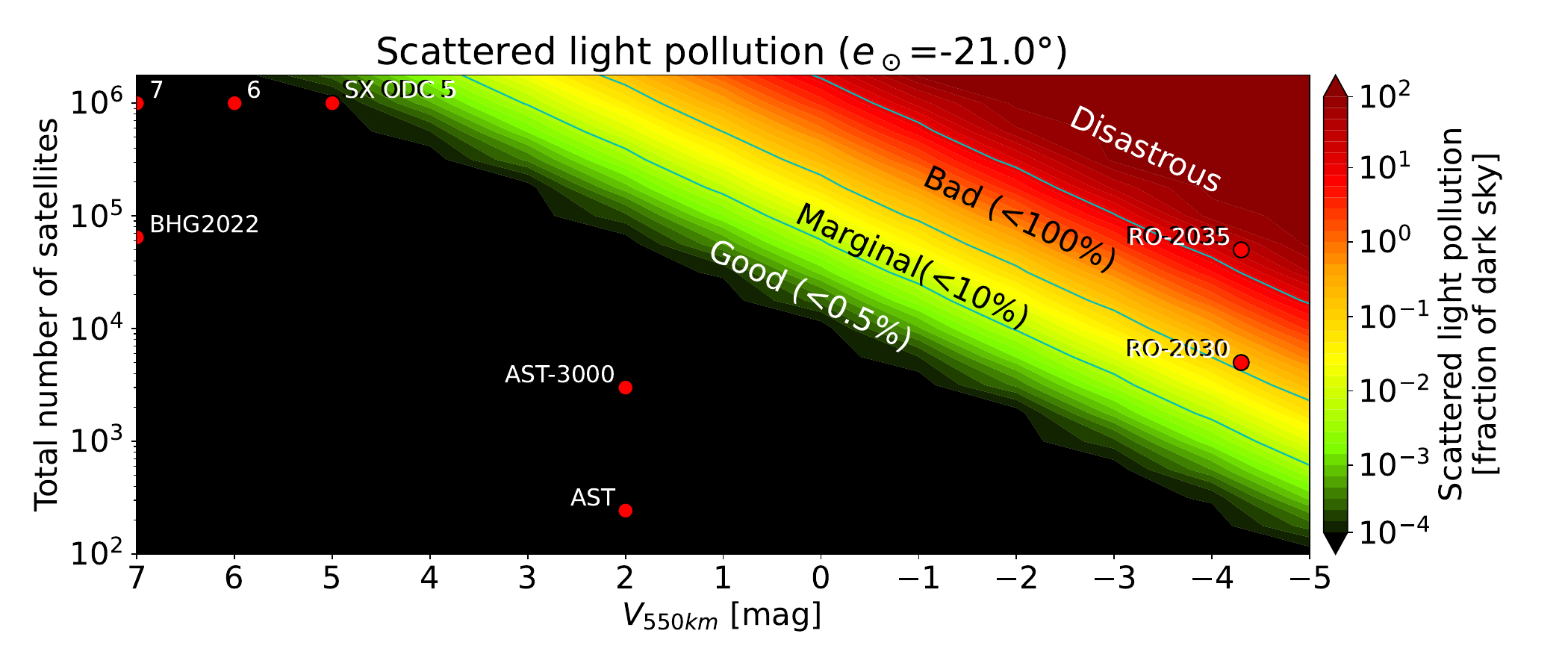}
    {\bf c}\includegraphics[width=12cm]{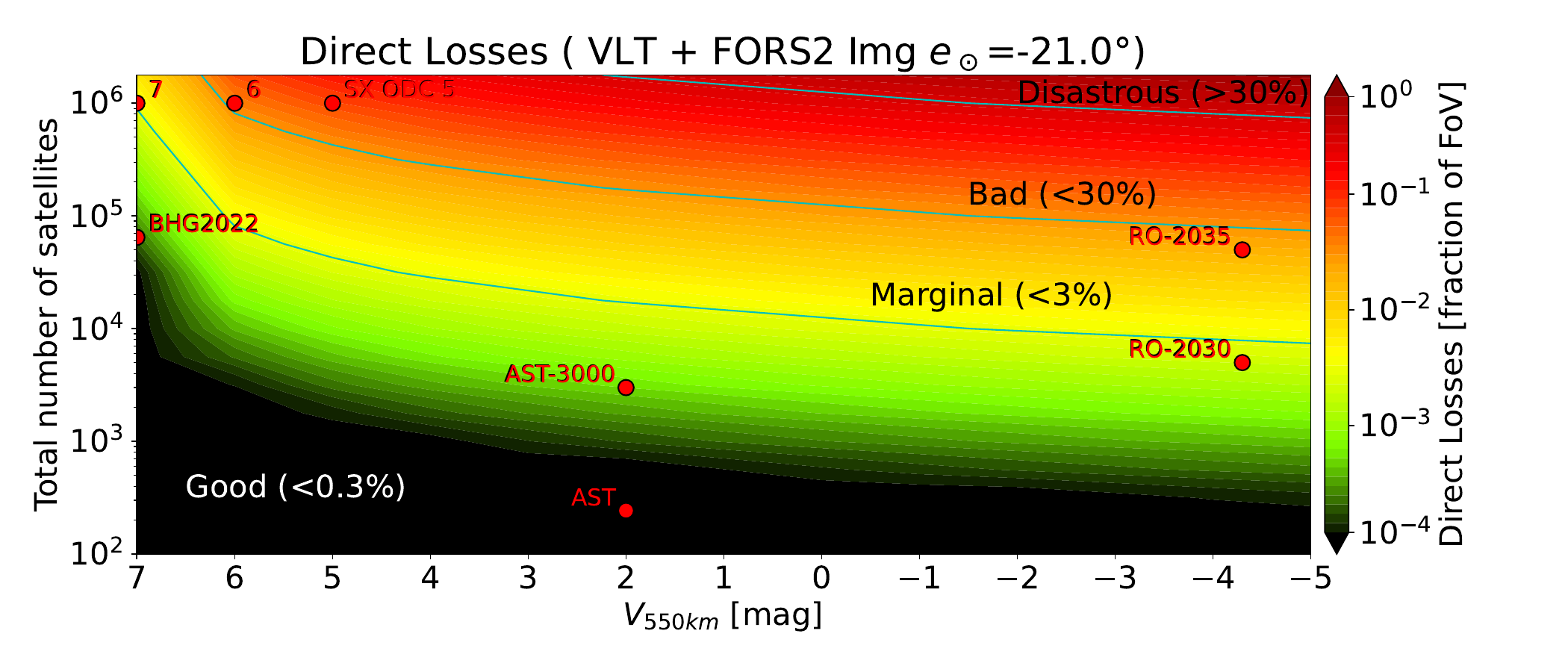}
    {\bf d}\includegraphics[width=12cm]{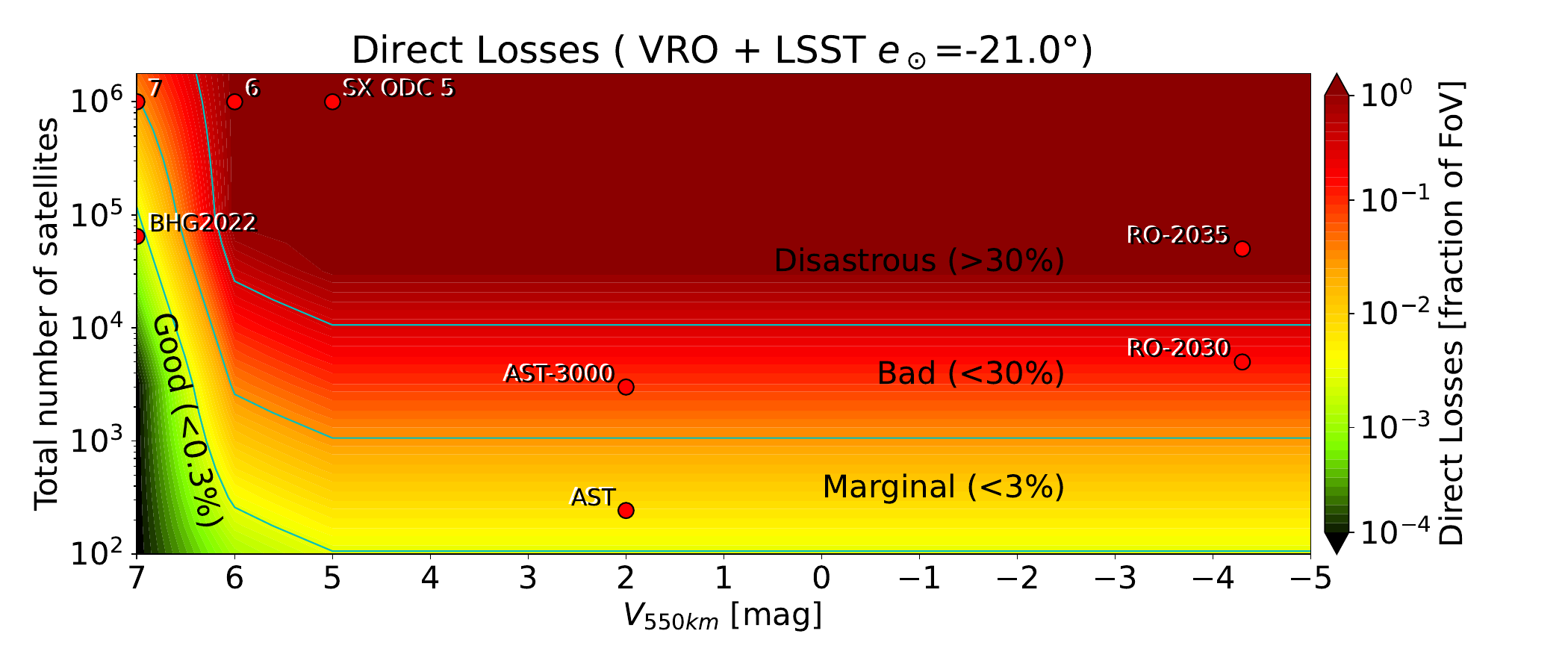}
    \end{minipage}
    \caption{Diffuse ({\bf a}) and scattered ({\bf b}) light pollution, and Field-of-View losses for FORS2 ({\bf c}, a traditional imaging camera) and LSST ({\bf d}, a saturating camera), as functions of the total number of satellites in orbit and their $V_\mathrm{550~km}$. 
    ~
    The light-pollution thresholds are 0.5\% (below the limit for an observatory), 10\% (generic IAU limit), and 100\%. The loss thresholds are 0.3\% (corresponding to one tenth of the technical losses), 3\% (equal to the technical losses), and 30\% (about 2--3$\times$ the technical and weather losses). The various constellations studied in this paper are indicated as dots.}
    \label{fig:generalization}
\end{figure*}

The previous sections, in which we evaluated in detail the effects of a series of constellations being considered for launch and operations, have revealed the importance of two main parameters: the total number of satellites and their magnitude $V_\mathrm{550~km}$. In order to characterize this dependence, we use the constellation from Sect.~\ref{Sect:SLOWGWAK}, with a realistic distribution of altitudes and inclinations. This constellation was scaled for a total number of satellites in the $10^2$ to $10^6$ range, and for $V_\mathrm{550~km}$ in the range from $-4$ (corresponding to the brightest objects in this study) to 7 (the IAU recommendation). 
Figure~\ref{fig:generalization} displays the results.

To keep the light pollution below the 1\% limit, the total number of satellites must be below $\sim 1000$ for $V_\mathrm{550~km}=2$, and below {\em a few tens} for $V_\mathrm{550~km}=-4$.
To keep the direct FoV losses below the technical downtime, the total number of satellites with $V_\mathrm{550~km}=7$ must remain below $\sim 10^5$.
Brighter satellites, even marginally, have a dramatic effect on the FoV losses for saturating instruments: just 2000 satellites with $V_\mathrm{550~km}\le 6$ would degrade the performance of the LSST camera by 3\%, and 20\,000 by 30\%. 

The effect of constellations is cummulative, it is therefore absolutely critical to take into account the total number of satellites in orbit.

\section{Discussion and conclusions}
We evaluated the impact of a range of plausible and proposed satellite constellations on astronomical observations. The FORS2 imager on ESO's VLT was used as a representative instrument for traditional cameras on large telescopes, while the LSST camera at the Rubin observatory was used as the archetype of instruments affected by saturation cross-talk, which can damage most of the field of view when satellites brighter than $V_\mathrm{550~km}<7$ cross the detector. Because the number of illuminated satellites depends on the solar elevation $e_\odot$ below the horizon, and therefore on local time and season, we adopted $e_\odot = -20^\circ$ to $-24^\circ$ as representative dark-time conditions, when most satellites are still illuminated. For critical cases, we also explored the full range from $e_\odot = 0^\circ$ (sunset or sunrise) to $e_\odot = -90^\circ$ (nadir), the latter corresponding to winter midnight at the VLT. Reported values were evaluated at zenith and at an elevation of $30^\circ$, thereby covering the range over which most astronomical observations are carried out.
The constellation architectures were taken from publicly available information when possible, and otherwise from standard, realistic assumptions.

We estimated the resulting losses using the analytical method of \citet{BHG22}, extended here to include diffuse light pollution (the contribution from satellites too faint to be detected as trails) and scattered light pollution (including Mie and Rayleigh scattering from all illuminated satellites). The Python code used for these calculations is publicly available on GitHub, and the relevant photometric and radiometric conversions are summarized in Appendix~\ref{Appendix}.

Our results lead to the following main conclusions about bright or very bright satellites, i.e. those with $V_\mathrm{550~km}<7$:
\begin{itemize}
    \item {\bf Bright satellites cause dramatic FoV losses for saturating instruments.}
    Constellations that do not respect the $V_\mathrm{550~km}>7$ limit will significantly degrade the performance of saturating instruments such as the Rubin Observatory LSST camera which suffers from saturation cross-talk, because of satellite brightness, the sheer number of satellites, or both.

    \item {\bf Bright satellites would dramatically affect the appearance of the night sky.}
    A large constellation such as SpaceX's Orbital Data Center, with $V_\mathrm{550~km}=5$, would place {\rm thousands} of satellites above naked-eye visibility threshold, comparable to the number of natural stars typically visible in a dark sky. The AST SpaceMobile constellation would add tens of bright objects ($V\sim 6$ to 2). The Reflect Orbital concept would produce more than 100 Venus-bright satellites in 2030 and more than 1000 in 2035, even assuming that none points its main beam directly at the observer. In  light-polluted areas, one could imagine a sky in which only artificial satellites are visible.

    \item {\bf Very bright satellites affect the sky background.} 
    Large (60 thousand satellites) to very large (one million satellites) constellations that respect the $V_\mathrm{550~km}>7$ limit have a negligible to modest effect on sky-background pollution. By contrast, even a moderate number of bright satellites --- such as 3000 BlueBird-like satellites with $V_\mathrm{550~km}=2$ --- would contribute up to 0.5\% of the dark-sky background as diffuse light, mainly affecting instruments such as spectrographs. The Reflect Orbital case is far more severe: its proposed 2030 constellation of 5000 very bright satellites would add scattered light equivalent to about 20\%--30\% of the natural sky background, corresponding to $V=21.8$ to 21.7~MpSA. The proposed 2035 configuration, with 50\,000 satellites, would raise the sky background by 200\%--300\%, to $V=20.8$ to 20.5~MpSA --- roughly a ``semi-suburban'' sky on the Bortle scale, and therefore {\em unsuitable} for dark-time observations.
\end{itemize}
    
The number of satellites also matters, even when they satisfy the $V_\mathrm{550~km}\ge 7$ recommendation:
\begin{itemize}
    \item A $\sim$60\,000-satellite set is manageable in terms of FoV losses (below the 1\% level).
    \item A mega-constellation of one million or more satellites would fundamentally alter observing conditions, ensuring that most exposures contain multiple trails during a large fraction of the night. The field-of-view losses then rise to 10\%--20\%, making satellites the dominant source of data loss, ahead of weather and technical downtime.
\end{itemize}

Using the \citet{BHG22} constellation set, we generalized these results as a function of both satellite brightness and total satellite number. The effect is cumulative, so the total population in orbit is the key parameter. For satellites that satisfy the $V_\mathrm{550~km}>7$ limit, the total number must remain below about 100\,000 if FoV losses are to stay below the level of technical downtime. For even slightly brighter satellites, the situation deteriorates rapidly for saturating instruments: only 2000 satellites with $V_\mathrm{550~km}\le 6$ would degrade LSST performance by 3\%, and 20\,000 would cause losses of 30\%. For satellites with $V_\mathrm{550~km}\le 5$, the tolerable population drops to only a few hundred if FoV losses are to remain below technical downtime.

In conclusion, the collection of constellations currently proposed for launch and operation, with over 1\,700\,000 satellites, many of them brighter or much brighter than $V_\mathrm{550~km}=7$, would have a devastating effect on astronomical observations.

This study addressed the direct effects of satellite trails crossing the field of view, and the indirect effects of diffuse and scattered light pollution, for optical astronomy. Satellite constellations also affect radio, millimetre, and sub-millimetre observations. Beyond astronomy, they raise broader concerns related to orbital crowding, space debris, and atmospheric pollution from launches and re-entries. These issues are important, but they lie beyond the scope of this paper.


\begin{acknowledgements}

This paper has been reviewed by two anonymous referees from the International Astronomical Union (IAU) Centre for the Protection of the Dark and Quiet Sky (CPS\footnote{\url{https://cps.iau.org}}). The author is grateful for their constructive comments. The author is also grateful to B.~M. Kioko (ESO), 
J.~C. Barentine (Dark Sky Consulting),
and C.~Walker (IAU CPS) for their helpful discussions and comments on the manuscript. 
Our SatConAnalytic python package relies on  
    NumPy\footnote{\url{https://numpy.org/}} \citep{numpy}, 
    Matplotlib\footnote{\url{https://matplotlib.org/}} \citep{Matplotlib:2007}, and
    Astropy\footnote{\url{https://www.astropy.org/}} \citep{astropy:2013, astropy:2018, astropy:2022}, a community-developed core Python package for Astronomy.
This paper used data from the 2008 Whole Heliosphere Interval (WHI) Solar Irradiance Reference Spectra (SIRS) described in Woods et al. 2009 (https://doi.org/10.1029/2008GL036373). These data were accessed via the LASP Interactive Solar Irradiance Datacenter (LISIRD) (https://lasp.colorado.edu/lisird/).   
\end{acknowledgements}

\bibliographystyle{aa} 
\bibliography{sky} 

\FloatBarrier
\appendix
\setcounter{section}{1}
\nolinenumbers

\begin{figure*}[ht!]
    \section*{Appendix A: Figure set}
    \centering
    {\bf a}\includegraphics[width=.48\linewidth]{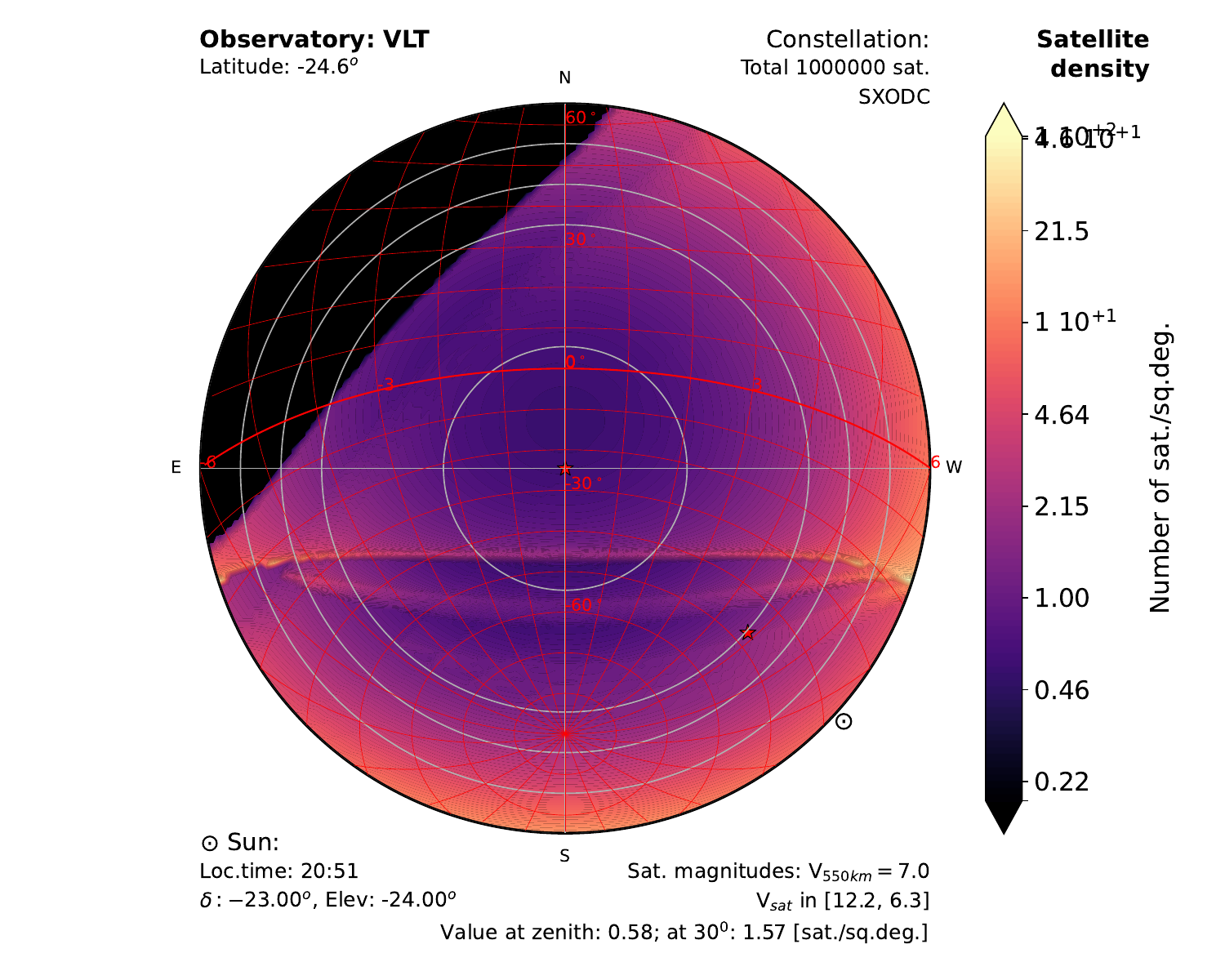}
    {\bf b}\includegraphics[width=.485\linewidth]{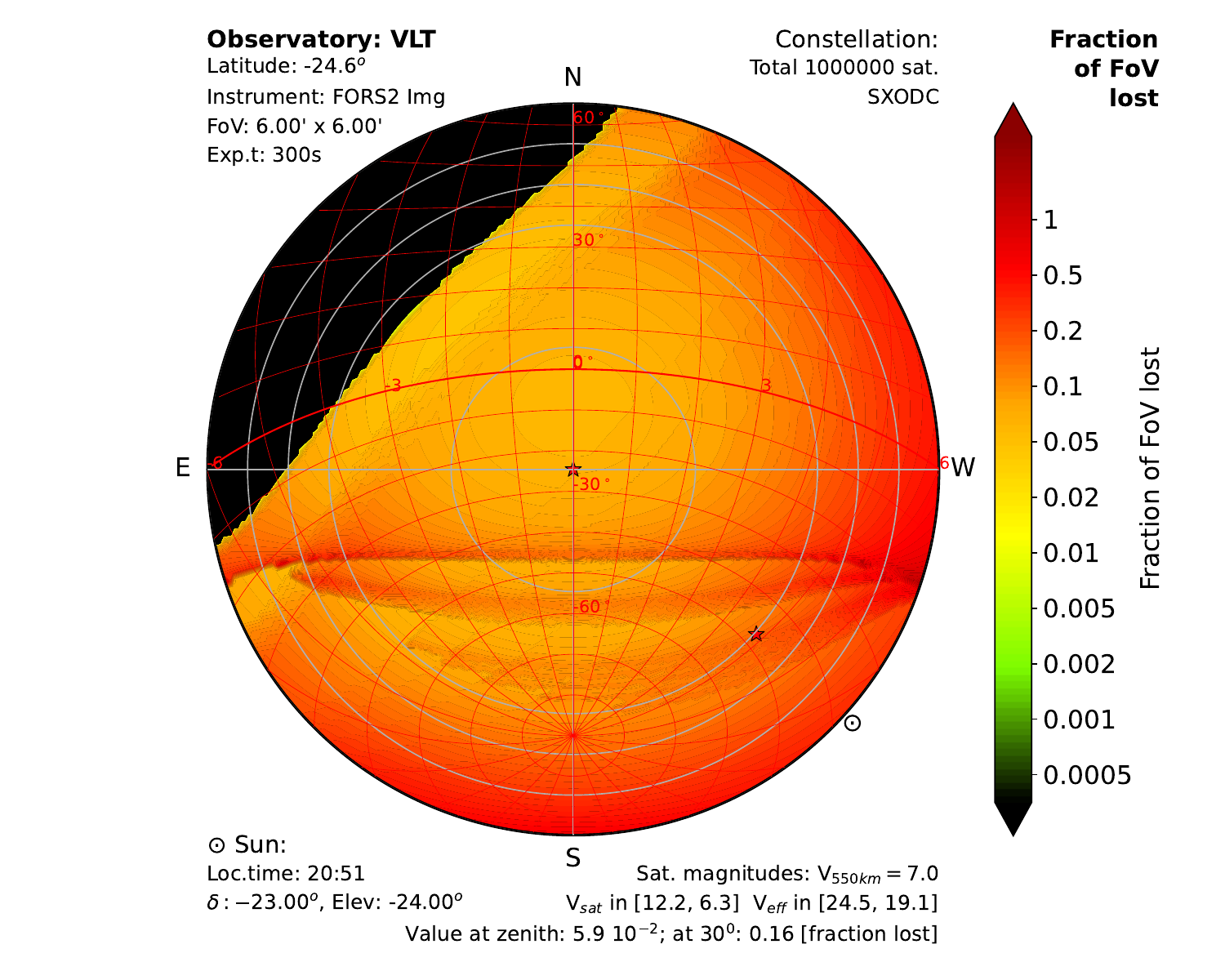}
    {\bf c}\includegraphics[width=.485\linewidth]{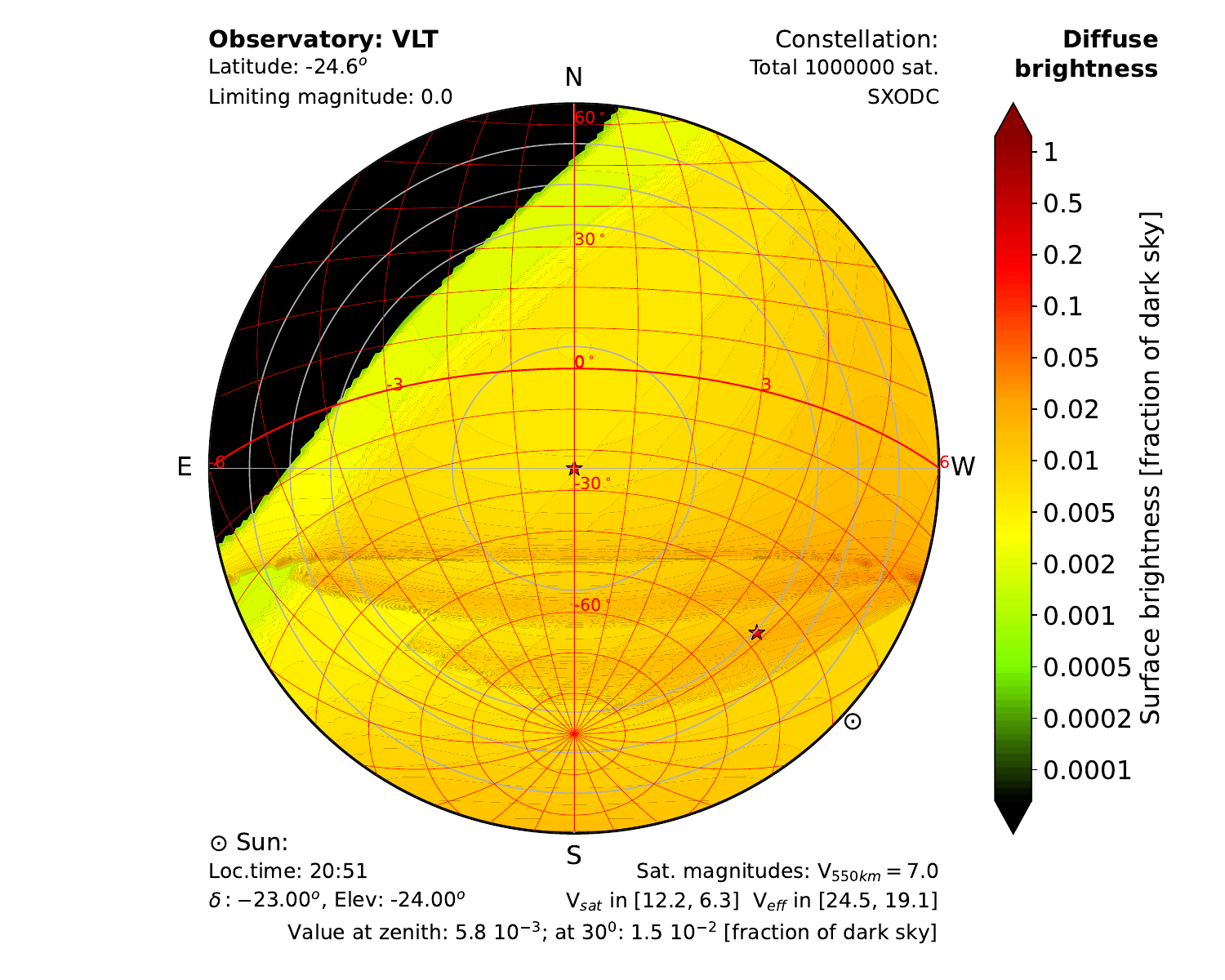}
    {\bf d}\includegraphics[width=.485\linewidth]{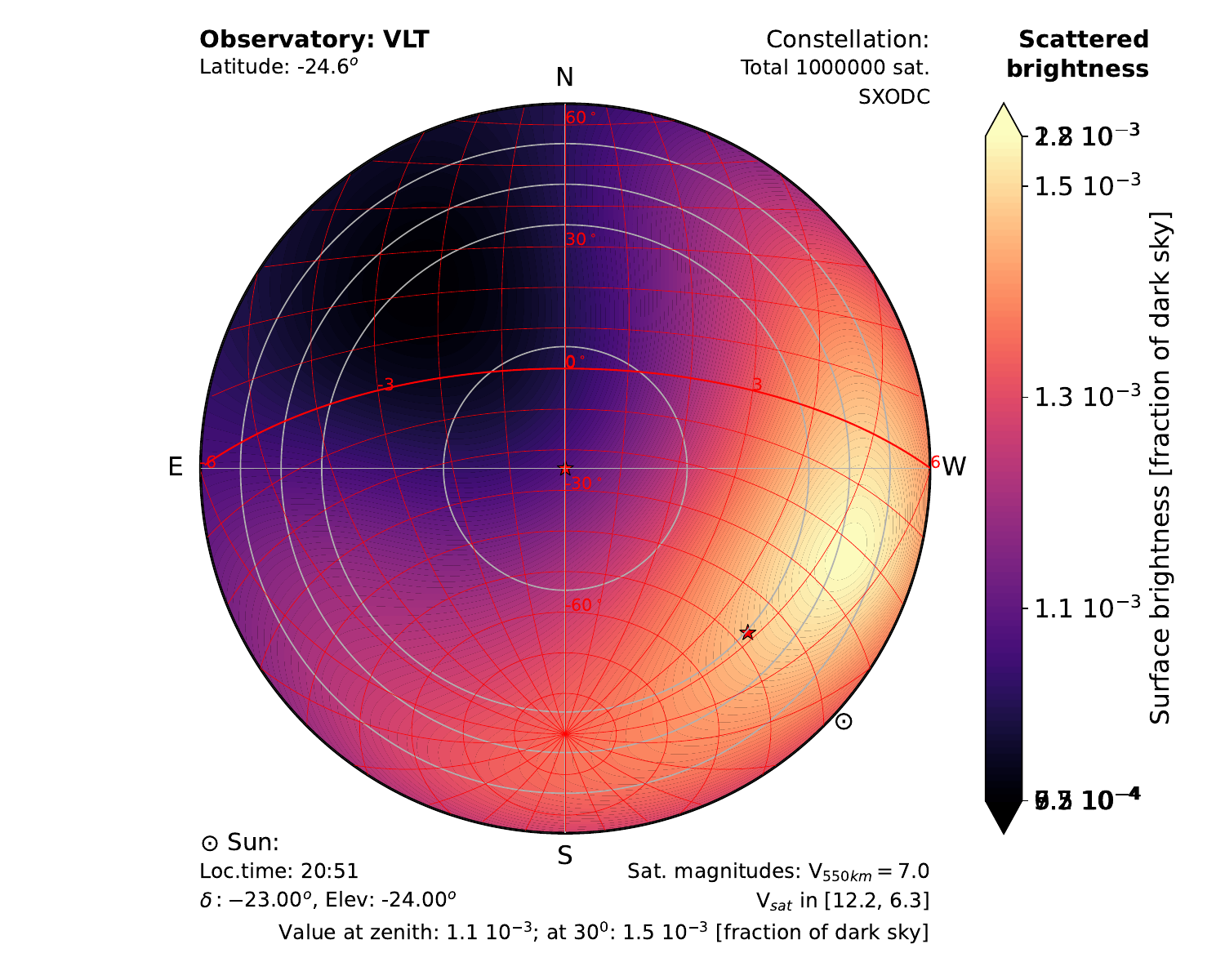}
    \caption{For one million satellites ($V_\mathrm{500~km}=7$), representing the SpaceX Orbital Data Center constellation:
    {\bf a} average satellite density [sat/sq.~deg.];
    {\bf b} fraction of the field of view lost to satellite trails in 300-second FORS2 exposures;
    {\bf c} diffuse sky brightness, as a fraction of the natural dark sky with $V=22$~MpSA;
    {\bf d} scattered sky brightness, as a fraction of the natural dark sky;
    See Fig.~\ref{fig:1Msat} for a general description of the sky maps.
    }
    \label{fig:ap.SXODC}
\end{figure*}

\begin{figure*}
    \centering
    {\bf a}\includegraphics[width=.485\linewidth]{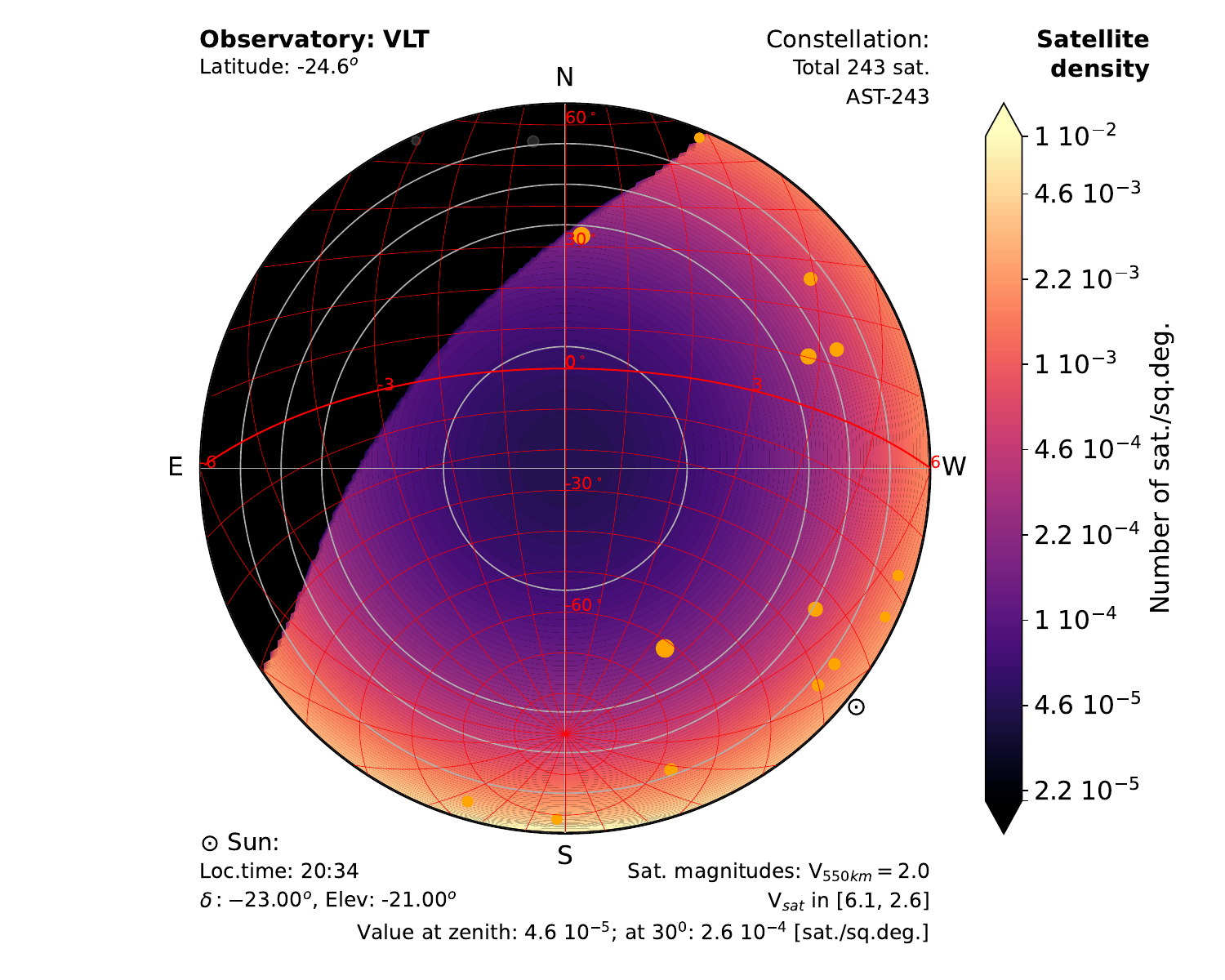}
    {\bf b}\includegraphics[width=.485\linewidth]{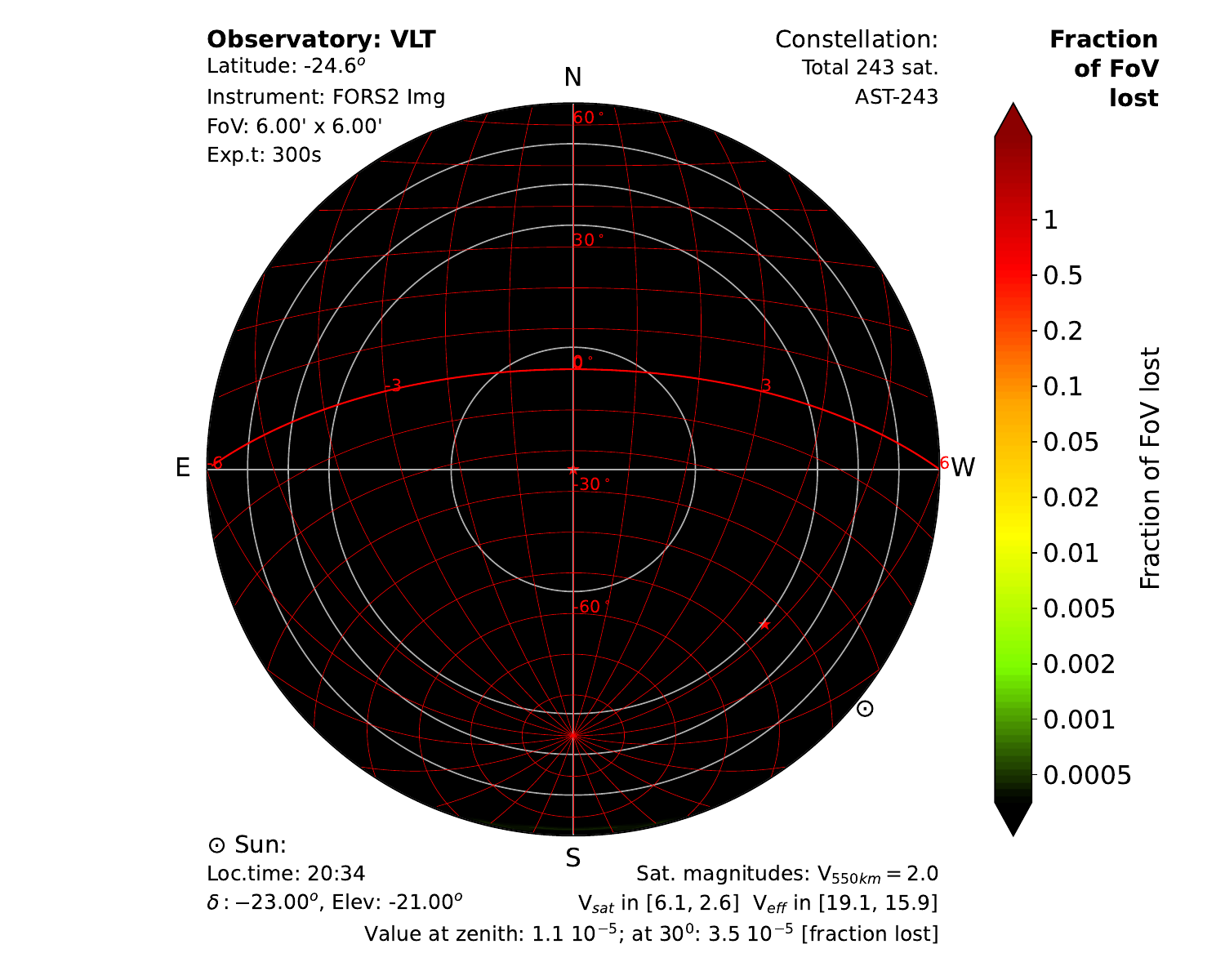}
    {\bf c}\includegraphics[width=.485\linewidth]{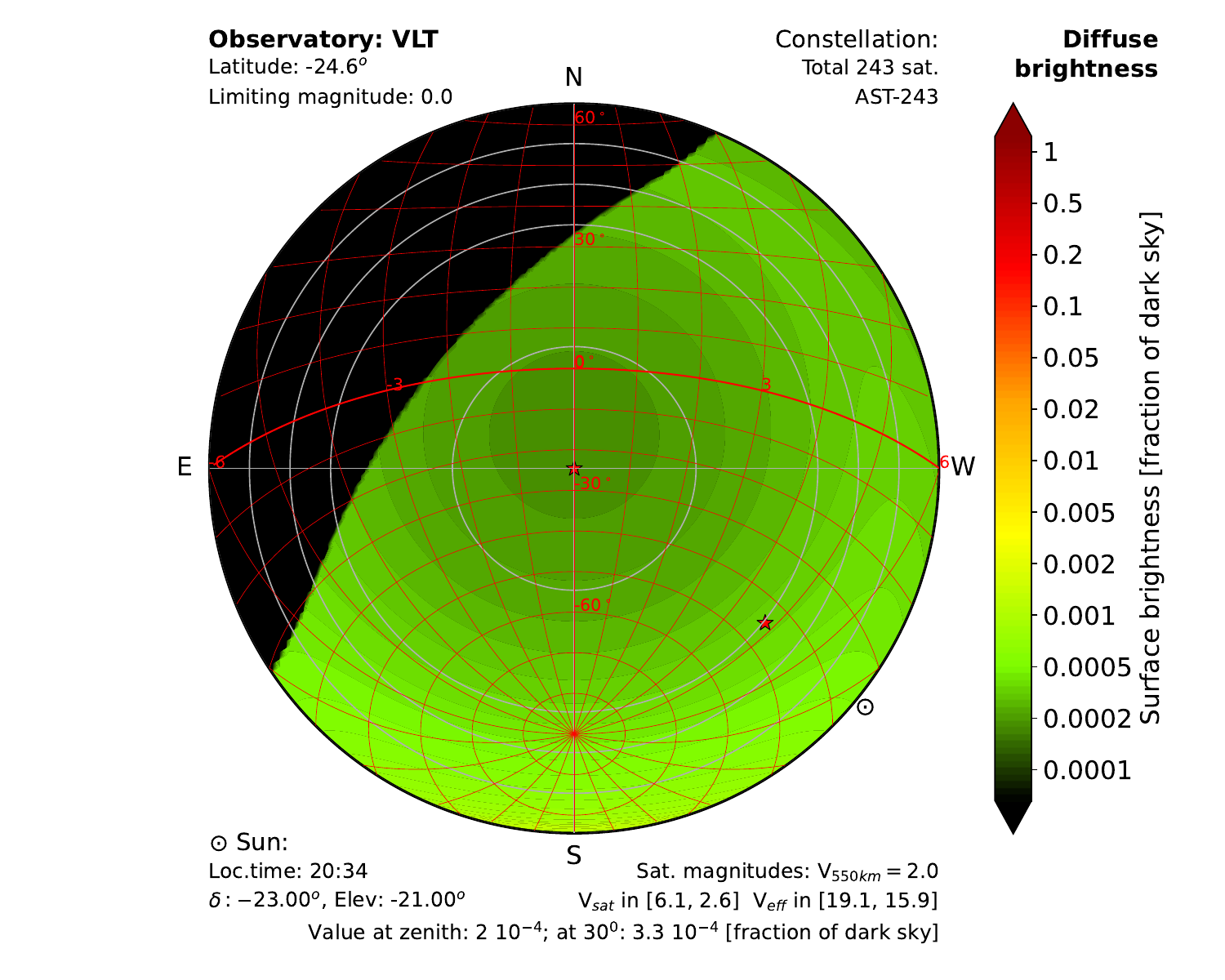}
    {\bf d}\includegraphics[width=.485\linewidth]{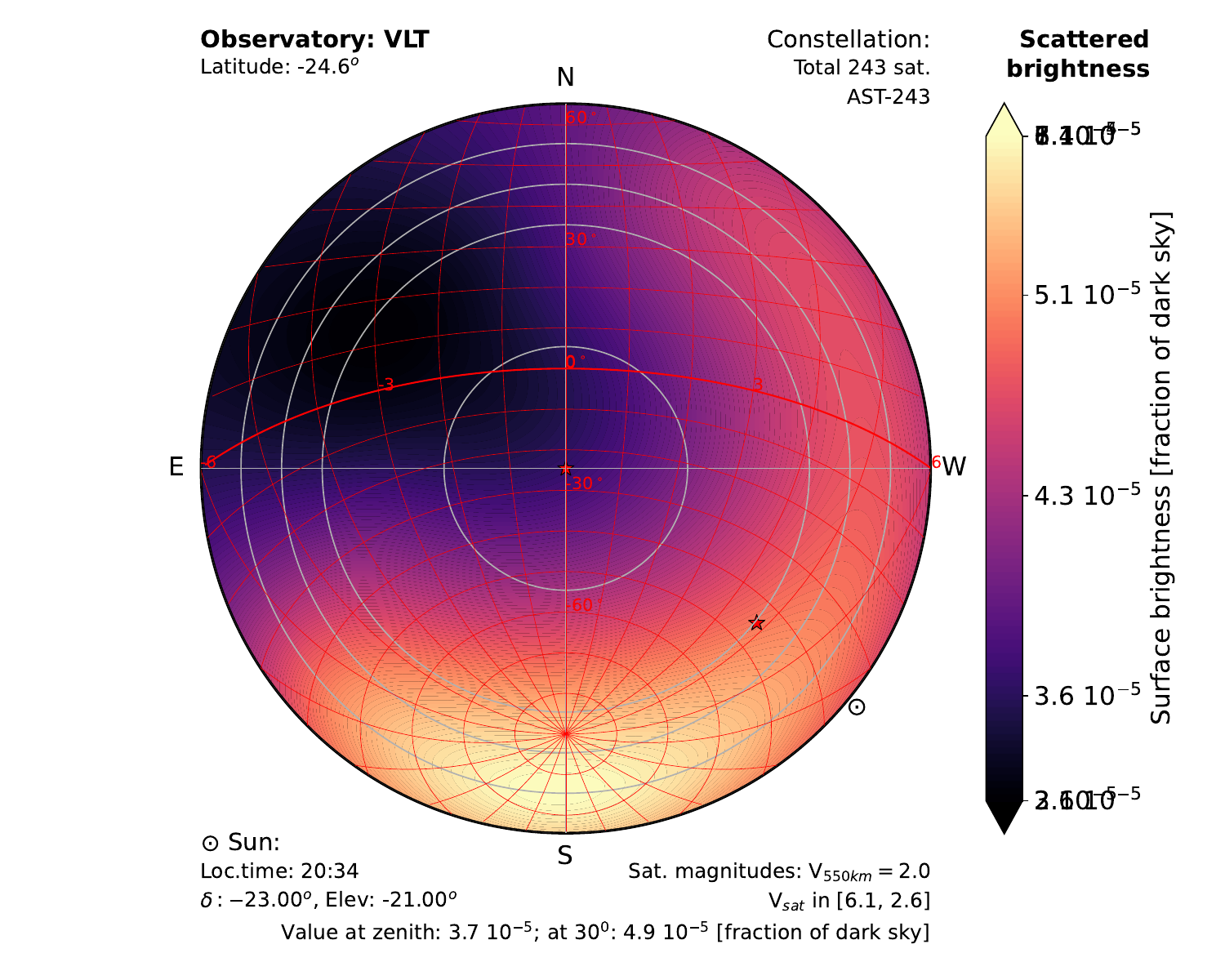}
    \caption{For 243 bright satellites ($V_\mathrm{500~km}=2$), representing BlueBird-like spacecraft in a constellation such as AST SpaceMobile:
    {\bf a} average satellite density [sat/sq.~deg.];
    {\bf b} fraction of the field of view lost to satellite trails in 15-second LSST exposures;
    {\bf c} diffuse sky brightness, as a fraction of the natural dark sky with $V=22$~MpSA;
    {\bf d} scattered sky brightness, as a fraction of the natural dark sky.
    For more details, see Fig.~\ref{fig:SLOWGWAK}.
    }
    \label{fig:ap.AST-243}
\end{figure*}

\begin{figure*}
    \centering
    {\bf a}\includegraphics[width=.485\linewidth]{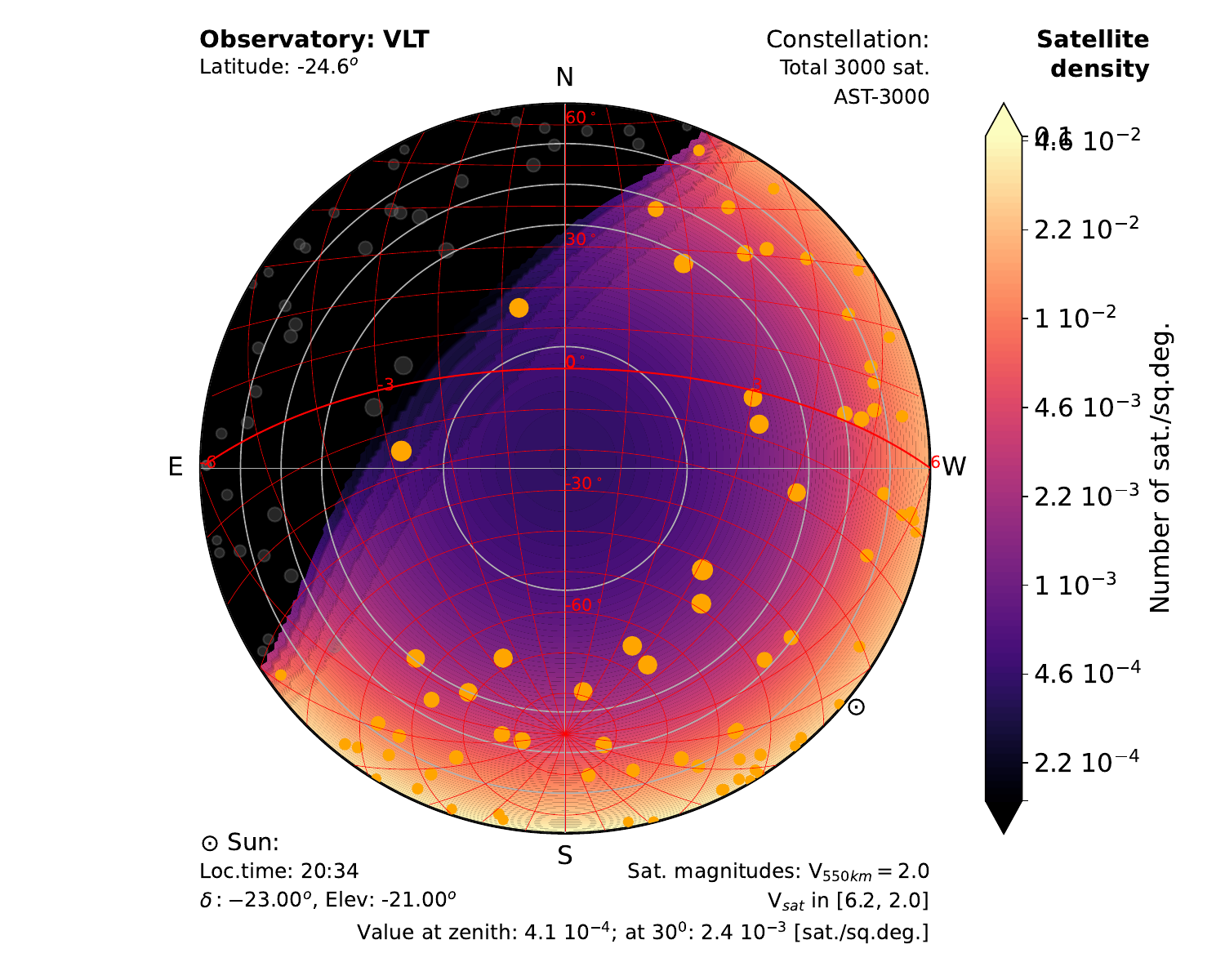}
    {\bf b}\includegraphics[width=.485\linewidth]{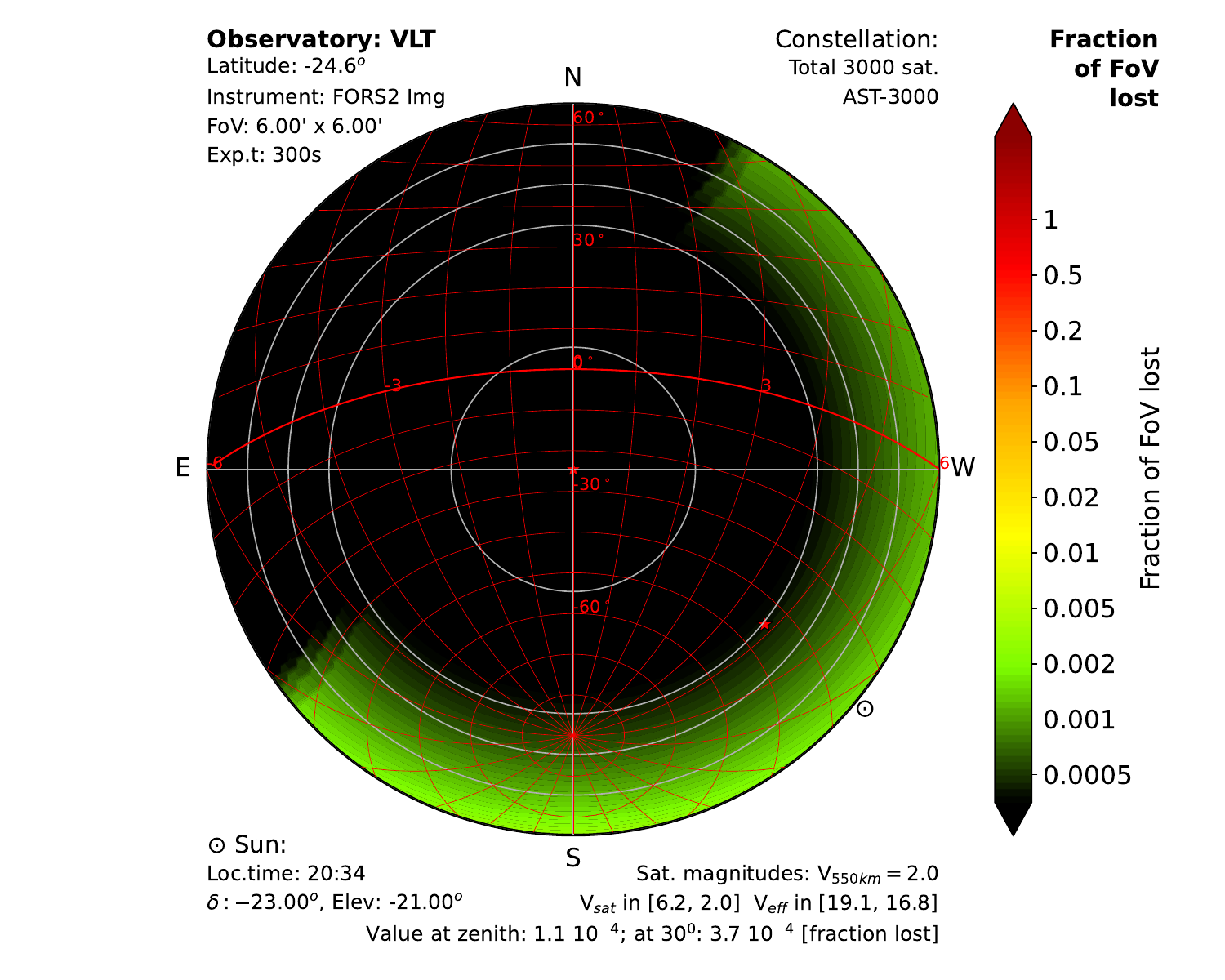}
    {\bf c}\includegraphics[width=.485\linewidth]{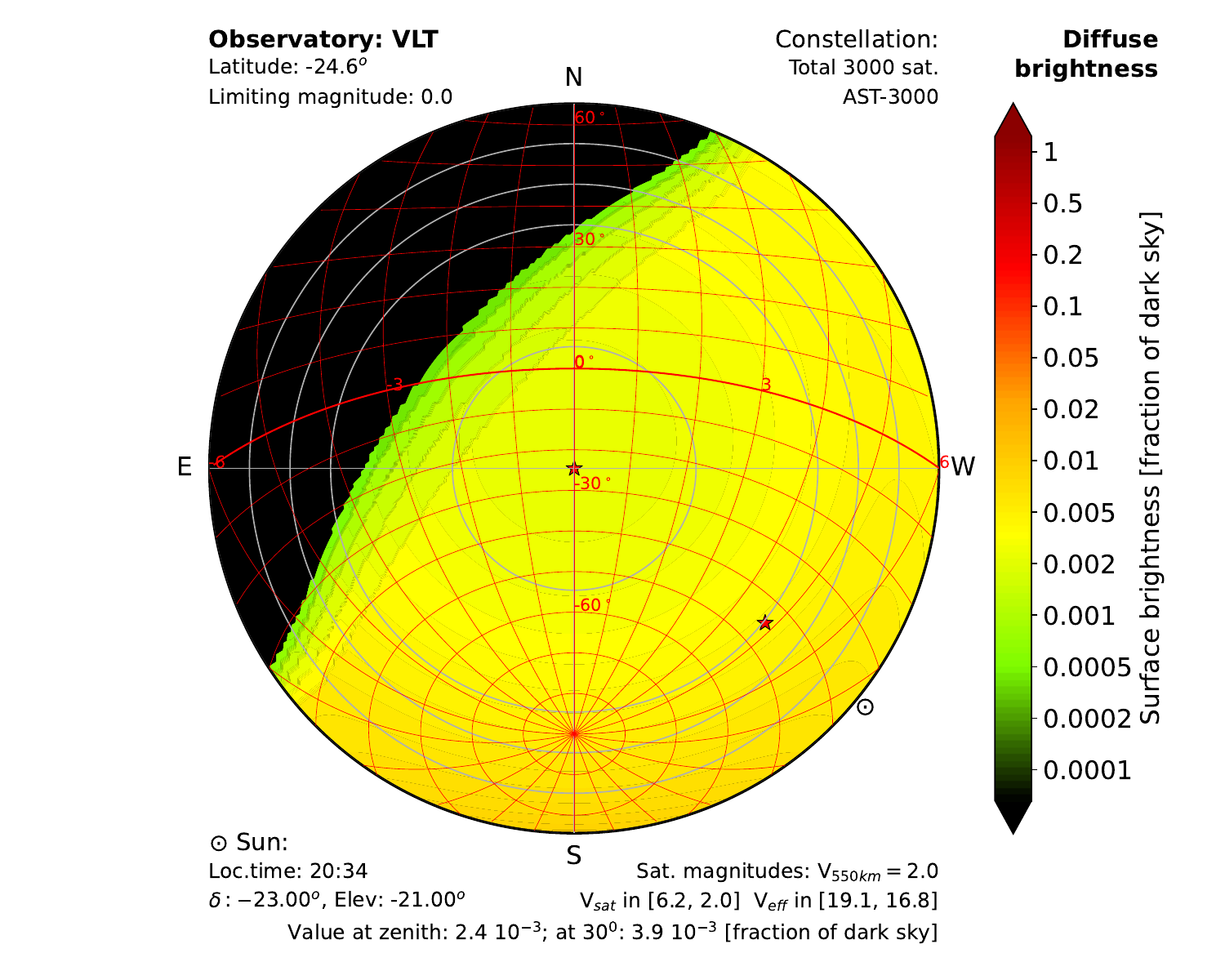}
    {\bf d}\includegraphics[width=.485\linewidth]{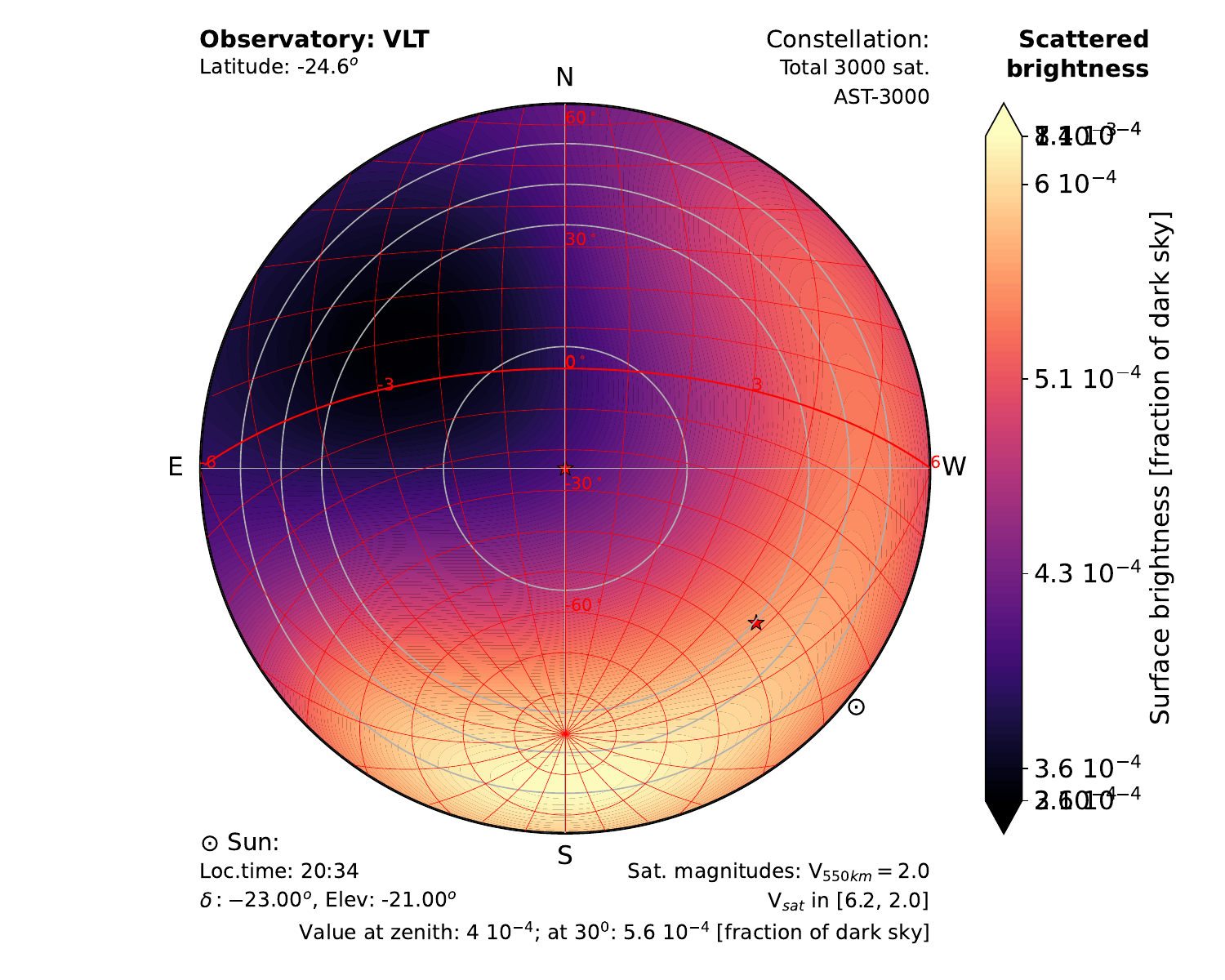}
    \caption{For 3000 bright satellites ($V_\mathrm{500~km}=2$), representing BlueBird-like spacecraft in a direct-to-cell constellation (code D3000 in the table):
    {\bf a} average satellite density [sat/sq.~deg.];
    {\bf b} fraction of the field of view lost to satellite trails in 15-second LSST exposures;
    {\bf c} diffuse sky brightness, as a fraction of the natural dark sky with $V=22$~MpSA;
    {\bf d} scattered sky brightness, as a fraction of the natural dark sky.
    For more details, see Fig.~\ref{fig:SLOWGWAK}.
    }
    \label{fig:ap.AST-3000}
\end{figure*}

\begin{figure*}
    \centering
    {\bf a}\includegraphics[width=.485\linewidth]{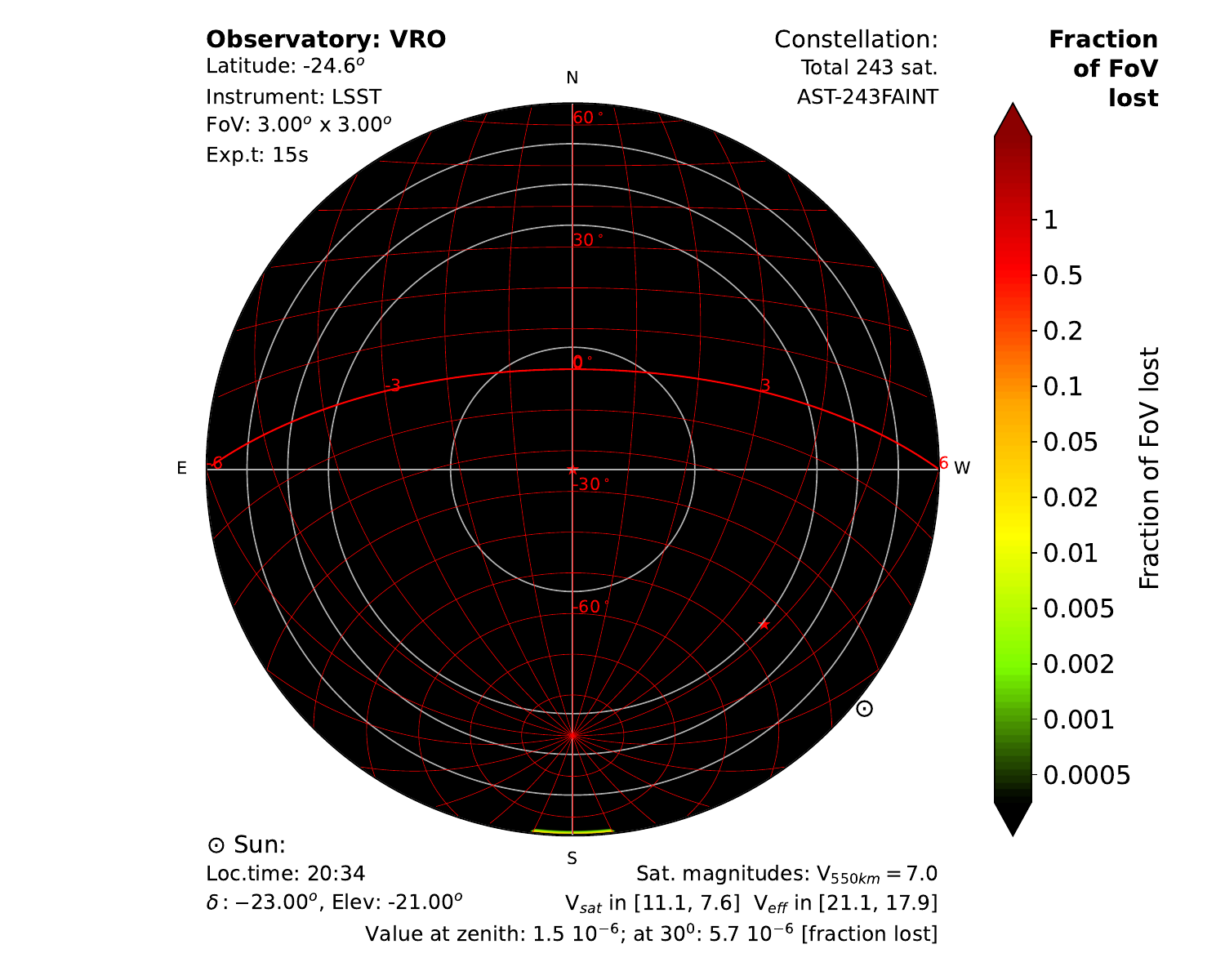}
    {\bf b}\includegraphics[width=.485\linewidth]{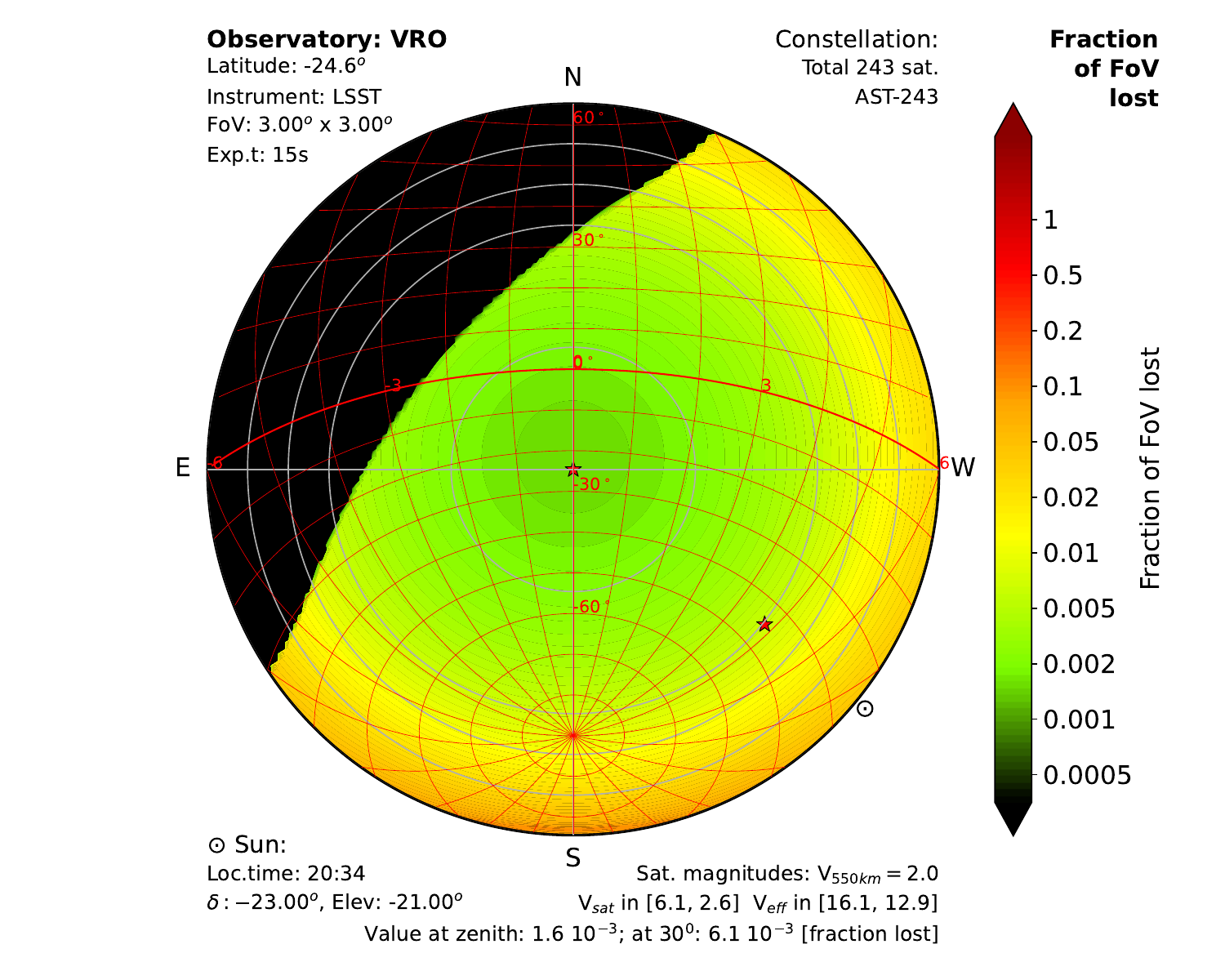}
    {\bf c}\includegraphics[width=.485\linewidth]{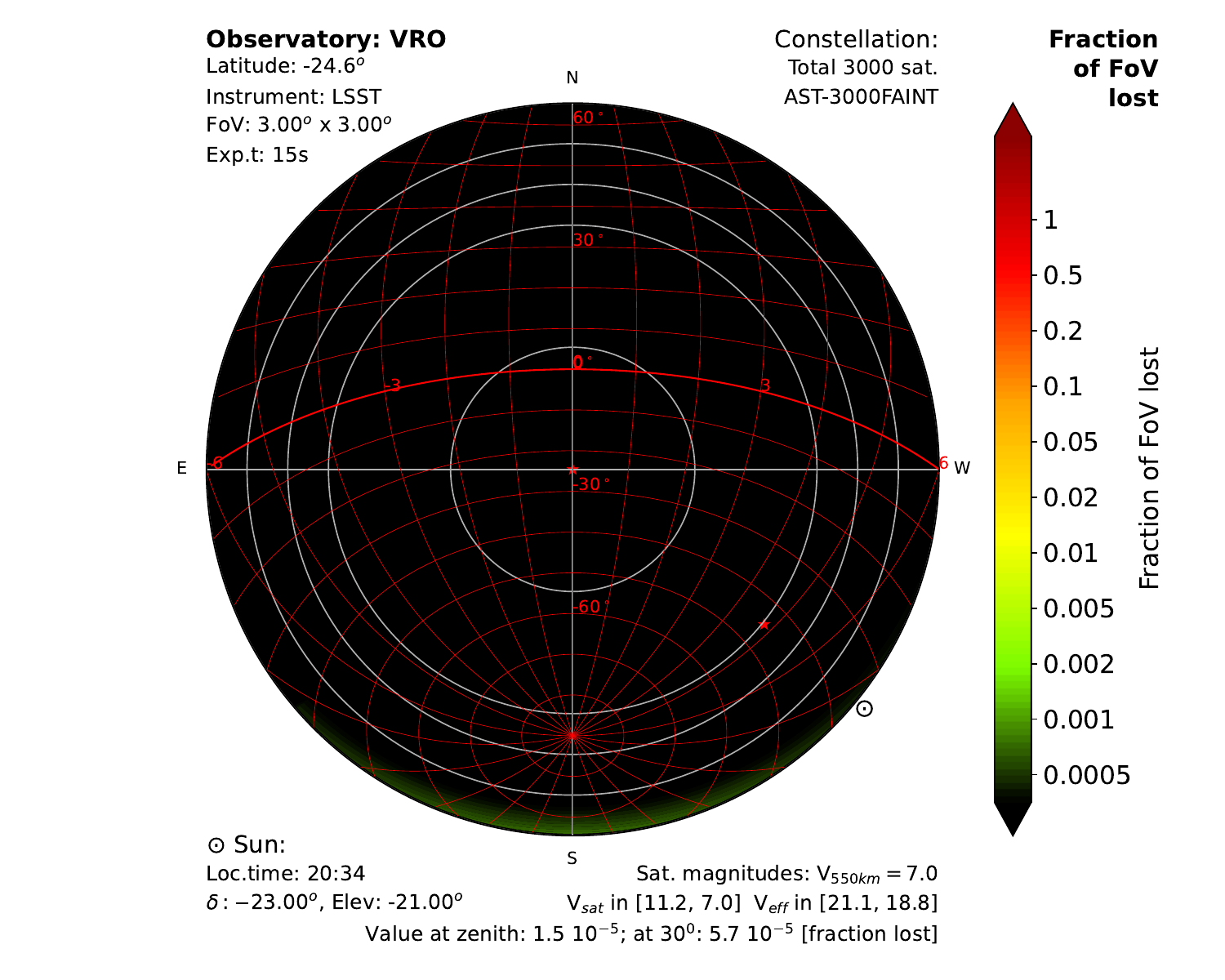}
    {\bf d}\includegraphics[width=.485\linewidth]{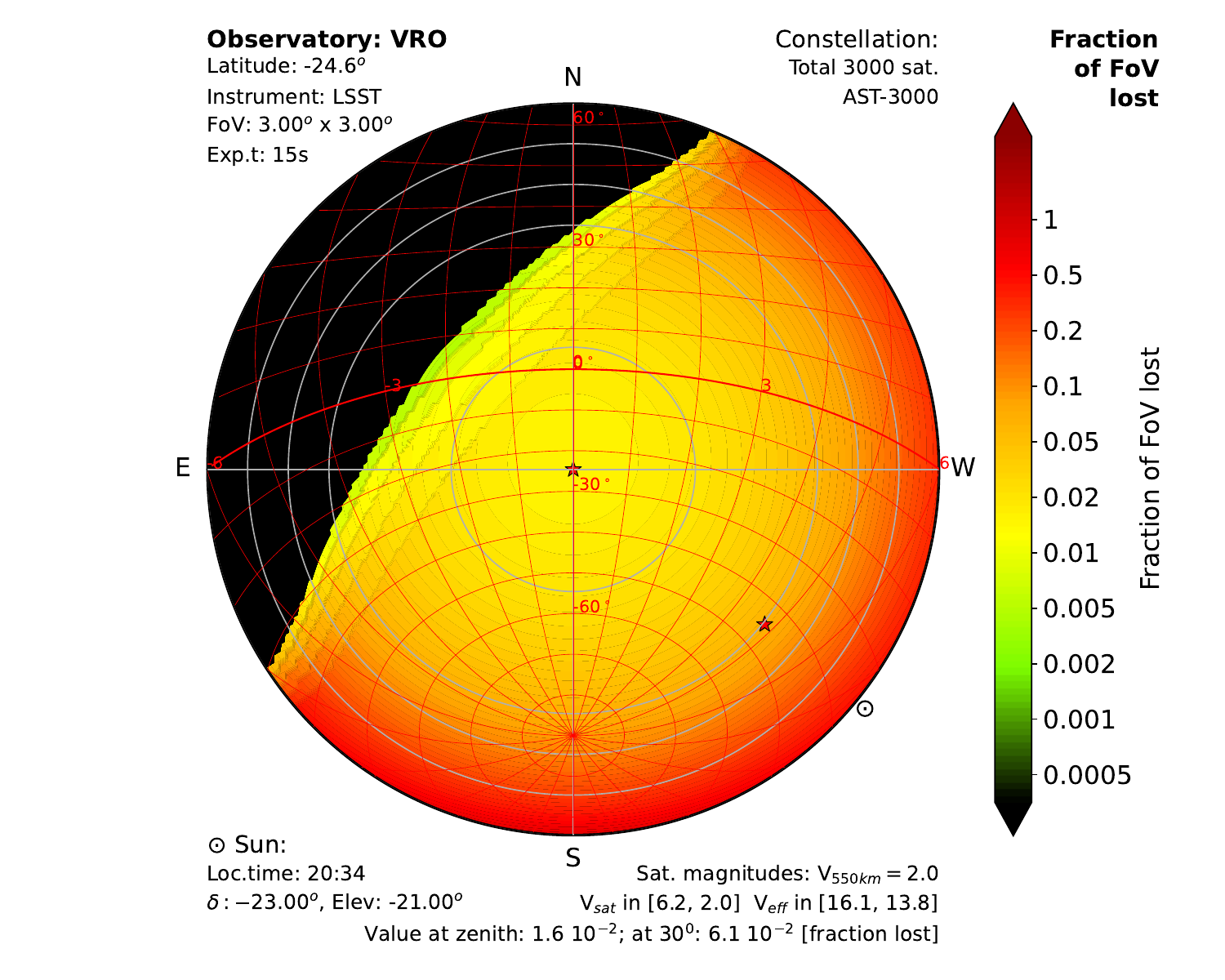}
    \caption{FoV losses for a saturating camera such as LSST, for constellations of 243 satellites (a, b, code AST in the table) and 3000 satellites (c, d, code D3000 in the table) representing BlueBird-like spacecrafts. In panels a and c, the satellites respect the $V<7$ limit; in panels b and d, they have the measured BlueBird brightness, $V=2$, that is, 100$\times$ brighter than the limit.
    For more details on these sky maps, see Fig.~\ref{fig:SLOWGWAK}.
    }
    \label{fig:ap.AST-faint-vs-bright}
\end{figure*}

\FloatBarrier


\section{Sky brightness unit conversion\label{Appendix}}
\begin{figure*}[t]
   \centering
   \includegraphics[width=\hsize]{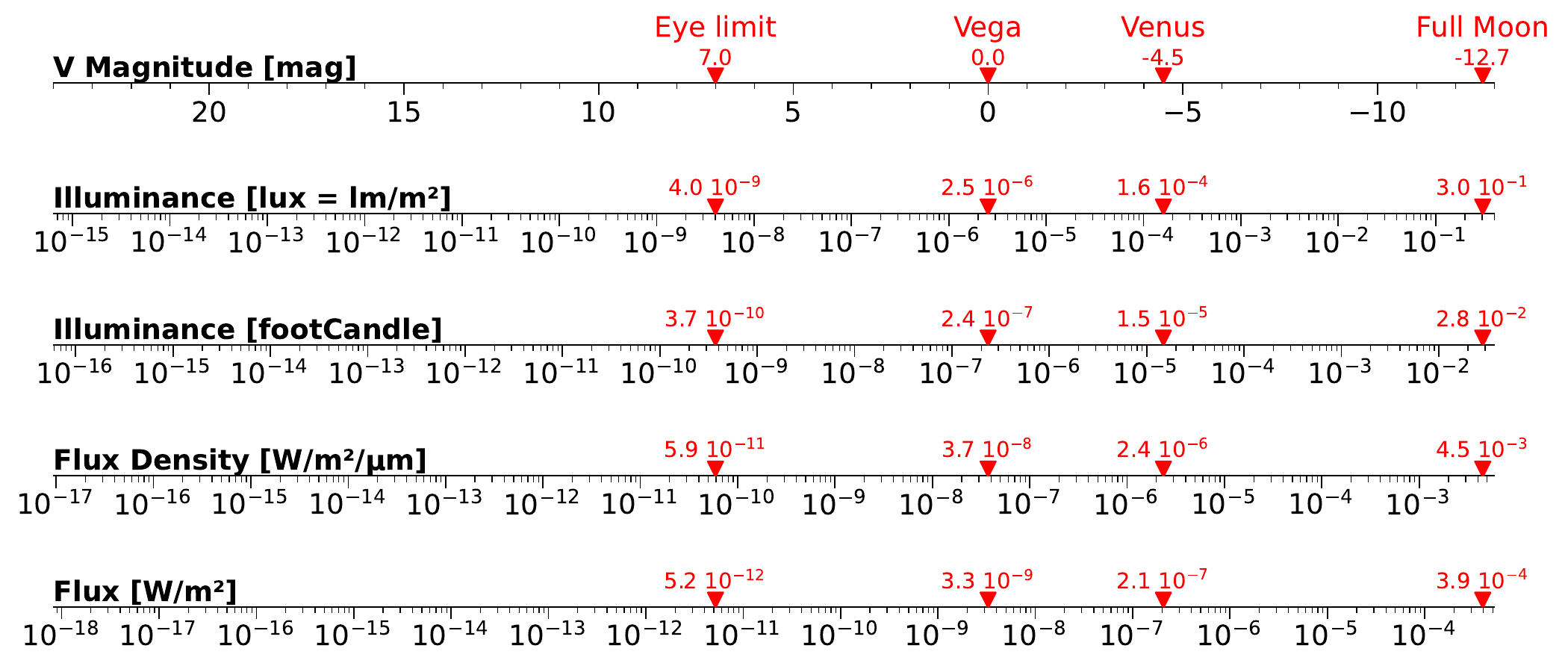}
   \caption{Conversion between magnitude, irradiance and illuminance in the $V$ filter for solar spectra.}
   \label{fig:abacusMag}
\end{figure*}

\begin{figure*}
   \centering
   \includegraphics[width=\hsize]{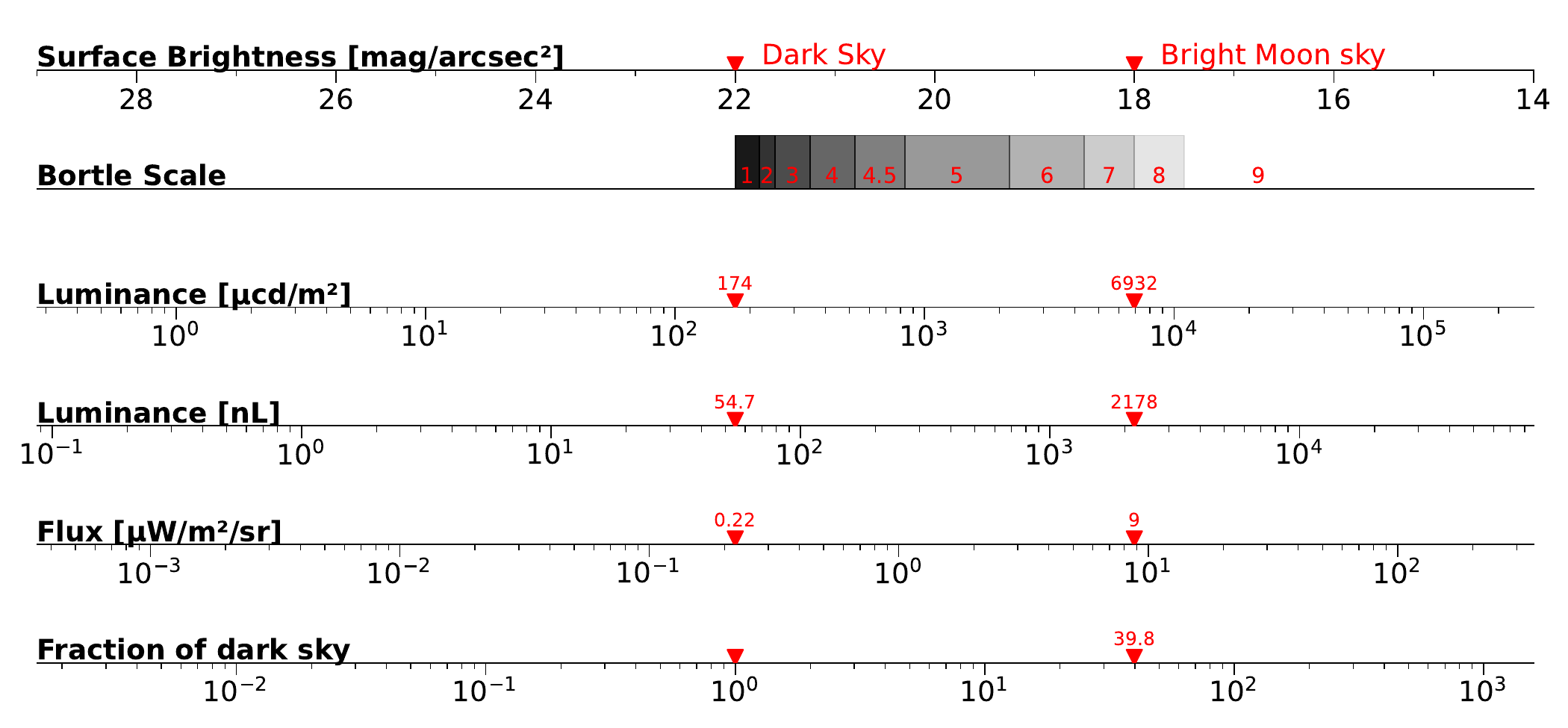}
   \caption{Conversion between surface brightness, radiance and luminance in the $V$ filter for solar spectra.}
   \label{fig:abacusSurfB}
\end{figure*}

This appendix summarizes the brightness quantities, units, and conversions used in this paper. The conversions are presented as nomograms in Fig.~\ref{fig:abacusMag} and \ref{fig:abacusSurfB}.

\subsection{Radiometric and photometric brightnesses}
Brightness can be expressed as power (in SI units, Watt), possibly per unit surface, per unit solid angle, and over the full spectrum or a restricted band. Such quantities are {\em radiometric}.

When the same quantities are weighted by the response curve of the human eye in photopic vision, as defined by the  International Commission on Illumination (CIE), they describe perceived brightness. The SI units for these quantities are called  (confusingly for an astronomer) {\em photometric}. In what follows, we will systematically specify whether “photometric” refers to the perception measurement or the branch of astronomy measuring the brightness of celestial objects.

Astronomical bandpasses are defined by standard photometric systems, such as Johnson--Kron--Cousins {\em U, B, V, R, I} and Sloan {\em u', g', r', i', z'}. The Johnson--Kron--Cousins {\em V} band is, by construction, a good match to the photopic eye response: it peaks near $\lambda \sim 555$~nm and extends roughly from 400 to 700~nm.

Radiometric measurements in $V$ can therefore be converted into perceptual photometric quantities. Vision peaks at a shorter wavelength ($\sim 500$~nm), but for this paper we neglect the differences between the various definitions of the $V$ filter and the exact CIE response curve.

\subsection{Illuminance, irradiance and magnitude}

The total flux of a celestial source through a $V$ filter, incident on a surface per unit area, is the {\em irradiance}, measured in \Wmm. It is obtained by integrating the spectral flux density $f(\lambda)$ (in \Wmm~\mum\m) over the filter response curve $V(\lambda)$:
\begin{equation}
    F_V = \int_0^\infty f(\lambda) ~ V(\lambda) d\lambda.         \label{Eq:irradiance}
\end{equation}
Alternatively, the flux density can be expressed per unit frequency; a customary non-SI unit is the Jansky, with $1~\mathrm{Jy} = 10^{-26}$~W~m\mm~Hz\m.

The {\em magnitude} of an object is the logarithmic counterpart of irradiance. It is defined as
\begin{equation}
    m_V = -2.5 \times \log_{10}\left( \frac{
    \int_0^\infty V(\lambda) ~ f(\lambda)
    }{
    \int_0^\infty V(\lambda) ~ f_o(\lambda)
    }                                                            \label{Eq:Mag}
 \right), 
\end{equation} 
where $f_0$ is a reference flux corresponding to $m=0$. The factor $-2.5\log_{10}$ preserves the historical magnitude scale going back to \cite{Hipparchus}. We work in the Vega system, in which Vega has $m=0$ in every filter \citep{JohnsonMorgan53}. Other systems exist, for example the AB system, where $f_0$ is constant and equal to 3631~Jy. In the $V$ band the AB--Vega offset is only $\sim 0.02$~mag \citep{Blanton+07}, negligible for our purpose.

Formally, Eqs.~\ref{Eq:irradiance} and~\ref{Eq:Mag} require the response curve $V(\lambda)$ and the source spectrum $f(\lambda)$. As we work with reflected solar light, we use $V_\odot = -26.75$ \citep{allen2000} and integrate the 2008 Whole Heliosphere Interval (WHI) Solar Irradiance Reference Spectra\footnote{\url{https://lasp.colorado.edu/lisird/data/whi_ref_spectra}} 
\citep{Woods09} over a generic Bessel $V$ transmission curve. This gives $F_{V,\odot} = 164$~\Wmm and therefore
\begin{equation}
    F_V = 164 \times 10^{-0.4 (m_V + 26.75 )}.     \label{Eq:mag2flux}
\end{equation}

The perceptual photometric counterpart of irradiance is the {\em illuminance}, whose SI unit is the lux (lx), equivalent to lumen per square metre (lm~m\mm). For monochromatic light at $\lambda = 555$~nm, where the eye is most sensitive, $1~\mathrm{lx} = 1.464\times10^{-3}$~W~m\mm by SI definition. A traditional non-SI unit is the foot-candle (fc), defined as the illuminance on the inside of a one-foot sphere produced by a one-candela source at its centre. Hence
\begin{equation}
    1~{\mathrm fc} \simeq 10.764~{\mathrm lx}.           \label{Eq:1fc}
\end{equation}

\cite{KS91} relate the magnitude of the Moon $m_V$ to its illuminance $I$, expressed in foot-candles, as
\begin{equation}
    I = 10^{-0.4 (m_V +16.57)}.                \label{Eq:Illuminance_fc}
\end{equation}

\subsection{Luminance and surface brightness}

For an extended source --- for example a nearby galaxy or the sky background --- brightness per unit solid angle is often more useful than total brightness.

The irradiance per unit solid angle is the {\em radiance}, in W~m\mm~sr\m. Astronomers, however, often use square arcseconds, with the conversion
\begin{equation}
    (1^{\prime\prime})^2 = \frac{\pi}{180 \times 3600} {\mathrm [sr]},  \label{Eq:arcsec2}
\end{equation}
and therefore often express radiance in W~m\mm~arcsec\mm, and spectral radiance in W~m\mm~arcsec\mm~\mum\m.

The logarithmic equivalent for extended sources is the {\em surface brightness}, expressed in magnitudes per square arcsecond (MpSA). Its definition is the same as Eq.~\ref{Eq:Mag}, replacing the spectral flux density by the spectral radiance $I(\lambda)$:
\begin{equation}
    M_V = -2.5 \times \log_{10}\left( \frac{
    \int_0^\infty V(\lambda) ~ I(\lambda)
    }{
    \int_0^\infty V(\lambda) ~ f_o(\lambda)
    }                                                            \label{Eq:surfBrg}
 \right). 
\end{equation} 

Using the same $V$-magnitude to flux relation as for irradiance (Eq.~\ref{Eq:mag2flux}), together with the square-arcsecond to steradian conversion (Eq.~\ref{Eq:arcsec2}), the relation between surface brightness $M_V$ and the radiance through the $V$ filter $I_V$, in \Wmm~sr\m, is
\begin{eqnarray}
         I_V  &=& \left(\frac{180.0 \times 3600.0}{\pi} \right) ^ 2 \times 164.346 \times 10^{-0.4  (M_V + 26.75)}  \\
           & = & 6.992\times 10^{12}   \times 10^{-0.4  * (M_V + 26.75)}. \label{Irrad}\\
\end{eqnarray}

Another commonly used non-SI unit is S$_{10}$, defined as the surface brightness corresponding to a 10th-magnitude star spread over one square degree. The conversion is
\begin{equation}
    1 ~\mathrm{S}_{10} = 27.78 \mathrm{MpSA}.
\end{equation}
The perceptual photometric counterpart is the {\em luminance}, expressed in candela per square metre (cd~m\mm), which we use here.
A common non-SI unit is the Lambert (L), defined as
\begin{equation}
    1 ~{\mathrm L} = \frac{10^4}{\pi} ~\mathrm{cd~m}^{-2} ~\simeq 3183.1 ~\mathrm{cd~m}^{-2}. \label{Eq:lambert}
\end{equation}

The conversion between radiometric and perceptual photometric units again depends on the source spectrum. For reflected sunlight, we use the calculations of \citet{Bara2019} for a $T=5500$~K blackbody in the $V$ filter (their Table~1 and Eq.~10):
\begin{equation}
    L_V = 10.987 \times 10^4 \times 10^{-0.4 M_V},      \label{Eq:Lv}
\end{equation}
with $M_V$ in MpSA and $L_V$ in cd~m\mm. 

Finally, \citet{Bortle2001} introduced a nine-class scale of sky darkness, from Class~1 for the darkest skies to Class~9 for inner-city skies. These classes have since been quantified in MpSA.

\end{document}